\documentclass[a4paper,11pt]{article}
\pdfoutput=1

\usepackage{jcappub}
\usepackage{tikz,xcolor,hyperref}
\usepackage{amsmath, amssymb, amsthm, graphicx, epsfig, fancyhdr,epsfig, slashed}
\usepackage[normalem]{ulem}
\usepackage{tikzsymbols}
\usepackage{natbib}
\usepackage{float}
\usepackage{makecell} 

\newcommand{\Trh}{T_\text{RH}}
\newcommand{\Tmax}{T_\text{max}}
\newcommand{\arh}{a_\text{RH}}

\newcommand{\aend}{a_\text{end}}
\newcommand{\br}{\text{Br}}

\newcommand{\rend}{\rho_\text{end}}
\newcommand{\rRH}{\rho_\text{RH}}

\def\be{\begin{equation}}
\def\ee{\end{equation}}
\def\beq{\begin{equation}\begin{aligned}}
\def\eeq{\end{aligned}\end{equation}}

\def\tev{\, {\rm TeV}}
\def\gev{\, {\rm GeV}}
\def\mev{\, {\rm MeV}}

\begin{document}
%%%%%%%%%%%%%%%%%%%%
\title{Gravity as a Portal to Reheating, Leptogenesis and Dark Matter}
%%%%%%%%%%%%%%%%%%%%%%%%%%%%%%%
\author[a,b]{Basabendu Barman,}
\author[c]{Simon Cléry,}
\author[d]{Raymond T. Co,}
\author[c]{Yann Mambrini,}
\author[d]{Keith A. Olive}
\affiliation[a]{\,Centro de Investigaciones, Universidad Antonio Nari\~{n}o\\Carrera 3 este \# 47A-15, Bogot{\'a}, Colombia}
\affiliation[b]{\,\,Institute of Theoretical Physics, Faculty of Physics, University of Warsaw,\\ ul. Pasteura 5, 02-093 Warsaw, Poland}
\affiliation[c]{\,Universit\'e Paris-Saclay, CNRS/IN2P3, IJCLab, 91405 Orsay, France}
\affiliation[d]{
 William I.~Fine Theoretical Physics Institute, 
       School of Physics and Astronomy,
            University of Minnesota, Minneapolis, MN 55455, USA
}
\emailAdd{basabendu88barman@gmail.com}
\emailAdd{simon.clery@ijclab.in2p3.fr}
\emailAdd{rco@umn.edu}
\emailAdd{yann.mambrini@ijclab.in2p3.fr}
\emailAdd{olive@physics.umn.edu}
%%%%%%%%%%%%%%%%%
\abstract{We show that a minimal scenario, utilizing only the graviton as an intermediate messenger between the inflaton, the dark sector and the Standard Model (SM), is able to generate {\it simultaneously} the observed relic density of dark matter (DM), the baryon asymmetry through leptogenesis, as well as a sufficiently hot thermal bath after inflation. 
We assume an inflaton potential of the form $V(\phi)\propto \phi^k$ about the minimum at the end of inflation.  
The possibility of reheating via minimal gravitational interactions has been excluded by constraints on dark radiation for excessive gravitational waves produced from inflation.
We thus extend the minimal model in several ways: i) we consider non-minimal gravitational couplings--this points to the parameter range of DM masses $M_{N_1} \simeq 2-10 $ PeV, and right-handed neutrino masses $M_{N_2} \simeq (5-20) \times 10^{11}$ GeV, and $\Trh \lesssim 3 \times 10^5$ GeV (for $k \le 20$); ii) we propose an explanation for the PeV excess observed by IceCube when the DM has a direct but small Yukawa coupling to the SM; and iii) we also propose a novel scenario, where the gravitational production of DM is a two-step process, first through the production of two scalars, which then decay to fermionic DM final states. In this case, the absence of a helicity suppression enhances the production of DM and baryon asymmetry, and allows a great range for the parameters including a dark matter mass below an MeV where dark matter warmness can be observable by cosmic 21-cm lines, even when gravitational interactions are responsible for reheating. We also show that detectable primordial gravitational wave signals provide the opportunity to probe this scenario for $\Trh\lesssim 5\times 10^6$~GeV in future experiments, such as BBO, DECIGO, CE and ET. 
}
%%%%%%%%%%%%%%%%%%
\begin{flushright}
  PI/UAN-2022-720FT, UMN--TH--4202/22, FTPI--MINN--22/26
\end{flushright}
\maketitle
%%%%%%%%%%%
\section{Introduction}
\label{sec:intro}
%%%%%%%%%%%
Since the first calculation of the mass of the Milky Way by Henry Poincaré in 1906 \cite{Poincare}, and the conclusion
that the {\it ``dark matter is present in much greater amount than luminous matter"} by 
Fritz Zwicky in 1933~\cite{Zwicky:1933gu},
the virial method has been used frequently to compute the amount of dark matter (DM) in the Universe. The presence of dark matter is generally deduced from its gravitational effects. The precise abundance of DM is obtained from observations of the anisotropy spectrum of the cosmological microwave background (CMB) \cite{Aghanim:2018eyx}. Initially, a neutrino, or more generally a heavy neutral lepton was thought to be an ideal dark matter candidate~\cite{Gunn:1978gr}. This candidate was assumed to interact weakly with the Standard Model (SM) and required a GeV scale mass to satisfy relic abundance constraints \cite{Hut:1977zn,Lee:1977ua}. Generalizations of the heavy neutrino DM candidate are referred to as WIMPs (weakly interacting massive particles). Perhaps the best studied WIMP
is the lightest supersymmetric particle in supersymmetric extensions of the SM \cite{Ellis:1983ew}. Other well studied candidates include those generated by a Higgs-portal \cite{Silveira:1985rk,McDonald:1993ex,Burgess:2000yq,Davoudiasl:2004be,Cline:2013gha,Han:2015hda,Djouadi:2011aa,Lebedev:2011iq,Mambrini:2011ik,Djouadi:2012zc,Casas:2017jjg}
or a $Z$-portal \cite{Arcadi:2014lta,Escudero:2016gzx}. However, these minimal constructions are now heavily constrained (see, for example, Refs.~\cite{Roszkowski:2017nbc,Arcadi:2017kky} for reviews), and even extensions to $Z'$-portal~\cite{Mambrini:2010dq,Lebedev:2014bba,Alves:2015pea,Alves:2016cqf,Arcadi:2017jqd} are in tension with the electroweak nature of dark matter.

The WIMP relic density is often determined by
thermal freeze-out. WIMPs are assumed to be in thermal equilibrium at temperatures higher than the mass, and the relic density is determined by the equilibrium density when DM annihilations can no longer keep up with the expansion rate of the Universe. 
It is also possible that particles with interactions much weaker than electroweak interactions and are never fully in equilibrium, but are nonetheless produced in the thermal bath after inflation.  An example of such a candidate is the gravitino \cite{nos,Ellis:1983ew,Khlopov:1984pf,Olive:1984bi}. More generally, these Feebly Interacting Massive Particles (FIMPs) have been proposed~\cite{McDonald:2001vt,Hall:2009bx,Mambrini:2013iaa} as an alternative to WIMPs (see
\cite{Bernal:2017kxu} for a recent review). In this framework, the DM candidate is never thermalized due its extremely
weak coupling to the SM, so weak that they evade the current accelerator constraints. 

In the early Universe, FIMPs can be produced
from either the decay or annihilation of states in the visible sector. When the SM temperature becomes smaller
than the typical mass scale of the interaction (i.e.~the maximum of the DM and the mediator mass), the generation
process becomes suppressed, leaving a constant comoving DM number density. Such a scenario is often referred as the freeze-in mechanism~\cite{Hall:2009bx}. In contrast to the ``WIMP-miracle" which produces the observed relic density with near weak-scale couplings and masses,
a ``FIMP-miracle" occurs when one considers renormalizable couplings of order 
$\sim\mathcal{O}(10^{-11})$ independent of the mass
of the DM. If a priori such couplings seem unatural, 
UV versions of the freeze-in mechanism may invoke
{\it effective} couplings, suppressed by a large mass scale
above the temperature of the thermal bath. 
This can be achieved 
via non-renormalizable operators~\cite{Elahi:2014fsa}, suppressed by a high mass
scale, e.g., in models
where the mediators between the visible sector
and the dark sectors are very massive.
This is the case
in unified theories like $SO(10)$ with a heavy $Z'$ gauge boson \cite{Mambrini:2013iaa,Mambrini:2015vna,Bhattacharyya:2018evo}, moduli fields \cite{Chowdhury:2018tzw}, high scale SUSY \cite{Benakli:2017whb,Dudas:2017rpa,Dudas:2017kfz,Dudas:2018npp,Kaneta:2019yjn}
or heavy spin-2 constructions \cite{Bernal:2018qlk}.
In other examples, freeze-in of DM may proceed via loops \cite{Kaneta:2019zgw,Heurtier:2019eou} or 4-body final states \cite{Mambrini:2022uol}. 
All of these scenarios are particularly interesting, as the DM yield is sensitive to the highest temperature
$\Tmax$ reached by the SM plasma~\cite{Chung:1998rq,Giudice:2000ex,Garcia:2017tuj,Garcia:2020eof,Garcia:2020wiy,Barman:2022tzk}, controlled by the dynamics of the inflaton decay. 

Even feebler interactions are possible when the only effective coupling at the UV scale is gravity.
Indeed, the minimal irreducible interaction that should exist between DM and the Standard Model (SM) is mediated by graviton exchange~\cite{Ema:2015dka,Garny:2015sjg,Garny:2017kha,Tang:2016vch,Tang:2017hvq,Ema:2016hlw,Bernal:2018qlk,Ema:2018ucl,Ema:2019yrd,Chianese:2020yjo,Chianese:2020khl,Redi:2020ffc,Mambrini:2021zpp,Barman:2021ugy,Haque:2021mab,Clery:2021bwz,Clery:2022wib,Ahmed:2022qeh,Ahmed:2022tfm}
which can lead to the observed amount of DM through the scattering of
the particles in the thermal bath or directly through the
gravitational transfer of the energy stored
in the inflaton condensate, as already been discussed in detail 
Refs.~\cite{Mambrini:2021zpp,Barman:2021ugy,Haque:2021mab,Clery:2021bwz,Clery:2022wib}. 

DM requires an extension to the SM, but it is not the only 
reason why an extension is necessary.
As is well known, the visible or baryonic matter content of the Universe is asymmetric. One interesting mechanism to produce the baryon asymmetry of the Universe (BAU) via the lepton sector physics is known as leptogenesis~\cite{Fukugita:1986hr}, where, instead of creating a baryon asymmetry directly, a lepton asymmetry is generated first and subsequently gets converted into baryon asymmetry by the $(B+L)$-violating electroweak sphaleron transitions~\cite{Kuzmin:1985mm}. In thermal leptogenesis \cite{Buchmuller:2002rq,Buchmuller:2003gz,Chankowski:2003rr,Giudice:2003jh}, the decaying particles, typically right-handed neutrinos (RHNs), are produced thermally from the SM bath. However, the lower bound on the RHN mass in such scenarios (known as the Davidson-Ibarra bound), leads to a lower bound on the reheating temperature $\Trh\gtrsim 10^{10}$ GeV~\cite{Davidson:2002qv} so that the RHNs can be produced from the thermal bath. 
One simpler alternative is the 
non-thermal production of RHNs~\cite{Giudice:1999fb,Asaka:1999yd,Lazarides:1990huy,Campbell:1992hd,Hahn-Woernle:2008tsk} originating from the decay of inflaton. This interaction is necessarily model dependent as it depends on the Yukawa interaction between the inflaton and the RHNs. 

In addition to providing the DM abundance, gravitational interactions can also be the source
of baryogenesis.
As shown in~\cite{Co:2022bgh}, it is possible to have a model-independent theory of non-thermal production of RHNs from inflation, once the inflaton potential is specified.\footnote{The simultaneous generation of gravitational DM and the baryon asymmetry was also discussed in~\cite{Bernal:2021kaj}. Our results differ,
as their choices of parameters are in conflict with the tensor-to-scalar ratio bound from Planck.}
The abundance of RHNs is calculated in the same manner as the dark matter abundance and 
can lead to observed BAU from the out-of-equilibrium CP violating decay of the RHNs, produced during the reheating epoch.

As noted above, the DM and RHNs may be produced through gravitational interactions emanating from the thermal bath or directly from the inflaton condensate. It has also been argued that the thermal bath itself may be generated from gravitational interactions \cite{Haque:2022kez,Clery:2021bwz,Co:2022bgh}. However, reheating the Universe from graviton exchange processes alone requires a steep inflaton potential during reheating, resulting in a low reheating temperature and a massive enhancement of tensor modes after inflation. Hence, the minimal scenario of gravitational reheating is excluded by an excessive generation of dark radiation in the form of gravitational waves (GWs) during BBN, as already noted in \cite{Figueroa:2018twl}. This limitation of minimal gravitational reheating is one motivation to introduce, as a natural generalization, non-minimal couplings of fields with gravity.

Motivated by these arguments, we derive a simultaneous solution for the DM abundance, the baryon asymmetry, and the origin of the thermal bath from purely gravitational interactions. In this sense, our scenario can be considered as the most minimal possible, since we do not introduce any new interactions for any process beyond the SM, except for gravity. The only new fields required are the dark matter candidate and the RHNs (which are anyway needed for the generation of neutrino masses). Our only model dependence comes from the choice of the particular inflaton potential. However we are mostly sensitive to the shape of the potential about the minimum after inflation. To be definite, we adopt the class of inflationary models called T-models~\cite{Kallosh:2013hoa}. But, as will be shown, even this dependence proves to be weak when it comes to combining the constraints of reheating, baryogenesis, and the dark matter relic density. We further show that the present framework can give rise to a detectable inflationary GW background, that in turn excludes the minimal gravitational reheating scenario which leads to an excess of the present-day GW energy density, in conflict with the BBN prediction. However, a large part of the parameter space still remain within the reach of several futuristic GW detection facilities.

The paper is organized as follows. After presenting our framework in Sec.~\ref{sec:framework}, we review the gravitational production from inflaton scattering and the thermal bath in Sec.~\ref{sec:grav-prod}, where
we also discuss the effect of non-minimal gravitational interactions. We derive the set of parameters (dark matter mass, RHN mass, and reheating temperature, $\Trh$, which simultaneously provide the correct relic density and BAU in Sec.~\ref{sec:result}. If the dark matter is not absolutely stable, we are able to propose an explanation for the PeV events observed at IceCube in the case of a long-lived candidate. Finally, we propose a novel scenario where the gravitational production is a two-step process passing through a scalar singlet which couples with the RHN sector in Sec.~\ref{sec:majoron}, before concluding in Sec.~\ref{sec:concl}.

%%%%%%%%%%%
\section{The framework}
\label{sec:framework}
%%%%%%%%%%%
If the metric is expanded around Minkowski space-time: $g_{\mu \nu}\simeq \eta_{\mu \nu}+\frac{2h_{\mu \nu}}{M_P}$, then the gravitational interactions are described by the Lagrangian~\cite{Choi:1994ax,Holstein:2006bh}
\beq
\sqrt{-g}\,\mathcal{L}_{\rm int}= -\frac{1}{M_P}\,h_{\mu \nu}\,\left(T^{\mu \nu}_{\rm SM}+T^{\mu \nu}_\phi + T^{\mu \nu}_{X} \right) \,,
\label{Eq:lagrangian}
\eeq
where $\phi$ is the inflaton and $X$ is a
particle which does not belong to the SM.\footnote{$M_P = (8 \pi G_N)^{-1/2} \simeq 2.4 \times 10^{18}$ GeV is the reduced Planck mass.} In the present context we consider $X$ to be a spin 1/2 Majorana fermion which
can be associated with the dark matter or a right-handed neutrino. 
The graviton propagator for momentum $p$ is
\begin{equation}
 \Pi^{\mu\nu\rho\sigma}(p) = \frac{\eta^{\rho\nu}\eta^{\sigma\mu} + 
\eta^{\rho\mu}\eta^{\sigma\nu} - \eta^{\rho\sigma}\eta^{\mu\nu} }{2p^2} \, .
\end{equation}
The form of the stress-energy tensor $T^{\mu \nu}_i$ depends on the spin of the field and, for Majorana spin-1/2 fermions, takes the form
\begin{equation}\label{eq:tmunu}
T^{\mu \nu}_{1/2} =
\frac{i}{8}
\left[ \bar \chi \gamma^\mu \overset{\leftrightarrow}{\partial^\nu} \chi
+\bar \chi \gamma^\nu \overset{\leftrightarrow}{\partial^\mu} \chi \right] -g^{\mu \nu}\left[\frac{i}{4}
\bar \chi \gamma^\alpha \overset{\leftrightarrow}{\partial_\alpha} \chi
-\frac{m_\chi}{2}\,\overline{\chi^c} \chi\right]\,,
\end{equation}
whereas for a scalar $\varphi$,
\begin{equation}
T^{\mu \nu}_{0} =
\partial^\mu \varphi \partial^\nu \varphi-
g^{\mu \nu}
\left[
\frac{1}{2}\partial^\alpha\varphi\,\partial_\alpha \varphi-V(\varphi)\right]\,.
\label{eq:tmunuphi}
\end{equation}

There are of course many possible scalar potentials $V(\phi)$ which can account for inflation. However, the calculations relevant in this paper are largely independent of the potential during inflation and depend only on the shape of the potential about the minimum. Without loss of generality, we will assume that $V(\phi)$ is among the class of $\alpha$-attractor T-models~\cite{Kallosh:2013hoa}
\begin{equation}
    V(\phi) \; = \;\lambda M_P^{4}\left|\sqrt{6} \tanh \left(\frac{\phi}{\sqrt{6} M_P}\right)\right|^{k} \, ,
\label{eq:Vatt}
\end{equation}
which can be expanded about the origin\footnote{Our discussion is general and not limited to T-models of inflation as the way we express the minimum of the potential is generic.}
\begin{equation}
    \label{Eq:potmin}
    V(\phi)= \lambda \frac{\phi^{k}}{M_P^{k-4}}\,; \quad \phi \ll M_P \, .
\end{equation}

In this class of models, inflation occurs at large
field values ($\phi > M_P$), and after the period of exponential expansion, the inflaton begins to oscillate about the minimum and the process of reheating begins. 
]The end of inflation may be defined when $\ddot a=0$ where $a$ is the cosmological scale factor. The inflaton field value at that time is given by \cite{Ellis:2015pla, Garcia:2020wiy} as 
\begin{align}
   \phi_{\rm end} &\simeq 
   \sqrt{\frac{3}{8}}M_P\ln\left[\frac{1}{2} + \frac{k}{3}\left(k + \sqrt{k^2+3}\right)\right].
   \label{phiend}
\end{align}
It is easy to show that at the end of inflation, 
the condition $\ddot a=0$ is equivalent to $\dot\phi_{\rm end}^2 = V(\phi_{\rm end})$
and thus the inflaton energy density at $\phi_{\rm end}$ is  $\rho_{\rm end}=\frac{3}{2}V(\phi_{\rm end})$.
The overall scale of the potential parameterized by the coupling $\lambda$, can be determined from the amplitude of the CMB power spectrum $A_S$,  \begin{align}
   \lambda &\simeq 
   \frac{18\pi^2 A_S}{6^{k/2}N^2} ,
\end{align}
where $N$ is the number of e-folds measured from the end of inflation to the time when the pivot scale $k_*=0.05~{\rm Mpc}^{-1}$ exits the horizon.
In our analysis, we use $\ln(10^{10}A_S)=3.044$ \cite{Planck:2018jri} and set $N=55$. This leads to an inflaton mass of $m_\phi\simeq 1.2\times 10^{13}$ GeV for $k=2$. More generically, $m_\phi\simeq 1.2\times 10^{13}$ is also the inflaton mass at the end of inflation for any larger $k$ when the full potential in Eq.~(\ref{eq:Vatt}) is used. While $N=55$ is appropriate for reheating temperatures of order $10^{12}$ GeV, for lower reheating temperatures (between $10 - 10^7$ GeV) $N = 45 -50$ \cite{egnov}. However, we have checked that our results are very insensitive to the value of $N$.

In addition to the inflationary sector and the SM,
neutrino masses and mixing require at least two (heavy) right-handed neutrino states for the seesaw mechanism~\cite{MINKOWSKI1977421,GellMann:1980vs,Yanagida:1979as,Mohapatra:1979ia,Schechter:1980gr,Schechter:1981cv}. One of these, if produced and remaining out-of-equilibrium until its decay, can produce a lepton asymmetry.
In order to account for the dark matter in a most economic way, we assume three RHNs, $N_i$, where for now,
the lightest of these, $N_1$ is decoupled from the other two and has a vanishing Yukawa coupling. 
Aside from the Yukawa couplings, the only couplings we consider between the SM, the RHNs, and the inflaton are gravitational of the form in Eq.~(\ref{Eq:lagrangian}). Needless to say, such interactions are unavoidable, and must be taken into account in any extensions beyond the SM.

As a concrete example, we consider the renormalizable interaction Lagrangian between the Majorana RHNs and the SM
\begin{align}\label{eq:RHN-lgrng}
& \mathcal{L}\supset-\frac{1}{2}\,M_{N_i}\,\overline{N_i^c}N_i-(y_N)_{ij}\,\overline{N}_i\,\widetilde{H}^{\dagger}\,L_j +{\rm h.c.}\,.
\end{align}
Here $H$ and $L$ are the SM Higgs and lepton doublet respectively.
Lepton number is clearly violated in this Lagrangian.\footnote{We consider the RHNs to be mass diagonal.}
For now, we assume that $(y_N)_{1i} = 0$ for all $i$ and that $N_1$ is stable. 
As a result, $N_1$ is a viable DM candidate.
Later, we will relax this condition and consider a metastable DM candidate with
$(y_N)_{1i}\neq 0$, allowing
for $N_1$ to decay into neutrinos that could be observed at IceCube. The preservation of the lepton asymmetry will provide a limit on $(y_N)_{1i}/M_{N_1}$. The other two RHNs, namely $N_{2,3}$ are assumed to be heavier and they participate in leptogenesis. 
 
We would like to remind the readers that there are three types of seesaw models, which differ by the properties of the exchanged heavy particles, e.g., 
\begin{itemize}
\item[(i)] Type-I: SM gauge fermion singlets
\item[(ii)] Type-II: SM $SU(2)_L$ scalar triplets
\item[(iii)] Type-III: SM $SU(2)_L$ fermion triplets. 
\end{itemize}
In the present case we are considering the Type-I scenario, which is evident from the Lagrangian in Eq.~\eqref{eq:RHN-lgrng}. The Type-I seesaw mechanism can indeed be realized with only
two active right-handed neutrino~\cite{Frampton:2002qc,Raidal:2002xf,Ibarra:2003up,Rink:2017zrf}. In this context, only the normal ($m_1<m_2<m_3$) and inverted ($m_3<m_2<m_1$) hierarchies are relevant, where  $m_i$ are the light neutrino masses. With only two RHN playing a role in the seesaw mechanism, we expect $m_1=0$ or $m_3=0$, depending on the hierarchy. Indeed, due to the reduced rank of the mass matrices (a $3\times 2$ Dirac matrix and a $2\times 2$ Majorana matrix) one neutrino remains massless, while the others acquire their light mass through the usual seesaw suppression of the order $m_i\sim \frac{(y_N)_{ii}^2\langle H \rangle^2}{M_{N_i}}$.

\begin{figure}[ht!]
\centering
\includegraphics[scale=1.5]{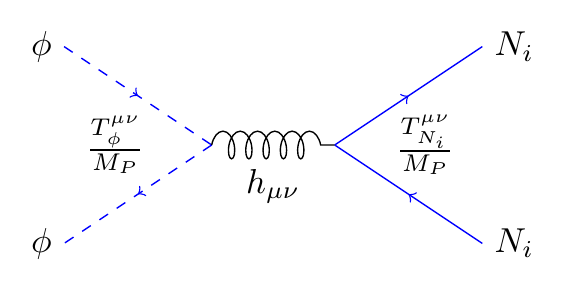}
\caption{\em Feynman diagram for the production of RHN through the gravitational scattering of the inflaton condensate. A similar diagram also exists with Standard Model particles in the initial state.}
\label{Fig:feynman}
\end{figure}

We further assume the absence of any direct coupling between the inflaton $\phi$ and the RHNs, such that there is no perturbative decay of the inflaton into the RHN final state; in other words we do not attribute a lepton number to the inflaton. Thus, the {\it only} possible production of the RHNs is the 2-to-2 gravitational scattering of the inflatons and of the particles in the radiation bath. As we will show, these production channels dominate in different regions of the parameter space.
In Fig.~\ref{Fig:feynman}, we show the $s$-channel exchange of a graviton obtained from the Lagrangian in Eq.~\eqref{Eq:lagrangian} for the production of RHNs from the inflaton condensate, to which we can add a similar diagram for the production of SM fields during the reheating process. Despite the a priori Planck reduced interactions, we will show that this framework is perfectly capable of simultaneously explaining the dark matter relic abundance and the observed baryon asymmetry, while also reheating the Universe. The Planck suppression due to graviton exchange is indeed partially compensated by the energy available in the inflaton condensate at the end of inflation.

%%%%%%%%%%%%%%%%%%%%%%%%%%%%%%%%%%%%%%%
%%%%%%%%%%%%%%%%%%%%%%%%%%%%%%%%%

%%%%%%%%%%%%%%%%%%%%%%%%%%%%%%
\section{Gravitational production of RHNs}
\label{sec:grav-prod}
%%%%%%%%%%%%%%%%%%%%%%%%%%%%%
In this section, we compute the gravitational interactions and resulting abundances of the dark matter candidate, $N_1$ as well as the abundance of the RHN neutrino responsible for leptogenesis.
We are particularly interested in the interactions of the type in Fig.~\ref{Fig:feynman} between the inflaton and  the $N_i$.
In addition, we are interested in the gravitational interactions between the inflaton and SM particles which make up the thermal bath. 
Furthermore, we will show that it is possible to produce the thermal bath assuming the absence of any inflaton decay mode leading to reheating.  We will quantify how such interactions can give rise to reasonable relic density and baryon asymmetry. The DM candidate $N_1$ can be produced during reheating from inflaton scattering $\phi \phi \rightarrow N_1 N_1$ as well as from the thermal bath (mediated by a massless graviton in both cases). 
For $(y_N)_{1i}=0$, $N_1$ couples only gravitationally, 
and the SM processes will not lead to its production. 
On the other hand, for the generation of the baryon asymmetry, we will cater to non-thermal leptogenesis, where the RHNs $N_{2,3}$ are too weakly coupled to reach thermal equilibrium. 
Hence they are predominantly produced only during 
reheating from gravitational inflaton scattering. 
To summarize, we consider the following production via graviton exchange
\begin{itemize}
\item $\phi\,\phi\to N_1\,N_1$, SM\,SM $\to N_1\,N_1$ for production of the DM candidate $N_1$.
\item $\phi\,\phi\to N_{2,3}\,N_{2,3}$ for production of $N_{2,3}$ that will lead to non-thermal leptogenesis. (Contributions from SM\,SM $\to N_{2,3}\,N_{2,3}$ are negligibly small.) 
\item $\phi \phi \rightarrow $ SM SM for the reheating process.
\end{itemize}
In the following subsections we determine the production rate of DM and RHNs from the processes listed above.

%%%%%%%%%%%%%%%%%%%%%%%%%%%%%%%
\subsection{Gravitational dark matter}
\label{sec:grav-dm}
%%%%%%%%%%%%%%%%%%%%%%%%%%%%%%%%
We start by computing the DM number density via 2-to-2 scattering of the bath particles, mediated by graviton exchange. In this case the interaction rate is given by~\cite{Bernal:2018qlk,Bernal:2021kaj,Barman:2021ugy,Clery:2021bwz}
\begin{align}\label{eq:dm-sm-rate}
& R_{N_i}^T = \frac{1}{2}\times\beta_{1/2}\,\frac{T^8}{M_P^4}\,,
\end{align}
where we have $\beta_{1/2} =11351 \pi^3/10368000 \simeq 3.4\times 10^{-2}$~\cite{Clery:2021bwz},
the explicit factor of $\frac{1}{2}$ accounting for the Majorana nature of $N_i$.
The evolution of RHN number density $n_{N_i}$ (with $i=1,2,3$) is governed by the Boltzmann equation
\begin{align}\label{eq:dm-beq}
& \frac{dn_{N_i}}{dt} + 3\,H\,n_{N_i} = R_{N_i}^T\,,
\end{align}
where $H=\dot a/a$ is the Hubble parameter. Defining the comoving number density as $Y_{N_i}=n_{N_i}\,a^3$ we can re-cast the Boltzmann equation as
\begin{align}\label{eq:dm-beq-comov}
& \frac{dY^T_{N_i}}{da}=\frac{a^2}{H}\,R_{N_i}^T\,, 
\end{align}
where $i=1$ for DM production, and the superscript
$T$ refer to the thermal source of production. In order to properly  capture the evolution of the inflaton and radiation energy density (and hence temperature) we solve the following set of coupled equations
\begin{align}\label{eq:BErho}
    &\frac{d\rho_\phi}{dt} + 3H\,(1+w_\phi)\, \rho_\phi = -(1+w_\phi)\,\Gamma_\phi\, \rho_\phi\,,
    \nonumber\\
    &\frac{d\rho_R}{dt} + 4H\, \rho_R = + (1+w_\phi)\,\Gamma_\phi\, \rho_\phi\,, 
\end{align}
where $w_\phi\equiv\frac{p_\phi}{\rho_\phi}=\frac{k-2}{k+2}$~\cite{Garcia:2020wiy} is the general equation of state parameter, $\rho_{\phi}$ the inflaton energy density, $\rho_R = \frac{\pi^2 g_*}{30}\,T^4\equiv c_* \,T^4$, and
$g_*$ is the effective number of degrees of freedom for 
the thermal plasma at the temperature $T$.
We recall that in our framework the potential $V(\phi)$ is proportional to $\phi^k$ [see, Eq.~\eqref{Eq:potmin}]. 

During reheating, the total energy density is dominated by the inflaton and we can approximate
the Hubble parameter by $H^2 \simeq \rho_\phi/3\,M_P^2$. In this case,  it is possible to analytically solve Eq.~\eqref{eq:BErho} and obtain 
\begin{equation}
\rho_\phi(a) = \rho_{\rm end} \left(\frac{a_{\rm end}}{a} \right)^\frac{6k}{k+2}\, .
\label{eq:rho_phi_a}
\end{equation}

Recall that we are assuming
that the radiation bath is produced {\it gravitationally} through inflaton scattering; namely, we do not rely on a specific
decay channel $\phi \rightarrow$ SM particles for reheating.
In this case, 
due to helicity conservation, 
the production of SM fermions from inflaton scattering is strongly suppressed by the mass of the fermions, whereas massless 
vectors are not produced because of the antisymmetry of $F_{\mu\nu}$.
However, scattering into scalars, especially Higgs scalars, 
is always allowed and
dominates the production rate. In \cite{Garcia:2020wiy}, the inflaton dissipation rate was parameterized as $\Gamma_\phi \propto \rho_\phi^l$. For a quartic interaction with constant coupling,
 $l = (3/k)-(1/2)$. However, for the effective gravitational coupling between the inflaton and 
 SM Higgs, the coupling is proportional to $m_\phi^2 \propto \rho_\phi^{(1-2/k)}$. 
 This leads to $l = (3/2)-(1/k)$. 
 More accurately, 
 expanding the potential energy in terms of the Fourier modes~\cite{Ichikawa:2008ne,Kainulainen:2016vzv,Clery:2021bwz,Co:2022bgh,Garcia:2020wiy,Ahmed:2022qeh}
\begin{equation}
V(\phi)=V(\phi_0)\sum_{n=-\infty}^{\infty} {\cal P}_n^ke^{-in \omega t}
=\rho_\phi\sum_{n = -\infty}^{\infty} {\cal P}_n^ke^{-in \omega t}\,,
\label{Eq:fourier}
\end{equation}
 the production rate of radiation is given by \cite{Clery:2021bwz, Clery:2022wib, Co:2022bgh}
\begin{equation}
   (1+w_\phi)\,\Gamma_\phi\, \rho_\phi = R^{\phi^k}_H \simeq \frac{N_h\rho_{\phi}^2}{16\pi M_P^4} \sum_{n=1}^{\infty}  2n\omega|{\mathcal{P}}^k_{2n}|^2 = \alpha_k M_P^5 \left(\frac{\rho_{\phi}}{M_P^4}\right)^{\frac{5k-2}{2k}} \, ,
   \label{Eq:ratephik}
\end{equation}
where $N_h = 4$ is the number of internal degrees of freedom for one complex Higgs
doublet and we have
neglected the Higgs bosons mass. 
The frequency of oscillations of $\phi$ is given by~\cite{Garcia:2020wiy}
\begin{equation}
\label{eq:angfrequency}
\omega=m_\phi \sqrt{\frac{\pi k}{2(k-1)}}
\frac{\Gamma(\frac{1}{2}+\frac{1}{k})}{\Gamma(\frac{1}{k})} \,,
\end{equation}
with 
$m_\phi^2= \frac{\partial^2V(\phi)}{\partial \phi^
2}$ being the inflaton mass squared.
The definition of $\alpha_k$ follows the analysis in~\cite{Co:2022bgh}, with the values given in Table~\ref{Tab:tablealphak}. 
For $l = (3/2)-(1/k)$ the results of \cite{Garcia:2020wiy} yields for the 
evolution of the radiation density
\begin{align}
\label{eq:rad-density-grav}
\rho_R(a) = 
& \rRH\,\left(\frac{\arh}{a}\right)^4\,\left[\frac{1-(\aend/a)^{\frac{8k-14}{k+2}}}{1-(\aend/\arh)^{\frac{8k-14}{k+2}}}\right]\,,
\end{align}
which can be obtained by combining Eqs.(\ref{eq:BErho}), (\ref{eq:rho_phi_a}) and (\ref{Eq:ratephik}).
The evolution in Eq.~(\ref{eq:rad-density-grav}) is valid when 
$\aend\ll a\ll \arh$ where $\aend$ marks the end of inflation (or the onset of reheating), while $\arh$ indicates the end of reheating defined as $\rho_\phi(\arh)=\rho_R(\arh)=\rRH$.
To obtain Eq.~(\ref{eq:rho_phi_a}), we have supposed $H \gg \Gamma_\phi$, which is valid for all $a$ because $\Gamma_\phi < H$ at the end of inflation and $\Gamma_\phi$ decreases faster than $H$ for all $k \ge 2$. As a result, we note that gravitational reheating is only possible for $\frac{6k}{k+2}>4$ [cf. Eq.~\eqref{eq:rho_phi_a}], {\it i.e.}, when $\rho_\phi$ redshifts faster than $\rho_R$. This limits our parameter space to $k > 4$. It is also important to ensure that a sufficiently large reheating temperature is attained to allow big bang nucleosynthesis to occur at $T \sim 1$ MeV as we will discuss in more detail in Sec.~\ref{sec:grav-reheat}.
%%%%%%%%%%%%%%%%%%%%
\begin{table}[htb!]
\centering
\begin{tabular}{|c|c|c|}
\hline
$k$ & $\alpha_k$ & $\alpha_k^{\xi}$ \\
\hline
6 &  0.000193 & $\alpha_k + 0.00766 \, \xi_h^2$ \\
\hline
8 & 0.000528 & $\alpha_k + 0.0205 \, \xi_h^2$ \\
\hline
10 & 0.000966 & $\alpha_k + 0.0367\, \xi_h^2$ \\
\hline
12 & 0.00144 & $\alpha_k +  0.0537 \, \xi_h^2$ \\
\hline
14 &  0.00192 & $\alpha_k +  0.0702 \, \xi_h^2$ \\
\hline
16 &  0.00239 & $\alpha_k +  0.0855 \, \xi_h^2$ \\
\hline
18 &  0.00282 & $\alpha_k +  0.0995 \, \xi_h^2$ \\
\hline
20 &  0.00322 & $\alpha_k +  0.112 \, \xi_h^2$ \\
\hline
\end{tabular}
\caption{Relevant coefficients for the gravitational reheating [cf.~Eq.~\eqref{Eq:ratephik} and Eq.~\eqref{ratexi}].}
\label{Tab:tablealphak}
\end{table}
%%%%%%%%%%%%%%%%%%%%%%%%%%%%%%%%%%%%%%%

Using Eqs.~(\ref{eq:dm-sm-rate}) and (\ref{eq:rad-density-grav}) and relating $T^8$ to $\rho_R^2$, 
we obtain the thermal rate of DM production 
\begin{align}\label{eq:inf-rate-grav}
R^T_{N_i} = \frac{1}{2}\times \beta_{1/2}\,\frac{\rRH^2}{c_*^2\,M_P^4}
& \left(\frac{\arh}{a}\right)^8\,\left[\frac{1-(\aend/a)^{\frac{8k-14}{k+2}}}{1-(\aend/\arh)^{\frac{8k-14}{k+2}}}\right]^2\,.
\end{align}
 The DM number density at the end of reheating can then be computed by integrating Eq.~\eqref{eq:dm-beq-comov}, leading to \begin{align}\label{eq:nT-grav}
& n^T_{N_i}(a_{\rm RH})=\frac{\beta_{1/2}\,(k+2)\,\rRH^\frac{3}{2}}{12\,\sqrt{3}M_P^3\,c_*^2}
\left(\frac{1}{1-r^{\frac{14-8k}{k+2}}}\right)^2\,
\nonumber\\ &
\left[ \frac{2(7-4k)^2}{(k+5)(k-1)(5k-2)}\,r^{\frac{10+2k}{k+2}}-\frac{9}{(k+5)} + \frac{18}{(5k-2)}\,r^{\frac{14-8k}{k+2}} - \frac{1}{(k-1)}\, r^{\frac{28-16k}{k+2}}\right]\, ,
\end{align}
where $r=\arh/\aend$. Since the gravitational reheating temperature is generally quite low as discussed in Sec.~\ref{sec:grav-reheat}, we can consider the limit $r\gg 1$ and the dominant term in the expression above is
\begin{equation}
    n^T_{N_i} (a_{\rm RH}) \simeq \frac{\beta_{1/2}\,(k+2)\,\rRH^\frac{3}{2}}{12\,\sqrt{3}M_P^3\,c_*^2} \frac{2(7-4k)^2}{(k+5)(k-1)(5k-2)}\,r^{\frac{10+2k}{k+2}}\,.
    \label{Eq:nthermal}
\end{equation}
The contribution of gravitational scattering of the particles in the primordial plasma to the DM relic abundance can then be determined using~\cite{book}
\begin{align}\label{eq:rel-SM}
& \Omega_{N_1}^T\,h^2 = 1.6\times 10^8\,
\frac{g_0}{g_{\rm RH}}\,\frac{n_{N_1}^T(\arh)}{\Trh^3}\,\frac{M_{N_1}}{\text{GeV}}\,,  
\end{align}
which gives
\begin{align}
\Omega_{N_1}^T\,h^2     & \simeq 1.6\times 10^8\times\frac{g_0\,\beta_{1/2}}{g_{\rm RH}}\,\times \frac{M_{N_1}}{\rm GeV}\, \frac{c_*^{-\frac{5}{6}-\frac{5}{3k}}\,(7-4k)^2\, (k+2)}{6\,\sqrt{3}\,(k+5)(k-1)(5k-2)}\left(\frac{\Trh}{M_P}\right)^{\frac{5k-20}{3k}}\left(\frac{\rho_{\rm end}}{M_P^4}\right)^{\frac{k+5}{3k}}\,,
     \label{Eq:omegathermal}
\end{align}
where $g_0=g_*(T_0)=43/11$ and $g_{\rm RH}=g_*(\Trh)=427/4$ are the number of relativistic degrees of freedom at present and at the end of reheating respectively. In (\ref{Eq:omegathermal}), we have used Eq.~(\ref{eq:rho_phi_a}) evaluated at $a = a_{\rm RH}$ to relate $r$, $\Trh$, and $\rho_{\rm end}$.

The DM candidate $N_1$ can also be produced directly
from inflaton scattering. In fact, particle production directly from inflaton scattering can be much more efficient than gravitational thermal scattering when $\Trh$ is low~\cite{Clery:2021bwz}.
As we shall see in the next section, the same process is also involved in the production of the baryon asymmetry. For the production of $N_1$ through the scattering of the inflaton condensate, we consider the time dependent 
oscillations of a classical inflaton field $\phi(t)$\footnote{As it has been pointed out in~\cite{Lozanov:2016hid,Lozanov:2017hjm,Maity:2018qhi}, for all potentials steeper than quadratic near the origin, the oscillating inflaton condensate may fragment due to self-resonance. The equation of state in that case approaches $w\to 1/3$ at sufficiently late times. If this occurs after $T = \Trh$, then this effect is not important since the inflaton energy would already be subdominant.  Furthermore, $N$ is predominantly produced at $\aend$ and so possible fragmentation at later times would not affect the calculation of the baryon asymmetry. Reheating may be affected; however, the exact time when $w$ transitions $1/3$ depends on $k$ and requires dedicated lattice simulations. Self-resonance becomes less efficient for larger $k$ as shown by the lattice results for $k$ up to 6~\cite{Lozanov:2016hid,Lozanov:2017hjm,Maity:2018qhi}, while most of our viable results are for $k > 6$. Performing such lattice simulations for larger $k$ is beyond the scope of the present analysis.}. The oscillating inflaton field with a time-dependent amplitude can be parametrized as
\begin{equation}
    \label{Eq:oscillation}
    \phi(t)= \phi_0(t)\cdot\mathcal{Q}(t) = \phi_0(t)\sum_{n=-\infty}^{\infty}\,{\cal Q}_n e^{-in \omega t}\,,
\end{equation}
where $\phi_0(t)$ is the time-dependent amplitude that includes the effects of redshift and $\mathcal{Q}(t)$ 
describes the periodicity of the oscillation. Furthermore, we assume a mass hierarchy $M_{N_{1,2,3}}<m_\phi$ 
such that the $s$-channel graviton mediated process 
(as shown in Fig.~\ref{Fig:feynman}) is kinematically allowed. 
Note that, since $N_1$ is effectively decoupled from $N_{2,3}$,
it does not necessarily need to be the lightest of the three. 
The production rate for $N_i$ from inflaton scattering mediated by gravity is given\footnote{Note the difference of factor 2 with~\cite{Clery:2021bwz}, comes from the Majorana nature of the RHNs.} by~\cite{Clery:2021bwz}
\begin{equation}
\label{eq:rateferm}
R^{\phi^k}_{N_i}=\frac{\rho_\phi^2}{4\pi M_P^4}\,\frac{M_{N_i}^2}{m_\phi^2}\,\Sigma_{N_i}^k \, ,
\end{equation}
where 
\begin{equation}
    \label{eq:ratefermion2}
    \Sigma_{N_i}^k = \sum_{n=1}^{+\infty} |{\cal P}_{2n}^k|^2\,
    \frac{m_\phi^2}{E_{2n}^2}\,
\left[1-\frac{4\,M_{N_i}^2}{E_{2n}^2}\right]^{3/2} \,,
\end{equation}
accounts for the sum over the Fourier modes of the inflaton potential, and $m_\phi^2=\lambda\, k\,(k-1)\,\left(\rho_\phi/(\lambda\,M_P^4)\right)^\frac{k-2}{k}$. 
Here $E_n = n \omega$ is the energy of the $n$-th inflaton oscillation mode and $M_{N_i}$ is the 
mass of the produced RHN.
Then, the number density of RHN is obtained by solving a Boltzmann equation analogous to 
that in Eq.~\eqref{eq:dm-beq-comov} as 
\begin{equation}\label{eq:dm-beq-inf}
\frac{dY^{\phi^k}_{N_i}}{da}=\frac{\sqrt{3}M_P}{\sqrt{\rho_{\rm RH}}}a^2\left(\frac{a}{a_{\rm RH}}\right)^{\frac{3k}{k+2}}R_{N_i}^{\phi^k}(a).
\end{equation} 
Integration of Eq.~\eqref{eq:dm-beq-inf}, leads to the following expression for the RHN density~\cite{Clery:2021bwz, Co:2022bgh}
\begin{equation}
n_{N_i}^{\phi^k}(a_{\rm RH})\simeq \frac{M_{N_1}^2\,\sqrt{3}\,(k+2)\,\rho_{\rm RH}^{\frac{1}{2}+\frac{2}{k}}}{24\,\pi\,k(k-1)\lambda^{\frac{2}{k}}\,M_P^{1+\frac{8}{k}}}
\left(\frac{\rho_{\rm end}}{\rho_{\rm RH}}\right)^{\frac{1}{k}}\,
\Sigma_{N_1}^k\,,
\label{Eq:num-den-inf}
\end{equation}
evaluated at the end of reheating. In order to obtain the DM relic abundance, one can again follow Eq.~\eqref{eq:rel-SM}, but now replacing $n_{N_1}^T(\arh)$ with $n_{N_1}^\phi(\arh)$, and obtain \cite{Clery:2021bwz}
\begin{eqnarray}\label{eq:omega-phi}
\frac{\Omega_{N_1}^{\phi^k} h^2}{0.12}&=&
\frac{\Sigma_{N_1}^k}{2.4^{\frac{8}{k}}}\frac{k+2}{k(k-1)}
\left(\frac{10^{-11}}{\lambda}\right)^{\frac{2}{k}}
\left(\frac{10^{40} {\rm GeV}^4}{\rho_{\rm RH}}\right)^{\frac{1}{4}-\frac{1}{k}}
\nonumber
\\
&\times&
\left(\frac{\rm \rho_{end}}{10^{64} {\rm GeV}^4}\right)^{\frac{1}{k}}
\left(\frac{M_{N_1}}{1.1\times 10^{7^+\frac{6}{k}} {\rm GeV}}\right)^3\,.
\end{eqnarray}
The total DM relic abundance is a sum of the gravitational contribution from thermal bath $(\Omega_{N_1}^T\,h^2)$ and from inflaton scattering $(\Omega_{N_1}^{\phi^k}\,h^2)$.

%%%%%%%%%%%%%%%%%%%%%%%%%%%%%%%
\subsection{Gravitational leptogenesis}
\label{sec:grav-lepto}
%%%%%%%%%%%%%%%%%%%%%%%%%%%%%%%%
Since $N_1$ is the stable DM candidate, in the present scenario the lighter of $N_{2,3}$ can undergo out-of-equilibrium decay to SM final states. We denote $N_2$ to be the lighter of these, and we must require that the mixing of $N_1$ and $N_2$ to be sufficiently small so as to prevent the decay of $N_2$ to $N_1$. For now, we take this coupling to be absent.  The resulting CP asymmetry from the decay of $N_2$ is of the form~\cite{Luty:1992un,Flanz:1994yx,Covi:1996wh,Buchmuller:2004nz,Davidson:2008bu}
\begin{align}\label{eq:cp}
& \epsilon_{\Delta L} = \frac{\sum_{\alpha}[\Gamma(N_2 \rightarrow l_{\alpha}+H)-\Gamma(N_2 \rightarrow \overline{l}_{\alpha}+H^{*})]}{\sum_{\alpha}[\Gamma(N_2\rightarrow l_{\alpha}+H)+\Gamma(N_2 \rightarrow \overline{l}_{\alpha}+H^{*})]}\, .
\end{align}
The resulting lepton asymmetry depends on the out-of-equilibrium abundance of $N_2$
as computed in the previous subsection. So long as $M_{N_2} \ll m_\phi$ and any kinematic suppression can be ignored, the number density of $N_2$ (at $a_{\rm RH}$) will be given by Eq.~(\ref{Eq:num-den-inf}) with the substitution $N_1 \to N_2$. The CP asymmetry can be expressed as~\cite{Buchmuller:2004nz,Kaneta:2019yjn,Co:2022bgh}
\begin{align}
& \epsilon_{\Delta L}\simeq \frac{3 \delta_{\rm eff}}{16\,\pi}\,\frac{M_{N_2}\,m_{\nu\,,\text{max}}}{v^2}\,,  
\end{align}
where $\langle H\rangle \equiv v \approx 174$ GeV is the SM Higgs doublet vacuum expectation value,  $\delta_{\text{eff}}$ is the effective CP violating phase in the neutrino mass matrix with $0\leq \delta_{\rm eff}\leq 1$,  and, we take $m_{\nu,\text{max}} = 0.05$ eV as the heaviest light neutrino mass. 

Here we are interested in non-thermal leptogenesis~\cite{Giudice:1999fb,Lazarides:1990huy,Campbell:1992hd,Asaka:2002zu,Senoguz:2003hc, Hahn-Woernle:2008tsk,Barman:2021tgt}. The gravitationally produced $N_2$ should not be thermalized into the bath for the consistency of the calculation. To check this, we note that the thermalization rate $\Gamma_{\rm th} \simeq y_{N_2}^2 T / 8 \pi$ decreases slower than the Hubble rate $H =\frac{\sqrt{\rho_\phi}}{\sqrt{3}M_P}$ based on Eqs.~(\ref{eq:rho_phi_a}) and (\ref{eq:rad-density-grav}). Thermalization is potentially dangerous until $T \simeq M_{N_2}$ when the $N_2$ out-of-equilibrium decay rate dominates over the thermalization rate. Using Eq.~(\ref{eq:rho_phi_a}), $\aend/a \simeq T/\Tmax$ based on Eq.~(\ref{eq:rad-density-grav}), and $y_{N_2} \simeq m_\nu M_{N_2} / v^2$, we find that $\Gamma_{\rm th}$ is always less than $H$ at $T=M_{N_2}$ in the parameter space of interest.\footnote{More precisely, $T/\Tmax = a_{\rm max}/a = (a_{\rm max}/\aend)(\aend/a) = ((6k-3)/(2k+4))^{((k+2)/(8k-14))} (\aend/a) \approx 3^{1/8} (\aend/a)$ for large $k$ \cite{Clery:2021bwz}.} Thus the resulting lepton asymmetry will not be suppressed by inverse decays. 

The produced lepton asymmetry is eventually converted to baryon asymmetry via electroweak sphaleron processes leading to
\begin{align}\label{eq:yb}
& Y_B = \frac{n_B}{s} =  \frac{28}{79} \epsilon_{\Delta L}\,  \frac{n^\phi_{N_2}(\Trh)}{s} \,,
\end{align}
where $n_{N_2}^\phi(\Trh)$ is the number density from Eqs.~(\ref{Eq:nthermal}) and (\ref{Eq:num-den-inf}) at the end of reheating and $s = 2 \pi^2 g_{\rm RH} \Trh^3 /45$
is the entropy density. The final asymmetry then becomes 
\begin{align}
&  Y_B \simeq 3.5\times 10^{-4}\, \delta_{\rm eff} \left(\frac{m_{\nu,\text{max}}}{0.05 \,\text{eV}}\right) \left(\frac{M_{N_2}}{10^{13}\,\text{GeV}}\right)\,\left.\frac{n_{N_2}^\phi}{s}\right|_{\Trh}\,,    
\label{Eq:baryonassym}
\end{align}
while the observed value, as reported by Planck~\cite{Aghanim:2018eyx}, is $Y_B^\text{obs}\simeq 8.7\times 10^{-11}$.

We note that the lepton asymmetry is not washed out because the lepton-number violating process involving the Yukawa scattering and the electroweak sphaleron processes are never in equilibrium at the same time.

%%%%%%%%%%%%%%%%%%%%%%%%%
\subsection{Gravitational reheating temperature}
\label{sec:grav-reheat}
%%%%%%%%%%%%%%%%%%%%%%%%%
In section \ref{sec:grav-dm}, we computed the energy density in radiation from purely gravitational process. However, to avoid conflict with the Big Bang nucleosynthesis (BBN), that requires the reheating temperature $\Trh\gtrsim 1$ MeV, one needs to consider\footnote{This requirement of having large $w_\phi$ can be relaxed with non-minimal gravitational couplings as discussed in~\cite{Co:2022bgh,Clery:2022wib}.} $w_\phi\gtrsim 0.65$~\cite{Haque:2022kez,Co:2022bgh}, or $k=\frac{2\,(1+w_\phi)}{(1-w_\phi)}\gtrsim 9$. This lower bound comes from the fact that, for higher $k$, the inflaton energy density redshifts faster so the transition to radiation domination is achieved sooner, at a higher temperature. 

The precise bound on $\Trh$ is in fact more involved especially for $k \leq 8$
for the following reason. 
As noted below Eq.~(\ref{eq:rad-density-grav}), the inflaton-dominated era ends when $\rho_\phi$ redshifts 
below $\rho_R$ as opposed to a complete transfer 
of the $\phi$ energy to radiation in conventional reheating by perturbative decays. Still, we define the reheating temperature $\Trh$ by $\rho_\phi(\Trh)=\rho_R(\Trh)$. The difference in the scale factor 
dependence between $\rho_\phi$ in Eq.~(\ref{eq:rho_phi_a}) and $\rho_R$ in Eq.~(\ref{eq:rad-density-grav}) increases with $k$. In other words, 
for smaller $k$, the inflaton energy density does not redshift significantly more than radiation. Thus, $\Trh$ for low $k$ needs to be substantially higher than $T_{\rm BBN} \approx 1$ MeV so that the inflaton energy density does not excessively modify the expansion rate of the universe at BBN. 
We can recast the BBN bound on the extra energy density in the form
of $\Delta N_\nu < 0.226$~\cite{Yeh:2022heq} into a bound on $\Trh$ as 
a function of $k$ using the following expression
\begin{align}
\label{eq:TRH_BBN}
 1 = \left. \frac{\rho_\phi}{\rho_R} \right|_{\Trh} = \left. \frac{\rho_\phi}{\rho_R} \right|_{T_{\rm BBN}} \frac{(a_{\rm BBN}/\arh)^{\frac{6k}{k+2}}}{(\Trh/T_{\rm BBN})^4} ,
\end{align}
with entropy conservation $g_*(\arh)\,\Trh^3\,\arh^3 = g_*(a_{\rm BBN})\, T_{\rm BBN}^3\,a_{\rm BBN}^3$ within the SM sector, as well as $\rho_\phi(T_{\rm BBN}) = \frac{7}{8}\,\left(\frac{4}{11}\right)^{\frac{4}{3}}\,\Delta N_\nu\,\frac{\pi^2}{30}\,T_{\rm BBN}^4$. The resulting bounds for each $k$ are $\Trh \lesssim 40 \mev~(k=6),\, 9\mev~(k=8),\, 5 \mev~(k=10),\, 4 \mev~(k=12)$ and $3 \mev~(14 \leq k \leq 20)$, which we will show as red-shaded regions in the subsequent figures concerning $\Trh$. We note that this estimate is in good agreement with a more rigorous treatment performed in Ref.~\cite{Co:2021lkc} using a BBN computing package for the case of kination (large $k$ limit).

The reheating temperature can be determined by solving the Friedmann equation \eqref{eq:BErho} for the radiation energy density. This yields \cite{Co:2022bgh}
\begin{align}
\rho_R(a) & \simeq \alpha_k\frac{k+2}{8k-14}\sqrt{3}M_P^4
\left( \frac{\rho_{\rm end}}{M_P^4}\right)^{\frac{2k-1}{k}}
\left(\frac{\aend}{a}\right)^4 \, ,
\end{align}
and evaluating this at $a_{\rm RH}$
we have
\begin{align}\label{eq:grav-trh}
& \Trh^4 = \frac{30}{\pi^2\,g_{\rm RH}}\, M_P^4\,\left(\frac{\rend}{M_P^4}\right)^{\frac{4k-7}{k-4}}\,\left(\frac{\alpha_k\,\sqrt{3}\,(k+2)}{8k-14}\right)^\frac{3k}{k-4}\, .    
\end{align}
From Eq.~\eqref{eq:grav-trh} we find $\Trh\simeq 60$ MeV for $k=10$ and $\rend\simeq(4.8 \times 10^{15}\,\text{GeV})^4$. Note that, due to the logarithmic dependence of $\phi_{\rm end}$ on $k$ in Eq.~(\ref{phiend}), $\rend$ changes very slowly with $k$ and remains approximately fixed to the above value.

\begin{figure}[htb!]
\centering
\includegraphics[width=0.495\linewidth]{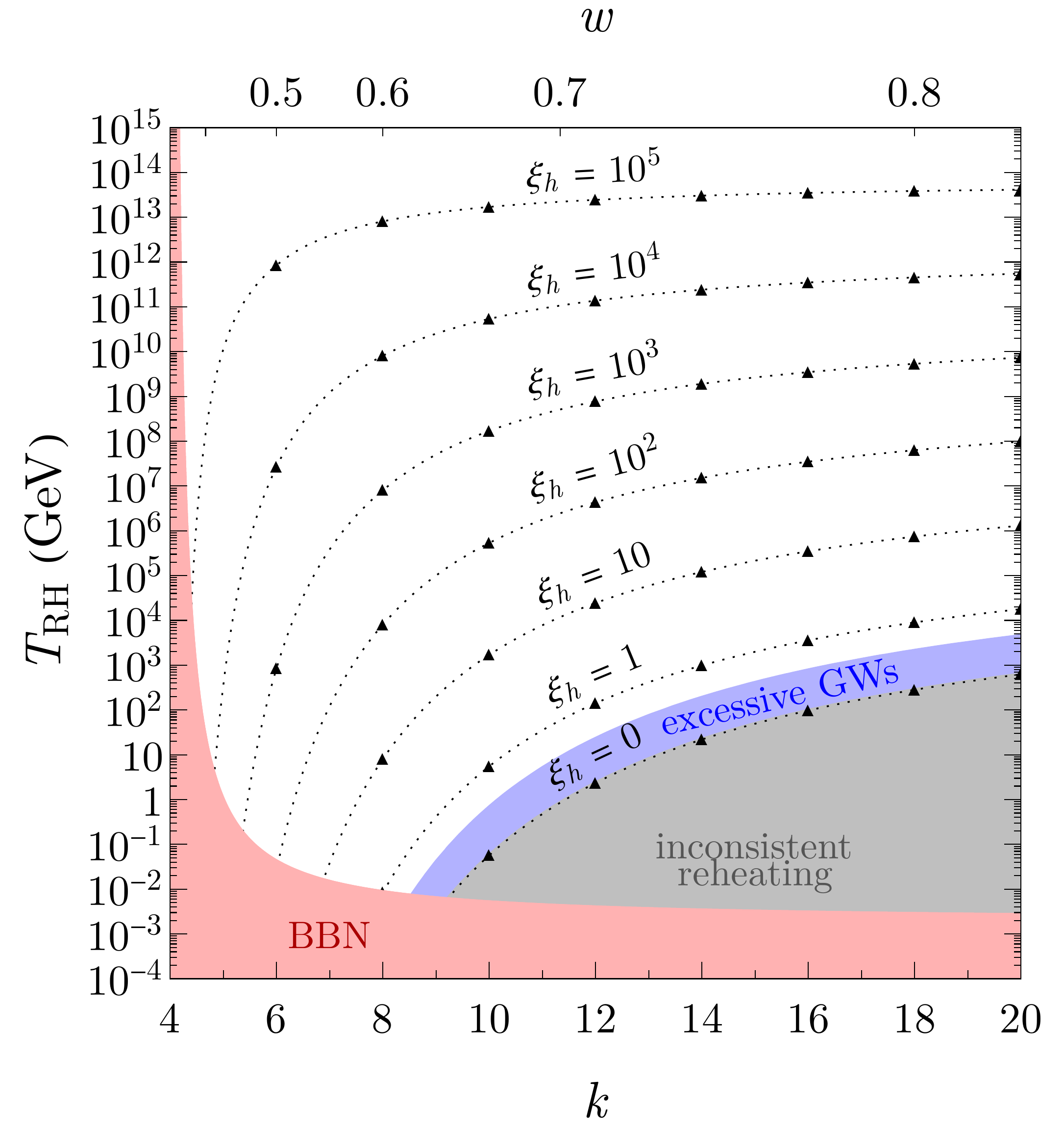}~~~
\includegraphics[width=0.495\linewidth]{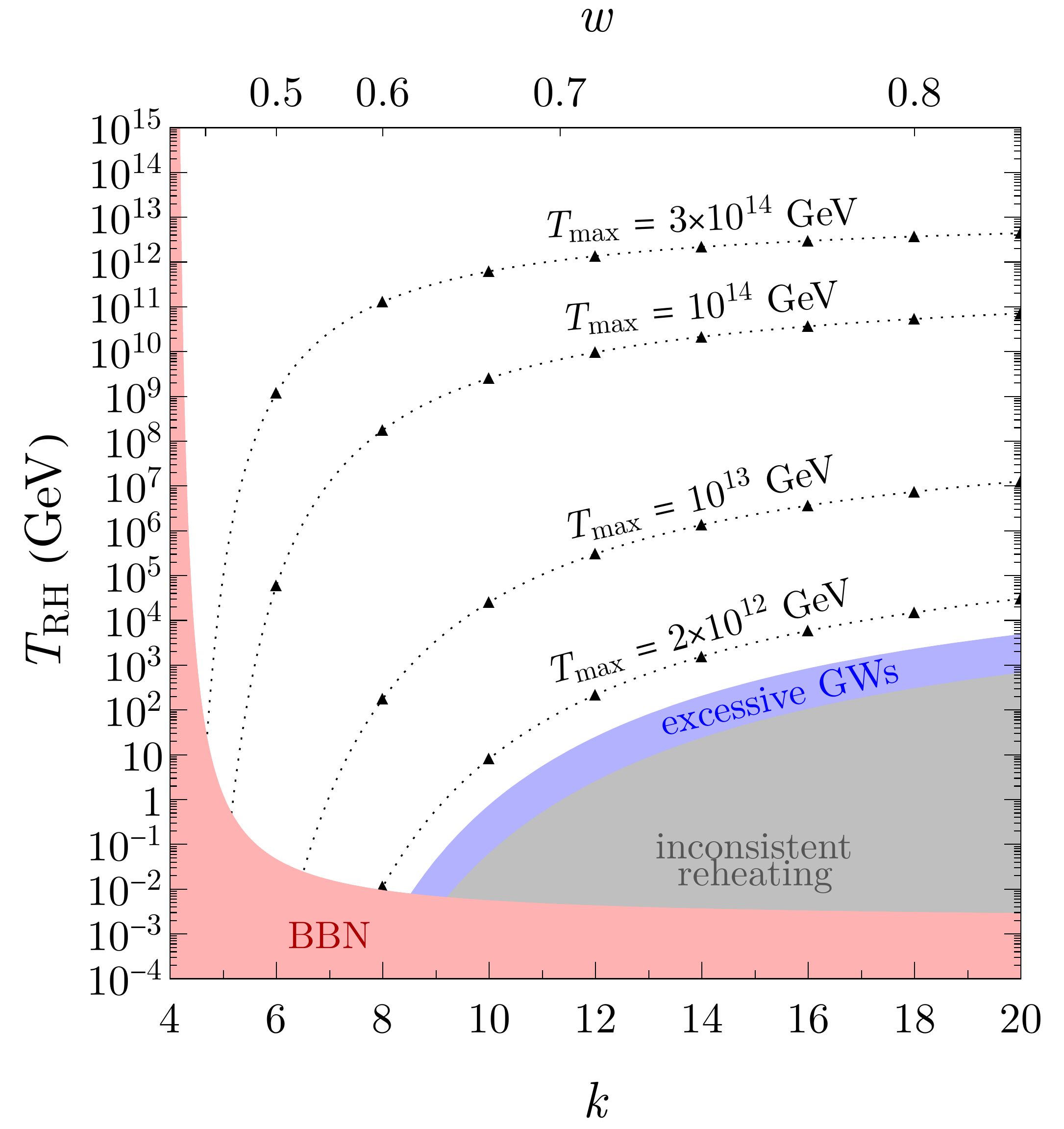}
\caption{\em Variation of $\Trh$ (left) and $\Tmax$ (right) as a function of $k$, for different choices of $\xi_h$. Triangles highlight the physical choices of even $k$. The red-shaded region is excluded by BBN because low reheating temperatures lead to an excessive inflaton energy density during BBN. The blue-shaded region is similarly excluded by BBN for excessive gravitational waves produced during inflation. The gray-shaded region is excluded as the lowest reheating temperature from gravitational reheating is that from minimal gravity (pure graviton exchange), i.e., $\xi_h = 0$.}
\label{fig:tmax/trh}
\end{figure}

In Fig.~\ref{fig:tmax/trh}, we show in the left panel the reheating temperature for minimal gravitational interactions by the curve labeled $\xi_h = 0$ (other values of $\xi_h$, non-minimal coupling of the Higgs to the Ricci scalar, are discussed in the next subsection). As one can see, $\Trh$ rises to $\simeq 1$ TeV, for $k=20$. This minimal case with $\xi_h = 0$ is excluded by excessive gravitational waves as dark radiation as will be discussed in Sec.~\ref{sec:grav-wave}, so a non-minimal coupling is ultimately required.  

%%%%%%%%%%%%%%%%%%%%%%%%%%%%%%%%%%%%%%%%%%%%%
%%%%%%%%%%%%%%%%%%%%%%%%%%%%%%%%%%%%%%%%%%%%%

At the start of the reheating process, 
the Universe quickly heats to a maximum temperature, $T_{\rm max}$.  As discussed in \cite{Clery:2021bwz}, the maximum temperature attained through purely gravitational processes is of order $10^{12}$ GeV, decreasing slightly with $k$. The maximum radiation density which determines $\Tmax$ was found to be \cite{Clery:2021bwz}
\begin{align}
\label{eq:tmax-grav}
\rho_\text{max}\simeq \sqrt{3}\,\alpha_k\,M_P^4\, \left(\frac{\rho_{\rm end}}{M_P^4}\right)^\frac{2k-1}{k}\,\frac{k+2}{12k-16}\,\left(\frac{2k+4}{6k-3}\right)^\frac{2k+4}{4k-7} \equiv c_*\, T_{\rm max}^4\,. 
\end{align}
Asymptotically at large $k$, $T_{\rm max} \approx 8 \times10^{11}$ GeV and  $\left(\frac{\Tmax}{\Trh}\right)_{k\gg4}\sim 
\left(\frac{1}{\alpha_{k\gg 4}}\right)^{1/2}\left(\frac{M_P^4}{\rho_{\rm end}}\right)^{1/2}\gg 1$. This represents a {\it minimal maximum} temperature, as other process such as decays (not considered here), may lead to a higher maximum temperature. 
The value of $\Tmax$ is shown in the right panel of Fig.~\ref{fig:tmax/trh}. For minimal gravitational interactions, 
corresponding to the simple exchange of a graviton, $\Tmax \simeq 10^{12} \gev$. In fact, as we noted before and will elaborate in Sec.~\ref{sec:grav-wave}, gravitational reheating with $\xi_h=0$ (graviton exchange) is already ruled out by the BBN bound on dark radiation in the form of GWs. 
Thus, in order to account for the reheating mechanism in a gravitational framework, it is necessary to introduce non-minimal couplings of fields to gravity. We compute the reheating and maximum temperatures for non-minimal gravitational interactions in the next subsection. The value of $\Tmax$ will be relevant when we discuss the DM warmness constraint, because DM is produced relativistically and predominantly at $\Tmax$.

%%%%%%%%%%%%%%%%%%%%%%%%%%%%
\subsection{Non-minimal gravitational production}
\label{sec:nonminimal}
%%%%%%%%%%%%%%%%%%%%%%
As pure gravitational particle production can be insufficient, we also consider the possibility that scalar fields have non-minimal couplings to gravity which generate effective couplings between these scalar fields and the RHNs. Thus, we allow both the inflaton $\phi$ and the Higgs field $H$ to be non-minimally coupled. We denote the complex Higgs doublet as $h$ throughout the following section. One can then write the action as 
\begin{equation}
    \mathcal{S}_J = \int\,d^4 x \sqrt{-\tilde{g}} \left[-\frac{M_P^2}{2}\,\Omega^2\, \widetilde{\mathcal{R}} +\widetilde{\mathcal{L}}_{\phi}  + \widetilde{\mathcal{L}}_{h} + \widetilde{\mathcal{L}}_{N_i} \right]\,,
    \label{eq:jordan}
\end{equation}
where
\begin{align}
& \widetilde{\mathcal{L}}_{\phi} = \frac{1}{2}\,\partial_\mu  \phi\,\partial^\mu \phi-V(\phi)\\
& \widetilde{\mathcal{L}}_{h} =  \partial_\mu  h\,\partial^\mu h^\dagger -V(h h^\dagger) \\
& \widetilde{\mathcal{L}}_{N_i} = \frac{i}{2}\,\overline{\mathcal{N}_i}\,\overleftrightarrow{\slashed{\nabla}}\,\mathcal{N}_i  - \frac{1}{2}\,M_{N_i}\,\overline{(\mathcal{N})^c}_i\,\mathcal{N}_i + \widetilde{\mathcal{L}}_\text{yuk}\\
& \widetilde{\mathcal{L}}_\text{yuk} = -y_{N_i}\,\overline{\mathcal{N}_i}\,\widetilde{h^\dagger}\,\mathbb{L}+\text{h.c.}\,,
\end{align}
where $\mathcal{N}\,,\mathbb{L}$ are the RHN and SM lepton doublet fields in Jordan frame.
The conformal factor $\Omega^2$ is given by
\begin{equation}
    \label{eq:conformalfact}
    \Omega^2\equiv 1 + \frac{\xi_{\phi} \,\phi^2}{M_P^2} + \frac{\xi_{h} \,|h|^2}{M_P^2} \, .
\end{equation}
It is convenient to remove the non-minimal couplings by performing the redefinition of the metric field via the usual conformal transformation from the Jordan frame to the Einstein frame,
\begin{equation}
    g_{\mu\nu} = \Omega^2\,\tilde{g}_{\mu\nu}\,,
\end{equation}
\begin{align}
    \label{eq:einstein0}
    & \mathcal{S}_E= \int d^4 x \sqrt{-g} \Biggl[-\frac{M_P^2\,\mathcal{R}}{2} + \frac{K^{ab}}{2}\,g^{\mu\nu} \partial_{\mu}S_a\,\partial_{\nu}S_b +  \frac{i}{2\,\Omega^3}\,\overline{N_i}\,\overleftrightarrow{\slashed{\nabla}}\,N_i-\frac{1}{\Omega^4}\,\left(\frac{M_{N_i}}{2}\,\overline{N_i^c}\,N_i+\mathcal{L}_\text{yuk}\right)  \nonumber\\&
    -\frac{3i}{4\Omega^4}\,\overline{N_i}\,\left(\overleftrightarrow{\slashed{\partial}}\,\Omega\right)\,N_i-\frac{1}{\Omega^4}\,\left(V_\phi+V_h\right) \Biggr]\,,
\end{align}
where we have used
\begin{equation}
    \sqrt{-\tilde{g}} \rightarrow \frac{\sqrt{-g}}{\Omega^4} 
\end{equation}
\begin{equation}
    \tilde{\slashed{\nabla}} \rightarrow \Omega \slashed{\nabla} -\frac{3}{2}\Omega^2 (\slashed{\partial}\Omega)  \,, 
\end{equation}
and the indices $a,b$ enumerate the fields $\phi$,  and the real components of $h$.
Then making spinor field redefinition $L\rightarrow\Omega^{3/2}L$, $N_i \rightarrow \Omega^{3/2}N_i$ and $\overline{N_i}\rightarrow \Omega^{3/2}\overline{N_i} $
we recover the following action with canonical kinetic term for the RHN
\begin{align}
    \label{eq:einstein}
    & \mathcal{S}_E= \int d^4 x \sqrt{-g} \Biggl[-\frac{M_P^2\,\mathcal{R}}{2} + \frac{K^{ab}}{2}\,g^{\mu\nu} \partial_{\mu}S_a\,\partial_{\nu}S_b -  \frac{1}{\Omega^4}\,\left(V_\phi+V_h\right)+\frac{i}{2}\,\overline{N_i}\,\overleftrightarrow{\slashed{\nabla}}\,N_i 
    \nonumber\\&
    - \frac{1}{2\,\Omega}\,M_{N_i}\,\overline{N_i^c}\,N_i + \frac{1}{\Omega}\,\mathcal{L}_\text{yuk} \Biggr]\,.
\end{align}
The kinetic function is given by
\begin{equation}
    \label{eq:kinfunc}
    K^{ab}  = 6\,\frac{\partial \log \Omega}{\partial S_a} \frac{\partial \log \Omega}{\partial S_b} + \frac{\delta^{ab}}{\Omega^2}\,.
\end{equation}
In what follows, we will be interested in the small-field limit
\begin{equation}
    \label{eq:smallfield}
    \frac{|\xi_{\phi}| \phi^2}{M_P^2},\,\frac{|\xi_{h}| |h|^2}{M_P^2}\ll 1\,.
\end{equation}
Since we necessarily consider large field values for the inflaton coming out of inflation, this constrains $\xi_\phi \ll 1$. 
However, there is no such constraint on $\xi_h$ which can take relatively large values. 

Expanding the kinetic and potential terms in the action in Eq.~\eqref{eq:einstein} in powers of $1/M_P^2$, we obtain a canonical kinetic term for the scalar fields, and deduce the leading-order interactions between scalars and the RHNs induced by the non-minimal couplings. Note that, the non-minimally coupled Yukawa interaction in Eq.~\eqref{eq:einstein} gives rise to a 3-to-2 (or 2-to-3) process with RHN in the final state above electroweak symmetry breaking temperature. These processes thus will be kinematically suppressed and have a subdominant contribution to the RHN yield.  The kinetic terms for the RHNs can be expressed in the form \begin{equation}
    \label{lag4point}
    \mathcal{L}_{\rm{non-min.}} \; = \; -\sigma_{hN_i}^{\xi}\, |h|^2\, \overline{N_i^c}N_i  - \sigma_{\phi N_i}^{\xi}\, \phi^2\,\overline{N_i^c}N_i \,,
\end{equation}
with 
\begin{equation}
    \label{appa:sigphiNi}
    \sigma_{\phi N_i}^{\xi} \; = \; \frac{M_{N_i}}{2M_P^2} \xi_{\phi}
\end{equation}
\begin{equation}
    \label{appa:sighNi}
    \sigma_{h N_i}^{\xi} \; = \; \frac{M_{N_i}}{2M_P^2} \xi_{h}\,.
\end{equation}
These non-minimal interactions open up additional channels~\cite{Clery:2021bwz} 
\begin{itemize}
    \item RHN production from inflaton scattering: $\phi\phi\to N_i\,N_i$ 
    \item RHN production from Higgs scattering: $hh^\dagger\to N_i\,N_i$
    \item Higgs production from  inflaton scattering: $\phi\phi\to hh^\dagger$. 
\end{itemize}
Interestingly, as can be seen from the interaction terms, the production of RHNs is systematically proportional to the mass of the fermion. Then, for the thermal production of RHNs, the production rate is
\begin{align}\label{eq:rTxiN}
&  R^{T,\xi}_{N_i}\simeq N_h\,\frac{\zeta(3)^2\,\xi_h^2}{32\,\pi^5}\,\frac{M_{N_i}^2\,T^6}{M_P^4}\,, 
\end{align}
where $\zeta(x)$ is the Riemann-zeta function. For both minimal and non-minimal gravitational couplings, the leading term in the production rate for scalar dark matter scales as $T^8$ \cite{Clery:2022wib}.
Similarly, the production rate for fermions
in minimal gravity also scales as $T^8$ as seen in Eq.~(\ref{eq:dm-sm-rate}). However, for non-minimal gravitational interactions, 
after the conformal rescaling to obtain canonical kinetic terms, there is no non-minimal coupling to the kinetic terms (in contrast to the scalars where this coupling is found in Eq.~(\ref{eq:einstein})). 
Instead, we are left with only the mass term coupled to $|h|^2$ and the thermal production rate is proportional only to $\propto M_{N_i}^2T^6$.  

Using the rate in Eq.~\eqref{eq:rTxiN} we obtain the number density at the end of reheating due to the non-minimal interaction as
\begin{align}
&  n_{N_i}^{T(\xi_h\neq0)}=  \left(\frac{\sqrt{3}\,N_h\,\zeta(3)^2\,\xi_h^2}{32\,\pi^5\,c_*^{3/2}}\,\frac{M_{N_i}^2\,\rRH}{M_P^3}\right)\,\frac{(k+2)\,\left(1-\left(\frac{\rend}{\rRH}\right)^{\frac{7}{3\,k}-\frac{4}{3}}\right)^{-3/2}}{72\,(5-4 k)\,\Gamma \left(\frac{29-20 k}{14-8 k}\right)}
\nonumber\\&
\times \Biggl[9 \sqrt{\pi } (5-4 k)\left(\frac{\rend}{\rRH}\right)^{1/k} \Gamma \left(\frac{4k-4}{4k-7}\right)
+ 4\,\left(\frac{\rend}{\rRH}\right)^{\frac{16 k^2+4 k+169}{21 k-12 k^2}}\,\Gamma \left(\frac{29-20 k}{14-8 k}\right)\,\mathcal{G}\Biggr]\,, 
\label{n-nonmin}
\end{align}
with
\begin{align*}
& \mathcal{G} = \left(\frac{\rend}{\rRH}\right)^{\frac{4 (k+30)}{3 k (4 k-7)}}\,\left[3(4k-5) \left(\frac{\rend}{\rRH}\right) ^{\frac{16 k^2+49}{3 k (4 k-7)}} - 6 \left(\frac{\rend}{\rRH}\right) ^{\frac{56}{3 (4 k-7)}}\right]\,{}_{2}F_{1}(...)\,,
\end{align*}
where ${}_{2}F_{1}\left(-\frac{1}{2},\frac{3}{4 k-7},\frac{4k-4}{4 k-7},\left(\frac{\rend}{\rRH}\right)^{\frac{7}{3 k}-\frac{4}{3}}\right)$ is the hypergeometric function.

For the inflaton scattering process $\phi\phi\rightarrow N_i\,N_i$, on the other hand, we find 
\begin{eqnarray}\label{eq:rphin}
    R^{\phi,\xi}_{N_i} &=& \frac{M_{N_i}^2\xi_{\phi}^2\phi_0^4\omega^2}{32 \pi M_P^4} \sum\limits_{n=1}^{\infty}(2n)^2|\mathcal{Q}^{(2)}_{2n}|^2\times  \sqrt{1-\frac{4M_{N_i}^2}{E_{2n}^2}}\,,
\end{eqnarray}
where we define $\phi_0=\left(\frac{\rho_\phi}{\lambda\, M_P^{4-k}}\right)^\frac{1}{k}$ and ${\cal Q}^{(2)}_n$ by
\begin{equation}
    \label{Eq:oscillation_2}
    \phi^2(t)=\phi_0^2(t)\cdot\mathcal{Q}^{2}(t)= \phi_0^2(t)\sum_{n=-\infty}^{\infty}\,{\cal Q}^{(2)}_n e^{-in \omega t}\,.
\end{equation} 
This rate is restricted by the small field limit that imposes a stringent bound on $\xi_{\phi}$ from  $\sqrt{|\xi_{\phi}|} \lesssim M_P/\langle \phi\rangle$. In particular, at the beginning of inflaton oscillations $\langle \phi \rangle \sim M_P$ and $| \xi_\phi | \lesssim 1$. When we compare this rate with the one due to inflaton scattering mediated by minimal gravitational interactions (Eq.~\eqref{eq:rateferm}), we obtain \begin{equation}
\frac{R^{\phi,\xi}_{\phi N_i}}{R^{\phi^k}_{N_i}} \approx \left[\frac{k\,(k-1)\,\xi_{\phi}\,(\omega/m_\phi)^2}{\sqrt{8}}\right]^2\,\frac{\,\sum\limits_{n=1}^{\infty}(2n)^2|\mathcal{Q}^{(2)}_{2n}|^2}{\sum\limits_{n=1}^{\infty}\frac{1}{(2n)^2}\,\left|\mathcal{P}^k_{2n}\right|^2}\,.
\end{equation} 
This ratio takes values between $\{32\, \xi_{\phi}^2, \, 242\, \xi_{\phi}^2\}$ for $k\in[6,20]$. Hence, the non-minimal contribution from inflaton scattering dominates over the graviton exchange for $\xi_{\phi}>\frac{1}{2\sqrt{8}}$ when $k=6$ and for $\xi_{\phi}>0.06$ when $k=20$. In what follows, we will neglect this contribution as it dominates for values of $\xi_{\phi}$ close to the small field limit, making the assumption of canonical kinetic terms of the fields invalid. 

The non-minimal coupling of Higgs bosons to gravity provides an additional channel to reheat the Universe through gravitational processes, with the following rate~\cite{Co:2022bgh}
\begin{equation}
   (1+\omega_\phi)\Gamma_\phi= R^{\phi,\xi}_H \simeq \frac{\xi_h^2N_h}{4\pi M_P^4} \sum_{n=1}^{\infty}  2n\omega \left|2\times{\mathcal{P}}^k_{2n}\rho_{\phi} + \frac{(n\omega)^2}{2}\phi_0^2|\mathcal{Q}_n|^2 \right|^2 = \alpha_k^{\xi} M_P^5 \left(\frac{\rho_{\phi}}{M_P^4}\right)^{\frac{5k-2}{2k}}\, ,
   \label{ratexi}
\end{equation}
where $\mathcal{Q}_n$ has been defined in Eq.~(\ref{Eq:oscillation})
and $\alpha_k^{\xi}$ is given in Table~\ref{Tab:tablealphak}.
If we solve Eq.~(\ref{eq:BErho}) for $\rho_R$,
the reheating temperature in the presence of the non-minimal coupling is then given by
\begin{align}\label{eq:non-minimal-trh}
& \left(\Trh^{\xi}\right)^4 = \frac{30}{\pi^2\,g_{\rm RH}}\, M_P^4\,\left(\frac{\rend}{M_P^4}\right)^{\frac{4k-7}{k-4}}\,\left(\frac{\alpha_k^{\xi}\,\sqrt{3}\,(k+2)}{8k-14}\right)^\frac{3k}{k-4}\,.  
\end{align}
The reheating temperature as a function of $k$ is shown in the left panel of Fig.~\ref{fig:tmax/trh} for several values of $\xi_h$.
The maximum temperature in this case is determined from 
\begin{align}\label{eq:tmax-xi}
& \rho_\text{max}^\xi\simeq \sqrt{3}\,\alpha_k^{\xi}\,M_P^4\, \left(\frac{\rend}{M_P^4}\right)^\frac{2k-1}{k}\,\frac{k+2}{12k-16}\,\left(\frac{2k+4}{6k-3}\right)^\frac{2k+4}{4k-7} \equiv c_*\,(\Tmax^{\xi})^4 \,.
\end{align}
Contours of $\Tmax^\xi$ in the $(k, \Trh)$ plane are shown in the right panel of Fig.~\ref{fig:tmax/trh}, where the appropriate values of $\xi_h$ taken from the left panel are used to calculate $\Tmax^\xi$. 

%%%%%%%%%%%%%%%%%%%%%%%%%
\subsection{Gravitational waves generated during inflation}
\label{sec:grav-wave}
%%%%%%%%%%%%%%%%%%%%%%%%%

In this section, we review the calculation of gravitational waves generated by quantum fluctuations during inflation, followed by a cosmological era where the the inflaton energy dominates and redshifts faster than radiation. This results in an enhancement of gravitational waves, which places a constraint from excessive gravitational waves as dark radiation and offers a gravitational wave signal with a distinctive spectrum. 

The ratio of the gravitational wave (GW) energy density to that of the radiation bath is given by~\cite{Saikawa:2018rcs}
\begin{align}
\frac{\rho_{\rm GW}}{\rho_R} = \frac{1}{32 \pi G \rho_R} \frac{k_{\rm GW}^2}{2} \mathcal{P}_T(k_{\rm GW})  ~~~{\rm with}~~~ \mathcal{P}_T(k_{\rm GW}) \equiv \frac{2 H_I^2(k_{\rm GW})}{\pi^2 M_P^2} \,,
\end{align}
where $k_{\rm GW}$ is the momentum mode of the GW, $\mathcal{P}_T$ is the primordial tensor power spectrum, $H_I(k_{\rm GW})$ is the Hubble scale during inflation when the mode $k_{\rm GW}$ exits the horizon, $T_{\rm hc}$ is the horizon-crossing temperature when the mode re-enters the horizon at $k_{\rm GW} = H(T_{\rm hc})$, and the factor of $1/2$ accounts for the time average of the rapidly oscillating metric perturbations. In our case, $\rho_{\rm GW}/\rho_R$ is redshift invariant up to the change of $g_*$ after $T=\Tmax$ because entropy is only efficiently produced at $T=\Tmax$ as discussed in Sec.~\ref{sec:grav-reheat}. Therefore, the final gravitational wave strength is given by $\Omega_{\rm GW} h^2 = \Omega_\gamma\,h^2\left(\rho_{\rm GW}/\rho_R\right)\times \left[g_{*s}^4({\rm eV})/g_*(T_{\rm hc}) g_*^3({\rm eV}) \right]^{\frac{1}{3}}$ where $\Omega_\gamma = \rho_{\gamma, 0}/\rho_{{\rm crit}, 0}$ is the fraction of the photon energy density today. Here $g_{*s}({\rm eV}) = \left[2+\frac{7}{8} \times 2 \times 3 \times\left(\frac{4}{11}\right)\right] \simeq 3.91$ and $g_*({\rm eV}) = \left[2+\frac{7}{8} \times 2 \times 3 \times\left(\frac{4}{11}\right)^{4 / 3}\right] \simeq 3.36$ denote the effective number of relativistic degrees of freedom relevant for the entropy density and the energy density, respectively.

\begin{figure}[htb!]
\centering
\includegraphics[width=0.495\linewidth]{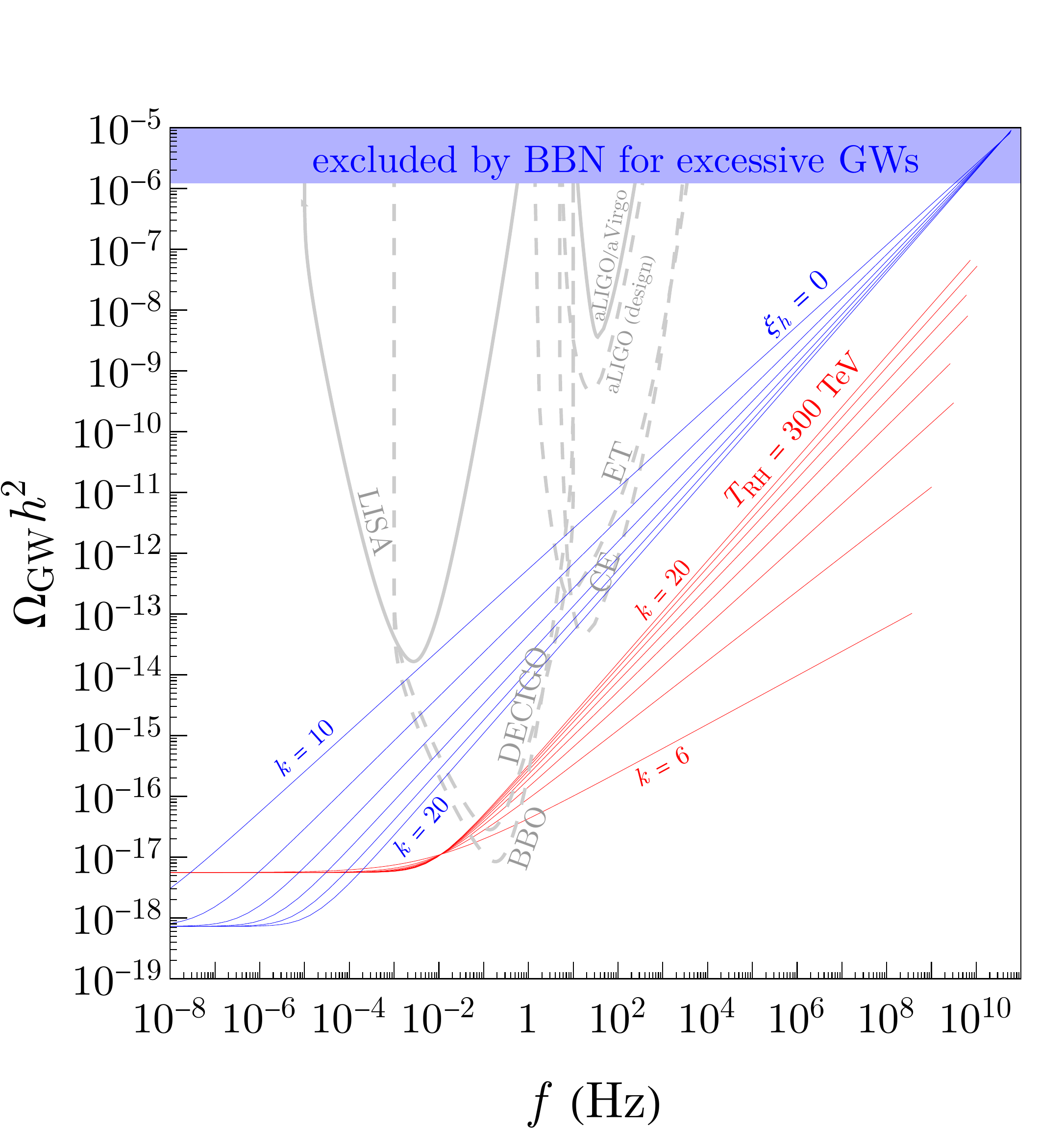}
\includegraphics[width=0.495\linewidth]{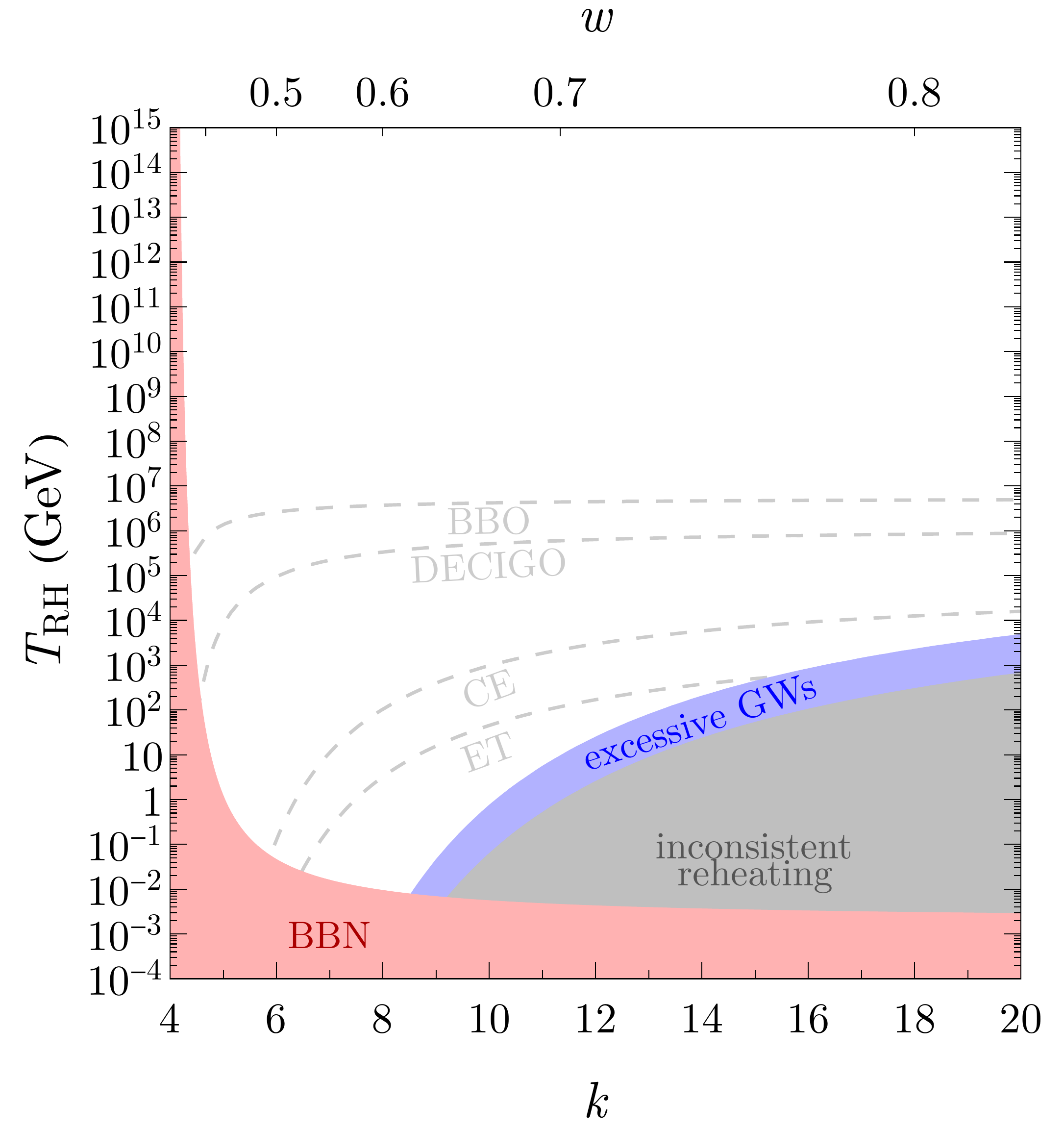}
\caption{\em Gravitational wave constraints and future prospects. Left: The blue-shaded region is excluded by BBN for excessive dark radiation. The blue and red curves fix $\xi_h = 0$ and $\Trh = 300 \tev$, respectively. Various curves of the same color use different values of $k$ as labeled and in increments of 2. The sensitivity of several future experiments as a function of frequency is also shown. Right: The blue region is excluded by BBN for the excessive GW energy as dark radiation. The regions below the gray dashed curves can be probed by the GW experiments as specified.}
\label{fig:GWs}
\end{figure}

As one can see, if horizon crossing occurs during radiation domination $k_{\rm GW}^2 = H^2(T_{\rm hc}) = \rho_R / (3 M_P^2)$, then the GW spectrum becomes scale invariant. On the other hand, if horizon crossing occurs during the inflaton-dominated era, the GW strength is enhanced by a factor of $\rho_\phi / \rho_R$ evaluated at $T_{\rm hc}$. As a result, the largest enhancement is for the mode that re-enters the horizon right after inflation at $\Tmax$. For minimal gravitational reheating $(\xi_h = 0)$, the enhancement in this mode is $\rend / \rho_R(\Tmax) \simeq (4-6) \times 10^{13}$ for $k\in[6,20]$, based on Eqs.~(\ref{eq:Vatt}) and (\ref{phiend}). This gives $\Omega_{\rm GW} h^2 \simeq (8-10) \times 10^{-6}$, which corresponds to the high frequency points of the blue curves, which fix $\xi_h = 0$, in the left panel of Fig.~\ref{fig:GWs}. These values are excluded by the BBN bound of $\Omega_{\rm GW} h^2 \simeq 1.3 \times 10^{-6}$~\cite{Yeh:2022heq}, shown by the blue-shaded region at the top. Therefore, the case with minimal gravitational interactions is excluded, which was previously pointed out by Ref.~\cite{Figueroa:2018twl}. The constraint is relaxed when $\Tmax$ is increased, e.g., by non-minimal gravitational interactions via $\xi_h$ [cf.~Eq.~\eqref{eq:tmax-xi}], because the GW energy density relative to that of radiation is smaller in this case. The blue region in the right panel of Fig.~\ref{fig:GWs} (and subsequent figures) shows the constraint in this non-minimal scenario, which excludes $\xi_h \lesssim 0.5$, as can be seen from the left panel of Fig.~\ref{fig:tmax/trh}. Alternatively, Ref.~\cite{Opferkuch:2019zbd} offers the solution where the inflaton energy is more efficiently transferred to radiation via a tachyonic growth of a new field.

%%%%%%%%%%%%%%%%%%%%%%%%%%%%%%%%%%%%%%%%%%%%%
%%%%%%%%%%%%%%%%%%%%%%%%%%%%%%%%%%%%%%%%%%%%%

In addition to setting a constraint, such enhanced gravitational waves offer an exciting signature to search for~\cite{Giovannini:1998bp}. The amount of enhancement depends on $\rho_\phi/\rho_R$ at the time of horizon crossing, implying that the GW spectrum depends on $k$ via $\rho_\phi$.
By analyzing modes that re-enter the horizon after $\Tmax$ and using $\rho_\phi \propto a^{-6k/(k+2)}$ from Eq.~(\ref{eq:rho_phi_a}), we find the GW spectrum scales with the frequency as $\Omega_{\rm GW}h^2 \propto f^{\frac{k-4}{k-1}}$, which is consistent with Ref.~\cite{Figueroa:2018twl}. 
The enhanced GW spectra are demonstrated in the left panel of Fig.~\ref{fig:GWs} for the different values of $k$ in the blue and red curves. The blue (red) curves correspond to the minimal scenario ($\Trh = 300 \tev$), and allow for $k$ to vary from 10 (6) to 20 in increments of 2 (for $\xi_h = 0$, $k < 10$ is excluded by BBN for low $\Trh$ according to Fig.~\ref{fig:tmax/trh}.) Here, the frequency is obtained by redshifting the initial momentum mode at $T_{\rm hc}$ to today's photon temperature $T_{\gamma,0}$ as
\begin{align}
f = \frac{k_{\rm GW}}{2\pi}  \frac{T_{\gamma, 0}}{T_{\rm hc}}  \left( \frac{g_*({\rm eV})}{g_*(T_{\rm hc})} \right)^{\frac{1}{3}}.
\end{align}
Therefore, by measuring the slope of $\Omega_{\rm GW}h^2$ as a function $f$, one can determine $k$ and thus reveal the shape of the inflaton potential energy near the minimum. Note that, the end-point frequencies for the red curves are different for different choices of $k$ (and different $\xi_h$), as for a given $k$ and $\Trh$, the maximum possible frequency is dictated by
\begin{align}
& f_\text{max} = \frac{H(\Tmax)}{2\pi}\,\frac{\aend}{a_0}\,,
\end{align} 
which for $\xi_h=0$ turns out to be $\simeq 7 \times 10^{10}$ Hz for all $k$, while modes with frequencies $f>f_\text{max}$ are never produced.  
In the right panel of Fig.~\ref{fig:GWs}, the regions below the gray dashed curves can be probed by the future gravitational wave observatories---BBO~\cite{Crowder:2005nr,Corbin:2005ny,Harry:2006fi}, DECIGO~\cite{Seto:2001qf,Kawamura:2006up,Yagi:2011wg}, CE~\cite{LIGOScientific:2016wof,Reitze:2019iox} and ET~\cite{Punturo:2010zz, Hild:2010id,Sathyaprakash:2012jk, Maggiore:2019uih}. Here we use the sensitivity curves derived in Ref.~\cite{Schmitz:2020syl}. Since these GW observatories probe frequencies that correspond to modes that exit the horizon early in the inflation period, we use the large-field asymptotic value of $V(\phi)$ in Eq.~(\ref{eq:Vatt}) to obtain $H_I$. In the left panel of Fig.~\ref{fig:GWs}, we illustrate the GW spectra in the red curves for a fixed $\Trh = 300 \tev$, which can be detected by DECIGO for $k \ge 8$ and by BBO for all $k \ge 6$. Remarkably, in the right panel, a large region in the parameter space with $\Trh < 5 \times 10^6 \gev$ can be probed by future GW detectors. We emphasize that this potential GW signal is generic for our model and applicable throughout this work, although we do not show these sensitivity curves in subsequent figures for clarity of presentation.
% 
%%%%%%%%%%%%%%%%%%%%%%%
\section{Results and discussion}
\label{sec:result}
%%%%%%%%%%%%%%%%%%%%%%%%
\subsection{A stable DM candidate}
%%%%%%%%%%%%%%%%%%%%%%%

As we have seen from the previous two subsections, for each value of $k$ and $\xi_h$ (in the non-minimal case), there is a unique value for $\Trh$. These are shown in the left panel of Fig.~\ref{fig:tmax/trh}. The gravitational thermal production of DM generally requires reheating temperatures much larger than can be obtained with $\xi_h = 0$. In this section, we will consider $\Trh$ and $k$ as free parameters and it should be understood that we are implicitly assuming that $\xi_h \ne 0$ and takes the necessary value to achieve a particular reheating temperature for a given value of $k$. For the production of DM, both minimal and non-minimal thermal contributions are included, whereas for the generation of a lepton asymmetry only minimal contributions from inflaton scattering are considered. 

\begin{figure}[htb!]
\centering
\includegraphics[width=0.44\linewidth]{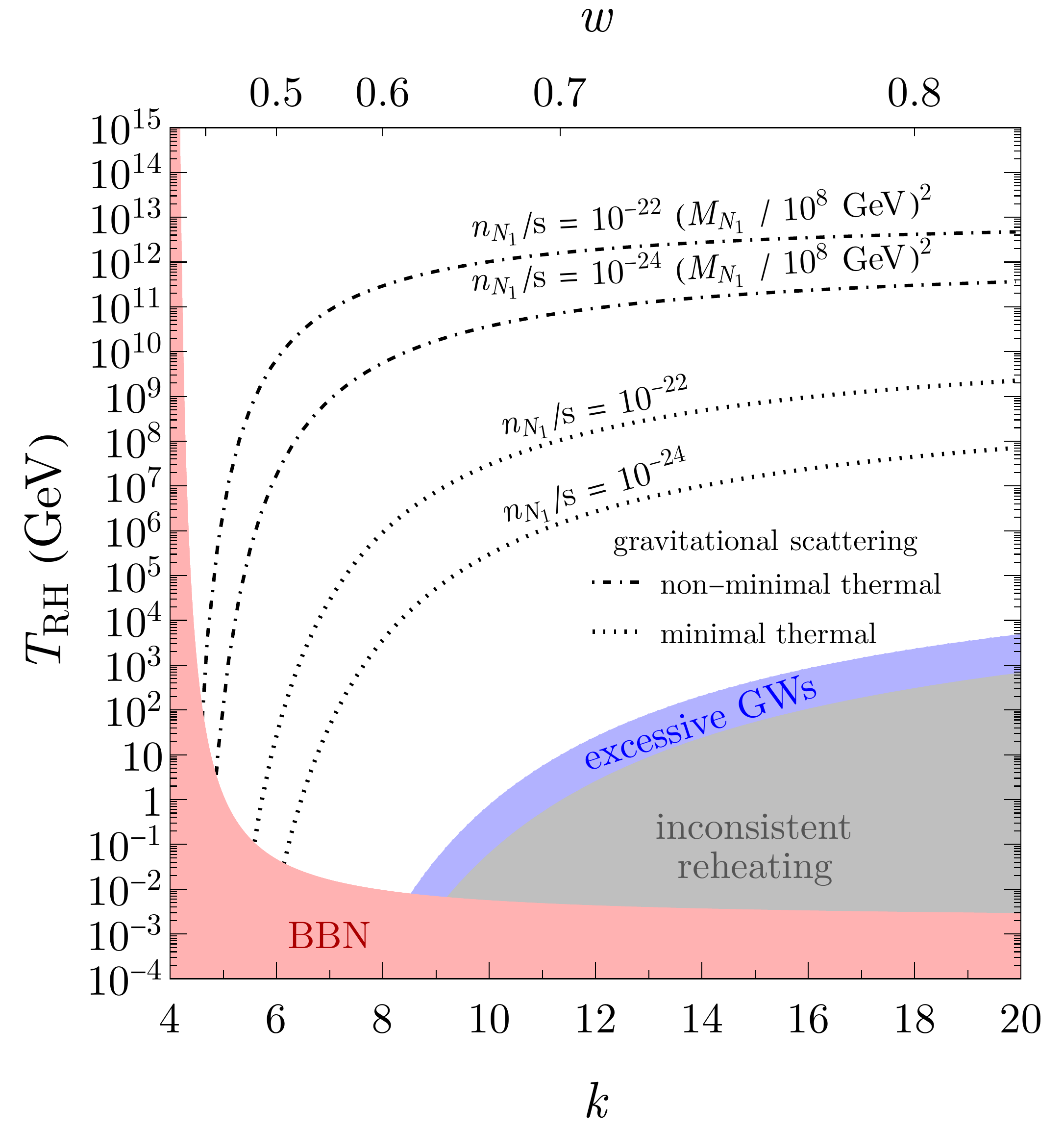}
\includegraphics[width=0.44\linewidth]{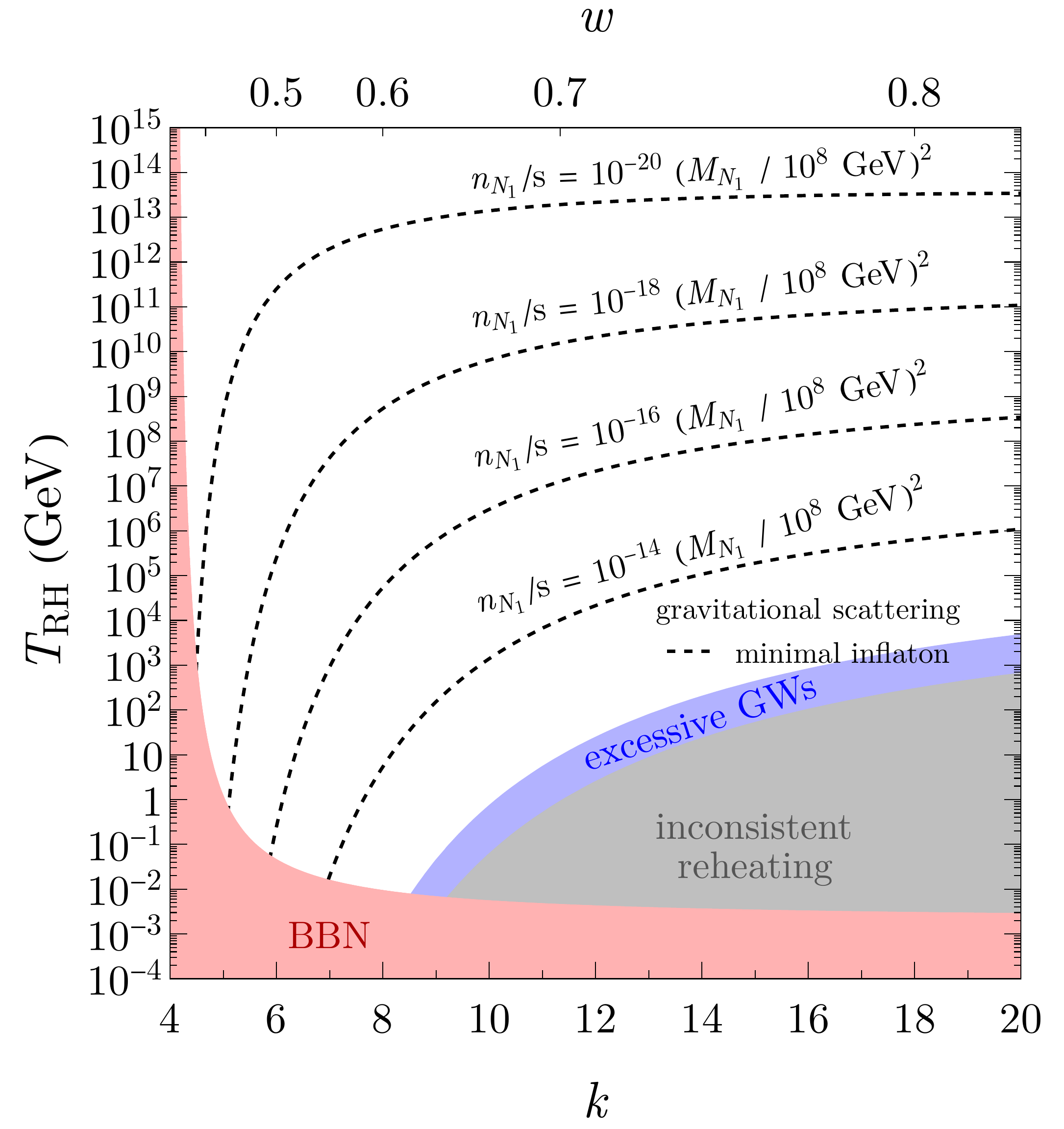}~\\[10pt]
\includegraphics[width=0.44\linewidth]{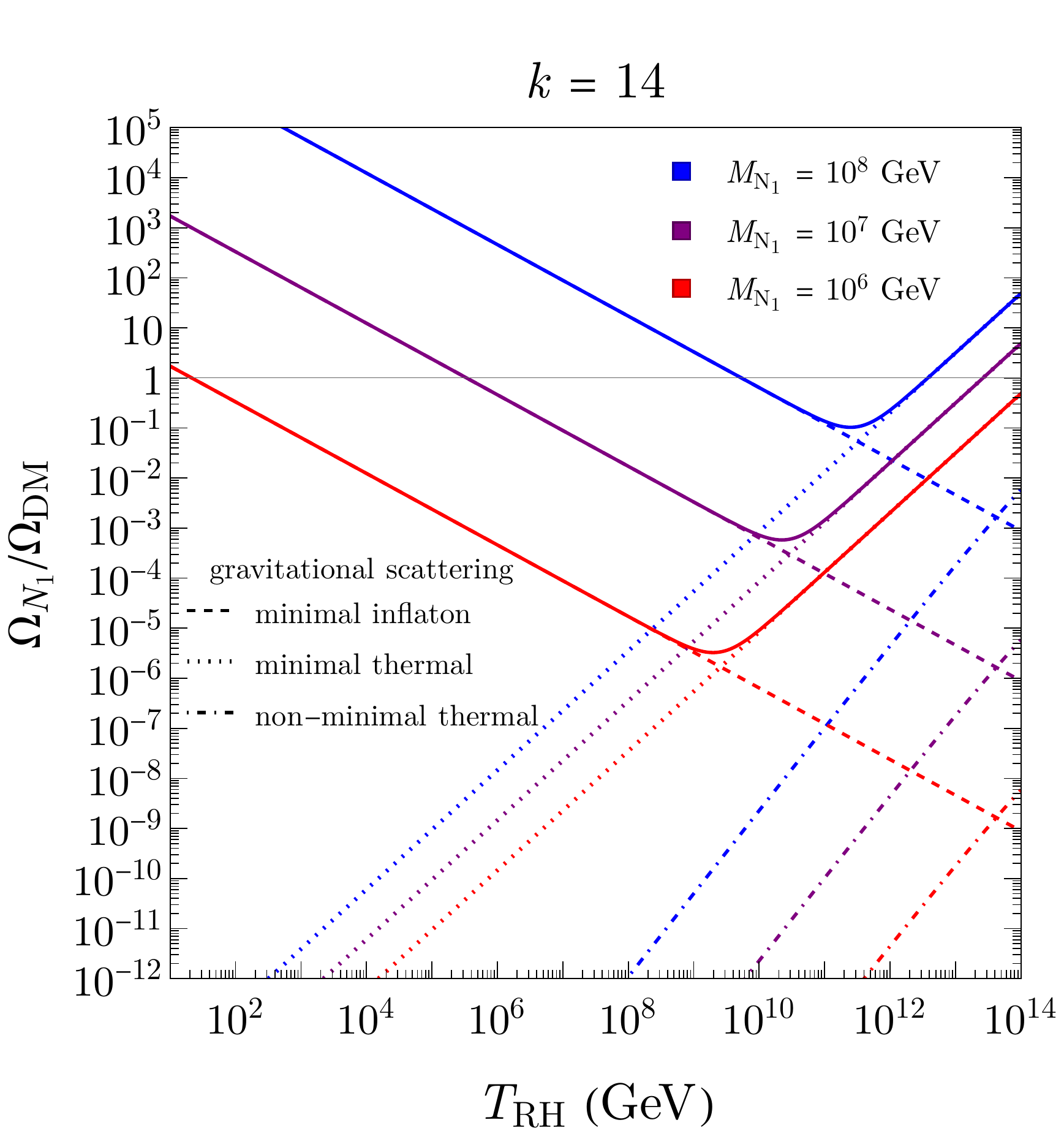}
\includegraphics[width=0.44\linewidth]{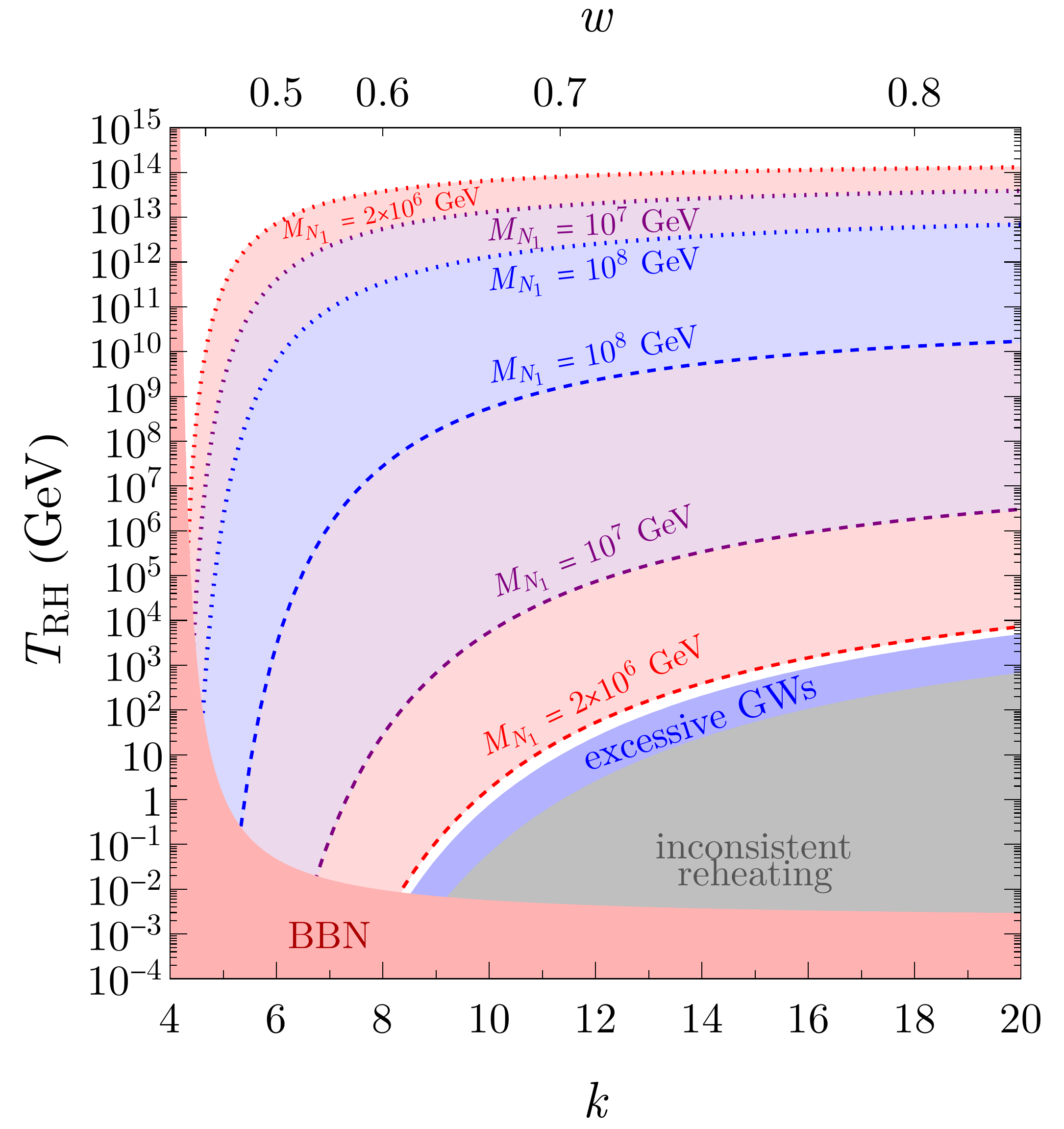}
\caption{\em Top left: Contours of fixed comoving number density, $n_{N_1}/s = 10^{-22}$ and $10^{-24}$ in the $(k, \Trh)$ plane. $M_{N_1} n_{N_1} / s \simeq 0.44$~eV is needed to explain the observed dark matter density, $\Omega_{N_1} h^2 = 0.12$. Dotted curves assume DM production solely from minimal gravitational scattering in the thermal bath. Dot-dashed curves correspond to non-minimal gravitational scatterings. The latter are scaled with $M_{N_1}^2$. Top right: Contours of $n_{N_1}/s = 10^{-20}, 10^{-18}, 10^{-16}$ and $10^{-14}$ each scaled with $M_{N_1}^2$ assuming DM production only from inflaton scattering. In both upper panels, the gray-shaded region is excluded as minimal gravitational interactions necessarily produce larger reheating temperatures. Low reheating temperatures shaded in red (blue) are excluded by BBN for an excessive inflaton (GW) energy. Bottom left: The total relic abundance $(\Omega_{N_1}^T+\Omega_{N_1}^{\phi^k}) h^2/0.12$ as a function of reheating temperature for three choices of DM masses $\{10^6$,\,$10^7$,\,$10^8\}$ GeV for fixed $k=14$. Individual contributions to the dark matter density are distinguished by line types as indicated. Bottom right: Coloured regions correspond to values of $(k,\Trh)$ with $(\Omega_{N_1}^T+\Omega_{N_1}^{\phi^k})h^2 \le 0.12$ for the three choices of $M_{N_1}$ used in the bottom left panel, and the lines styles indicate the dominant contribution.
}
\label{fig:thermrelic}
\end{figure}

The results presented in this section depend on the underlying class of inflationary models. As noted earlier, we consider T-models of inflation \cite{Kallosh:2013hoa}  for which we have determined $\lambda$ and $\rend$. As discussed above, there are two contributions to the DM relic density: from gravitational scattering within the newly formed primordial plasma and directly from inflaton scattering. These two contributions are presented separately in the upper two panels of Fig.~\ref{fig:thermrelic}. In the upper left panel, we show two contours of the yield, $n_{N_1}/s = 10^{-22}$ and $10^{-24}$, for both minimal gravitational interactions (dotted curves) using Eq.~(\ref{eq:nT-grav}) and non-minimal interactions (dot-dashed curves) using Eq.~(\ref{n-nonmin}).\footnote{Note that minimal gravitational interactions ($\xi_h = 0$) are not actually possible at these reheating temperatures which require $\xi_h \ne 0$. } Note the latter yield is proportional to $M_{N_1}^2$ as shown by the contour labels, and we have normalized these contours by choosing $M_{N_1} = 10^8$ GeV. Also note, $M_{N_1} n_{N_1} / s \simeq 0.44$~eV is needed to explain the observed dark matter density, $\Omega_{N_1} h^2 = 0.12$. For $k>4$ and minimal gravitational interactions, the relic density increases with reheating temperature, $n_{N_1}/s \sim \Trh^{\frac{5k-20}{3k}}$. The scaling of $n_{N_1}/s$ for non-minimal interactions is more complicated but also increases with $\Trh$.

%%%%%%%%%%%%%%%%%%%%%%%%%%%%%%%%%%%%%
%%%%%%%%%%%%%%%%%%%%%%%%%%%%%%%%%%%%%%%

In the upper right panel of Fig.~\ref{fig:thermrelic}, we provide four contours of the yield, $n_{N_1}/s$, produced from inflaton scattering, which also scales as $M_{N_1}^2$. The gravitational production process from inflaton scattering is complementary to the thermal production process just discussed. Recall that we are assuming $\xi_\phi$ is small enough that non-minimal scattering processes can be ignored. In this case,  from Eq.~\eqref{Eq:num-den-inf}, we see that $n_{N_1}/s\sim  \Trh^{-1+\frac{4}{k}}$ and for $k>4$, the relic density decreases with increasing $\Trh$. Recalling that $M_{N_1} n_{N_1} / s \simeq 0.44$~eV is needed to explain the observed DM density, $\Omega_{N_1} h^2 = 0.12$, we find indeed that higher reheating temperatures require 
{\it lighter} DM candidates to fit with the relic abundance constraint.

Combining the two constraints shown in the top panels of Fig.~\ref{fig:thermrelic} we see that for a given $k$ and $M_{N_1}$, there are both upper (from thermal scattering) and lower (from inflaton scattering) limits to $\Trh$ so as to avoid exceeding the observed cold DM abundance. 
The resulting relic density as a function of $\Trh$ is shown in the bottom left panel of Fig.~\ref{fig:thermrelic}, where we show the total relic abundance $(\Omega_{N_1}^T+\Omega_{N_1}^{\phi^k})\,h^2$ relative to the observed abundance for a fixed $k=14$ and three choices of the DM mass $M_{N_1}=\{10^6$,\,$10^7$,\,$10^8\}$ GeV.
We clearly see that the desired relic density $(\Omega_{N_1} = 0.12)$ is obtained {\it twice}: (i) at a lower reheating temperature, where inflaton scattering dominates, and (ii) for a higher reheating temperature, when we are in the thermal production regime. The allowed region corresponds to the parameter space at or {\it below} the line $\Omega h^2/ 0.12 = 1$ in the bottom left  panel of Fig.~\ref{fig:thermrelic}. For $M_{N_1} > 3\times 10^8$ GeV, there are no values of ($\Trh$, $k$) that result in an acceptable density of DM, and the allowed range in $\Trh$ is larger with lighter DM. This is understandable as, the thermal relic requiring a {\it larger upper bound} on $\Trh$ for lighter DM, while the inflaton scattering requires
a {\it smaller lower bound} on $\Trh$ for lighter DM. 

A two-dimensional version of the lower left panel of Fig.~\ref{fig:thermrelic}, over a range in $k$, is shown in the lower right panel. Low values of $\Trh$ are excluded by BBN. Once again, the gray-shaded region in the lower right corner of this panel is also excluded since minimal gravitational interactions would produce a reheating temperature larger than the values in that region. Within each shaded band (the color corresponds to a specific choice of $M_{N_1}$), the total relic density is below the observed DM density. The observed value is reached on the border of the colored bands. For DM of masses very close to $1$ PeV, there exists a viable parameter space for $k \ge 9$ (along the boundary of the excessive GWs region), requiring $\xi_h \simeq 0.5$.  For larger masses, the range in $k$ extends to lower values, and higher reheating temperatures are possible and require larger non-minimal coupling to gravity.

\begin{figure}[htb!]
\centering
\includegraphics[width=0.495\linewidth]{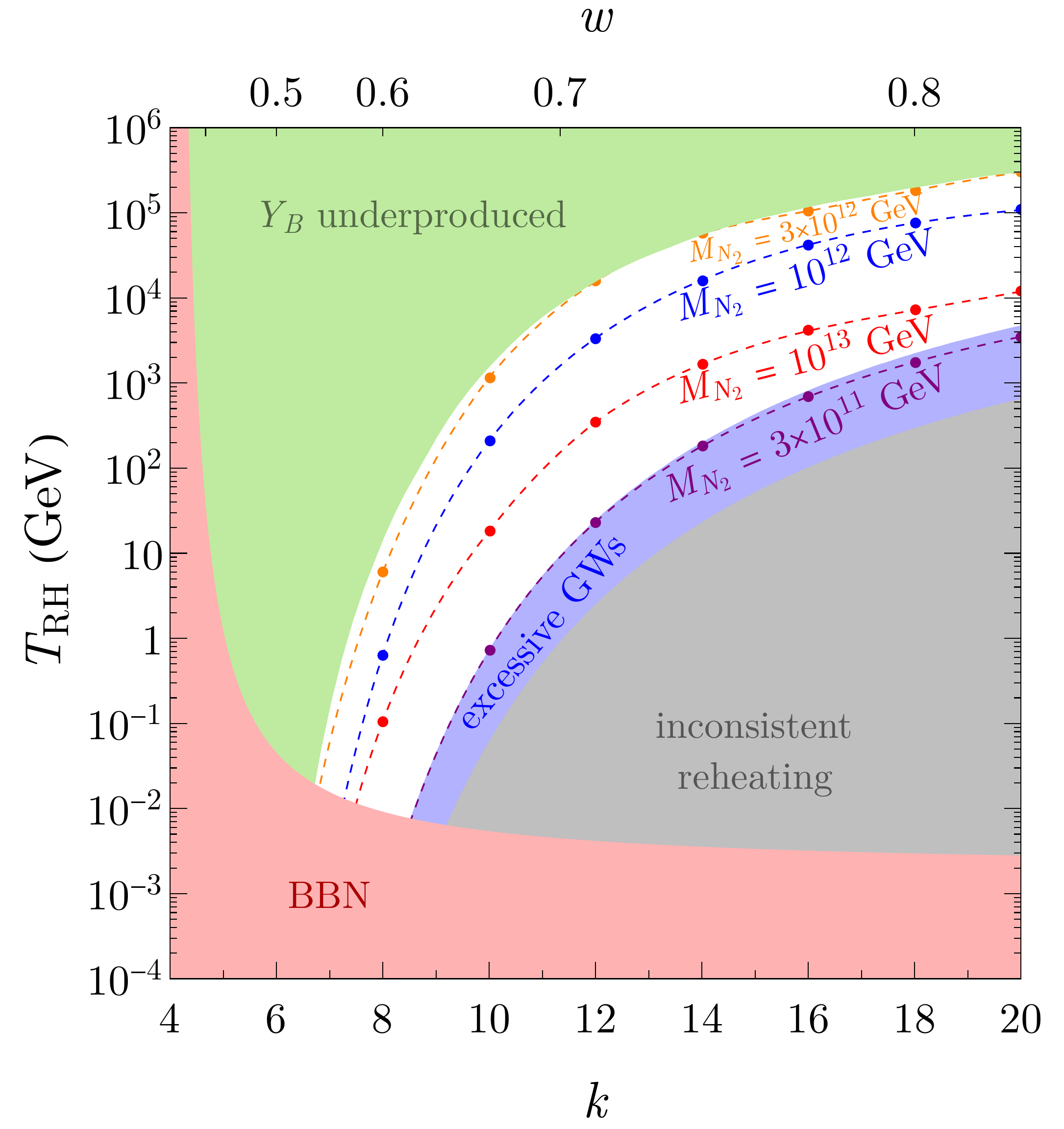}
\caption{\em Contours of $M_{N_2}$ corresponding to the observed baryon asymmetry [cf.~Eq.~\eqref{Eq:baryonassym}] in the $(k,\Trh)$ plane. The red-shaded region correspond to the lower bound on $\Trh$ from BBN, and the green region leads to underproduction of $Y_B$ due to the kinematic suppression in inflaton scattering when $M_{N_2}$ approaches $m_\phi$.
}
\label{fig:asym}
\end{figure}

Having identified the regions of the $(k,\Trh)$ parameter space with a suitable DM density, we turn to the production of the baryon asymmetry through gravitationally induced leptogenesis.  This analysis was performed in
\cite{Co:2022bgh} and therefore we only briefly summarize the results found there. We note, however, Ref.~\cite{Co:2022bgh} neglected the kinematic suppression in Eq.~(\ref{eq:ratefermion2}) to maintain the model independence of the analysis, though this effect is included in the present work. In Fig.~\ref{fig:asym}, we show contours of some benchmark values of the mass of $N_2$ that reproduce the observed baryon asymmetry $Y_B^\text{obs}$. We find that the gravitational contribution to the baryon asymmetry is essentially entirely due to inflaton scattering rather than the thermal particles in the SM bath.  
Since minimal gravitational interactions are excluded by excessive GWs, non-minimal interactions are required to produce a sufficiently large thermal bath so that GW fractional energy is consistent with BBN. Leptogenesis via $N_2$ is therefore possible above the border of the blue-shaded region in Fig.~\ref{fig:asym}, indicating a mass $M_{N_2} \gtrsim 3 \times 10^{11}$ GeV is required. 
Larger values of $M_{N_2}$ can produce the correct asymmetry so long as $\xi_h > 0$. Nonetheless, when $M_{N_2} \gtrsim 3 \times 10^{12} \gev$, the baryon asymmetry starts to become suppressed for the following reason. The inflaton mass obtained from Eqs.~(\ref{eq:Vatt}) and (\ref{phiend}), $m_\phi \simeq 1.2 \times 10^{13} \gev$ across all $k$ values, is no longer much larger than $M_{N_2}$ and the kinematic suppression in Eq.~(\ref{eq:ratefermion2}) becomes important. This explains the existence of the green region as well as why the curve for $M_{N_2} = 10^{13} \gev$ is at a lower $\Trh$ than that for $M_{N_2} = 3 \times 10^{12} \gev$.
Once again, the bottom red region is forbidden by BBN because of an excessive inflaton energy density during BBN. 
In summary, we observe that, saturating the bound on GWs from BBN, together with the right DM abundance and successful leptogenesis requires $\xi_h \simeq 0.5$, $M_{N_2}\simeq 3 \times 10^{11}$ GeV and $M_{N_1}\simeq 10^6$ GeV. As discussed above, this parameter space can be extended, allowing larger values $\{M_{N_1},M_{N_2}\}$  if one considers stronger non-minimal gravitational couplings by $\xi_h \gtrsim 0.5$, thus allowing a larger reheating temperature [cf.~Eq.~\eqref{eq:non-minimal-trh}]. 

%%%%%%%%%%%%%%%%%%%%%%%%%%%%%%%%%%%%%%
%%%%%%%%%%%%%%%%%%%%%%%%%%%%%%%%%

\begin{figure}[htb!]
\centering
\includegraphics[width=0.495\linewidth]{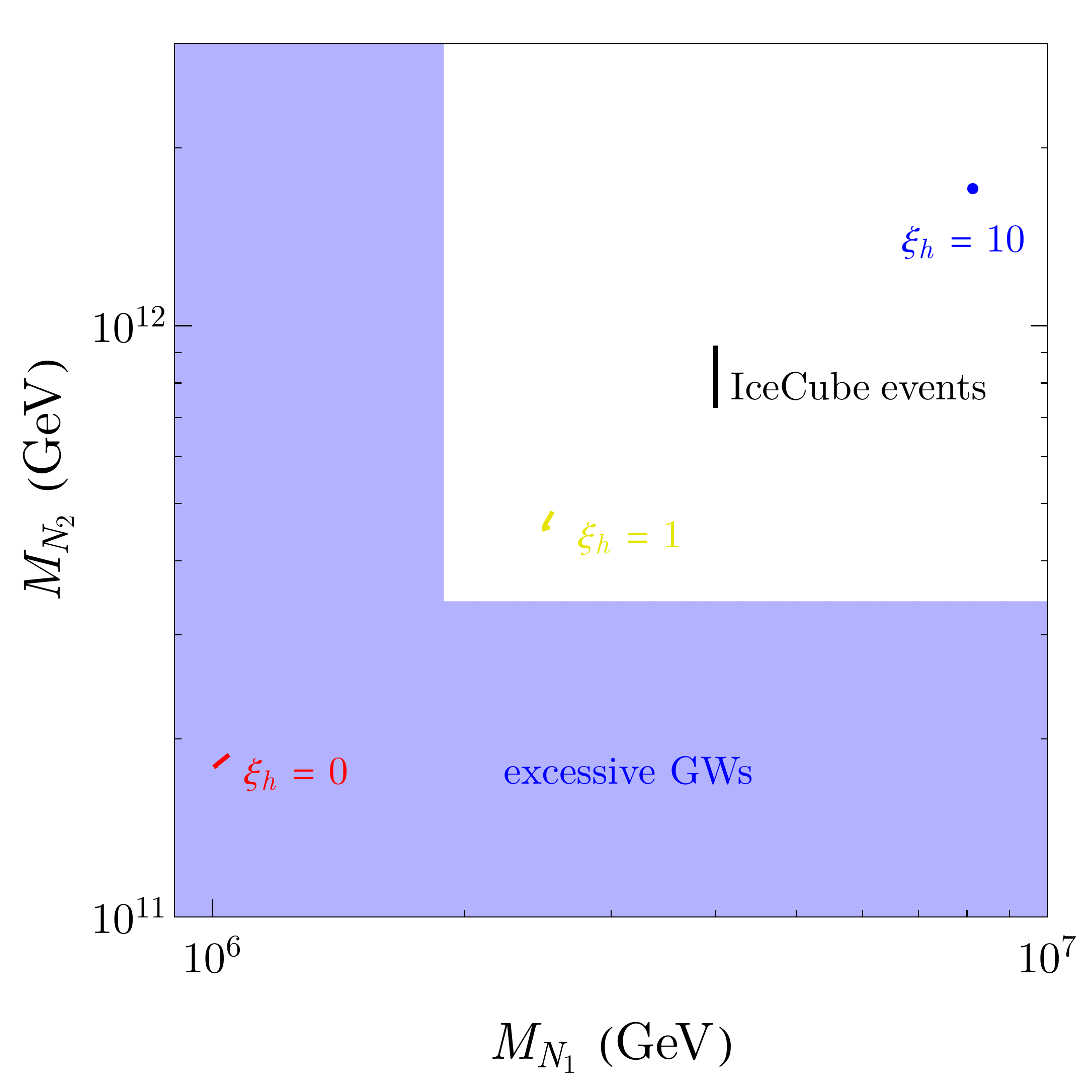}~~
\includegraphics[width=0.495\linewidth]{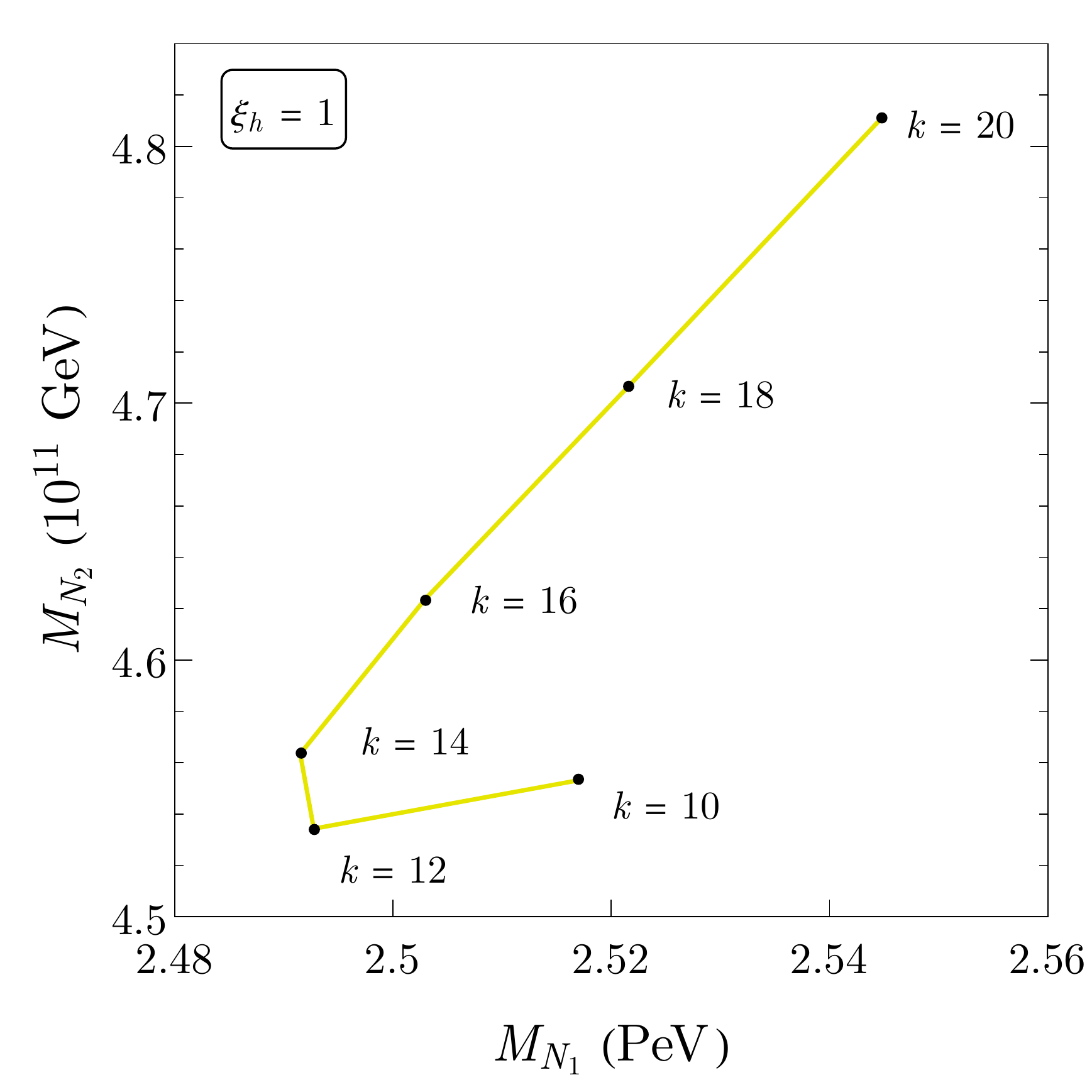}
\caption{\em Viable parameter space in the ($M_{N_1}, M_{N_2}$) plane for which gravitational interactions are responsible for  
the observed DM relic abundance (in $N_1$), the baryon asymmetry (produced from $N_2$ decays), and reheating for $k\in [6,\,20]$. In the left panel, different colours correspond to $\xi_h=\{0,\,1,\,10\}$ diagonally from bottom left (red) to top right (blue). The vertical black segment indicates the range in $M_{N_2}$ for $M_{N_1} = 4$ PeV for the range in $k$ considered, where the connection to the IceCube high-energy neutrino excess will be discussed in the next subsection. In the right panel, we magnify the parameter space for a fixed non-minimal coupling $\xi_h=1$. The dots correspond to even values of $k$ as indicated.
}
\label{fig:allowed}
\end{figure}

Combining our preceding analyses, it is possible, for a given $V(\phi)$, to constrain the ($M_{N_1}$, $M_{N_2}$, $\xi_h$) parameter space  by requiring leptogenesis, DM production and reheating to have a common gravitational origin. Indeed, for a given $k$ and DM mass $M_{N_1}$, the temperature $\Trh$ can be determined by the relic abundance constraint. In turn, $\Trh$ determines the value of $\xi_h$ needed to reheat the Universe, as well as the value of $M_{N_2}$ which gives the desired baryon asymmetry through leptogenesis. To illustrate our result, we project the viable parameter space in the ($M_{N_1}$,$M_{N_2}$) plane in Fig.~\ref{fig:allowed} for different values of $\xi_h$, allowing $k$ to vary within $k\in [6,\,20]$. In each coloured line segment, gravitational interactions are responsible for the observed DM relic abundance, the baryon asymmetry and reheating.  Different coloured slanted line segments in this figure correspond to different choices of the non-minimal coupling $\xi_h$, with $\xi_h=0$ being ruled out from overproduction of GWs. 
The maximum possible value for $\xi_h$ is around 13.5, above which the mass $M_{N_2}$ necessary to reproduce the observed baryon asymmetry gets too close to $m_\phi$ and kinematic suppression becomes significant, as can be seen from Fig.~\ref{fig:asym}. Note that for each $\xi_h$, the allowed parameter space satisfying all the constraints, is rather restricted. This is better seen from the right panel figure, where we have zoomed in to the $\xi_h=1$ case. Interestingly, this shows that the viable parameter space is approximately independent of $k$, while $k=6$ and $8$ are excluded by BBN as can be seen from the left panel of Fig.~\ref{fig:tmax/trh}.

%%%%%%%%%%%%%%%%%%%%%%%%%%%%%%%%%

%%%%%%%%%%%%%%%%%%%%%%%%%%%%%%%%%%%%%
\subsection{The case for a decaying gravitational DM \& IceCube events}
\label{sec:dm-decay}
%%%%%%%%%%%%%%%%%%%%%%%%%%%%%%%%%%%%%%
Until now, we have assumed that the DM candidate, $N_1$, is absolutely stable. If it is not, and $N_1$ has non-zero Yukawa components, $y_{1i}$, $N_1$ can decay to SM final states. In this case, one necessary (but not sufficient) constraint on the DM mass and Yukawa coupling arises from the requirement of having a lifetime larger than the age of the Universe $\tau_{N_1} \gtrsim \tau_\text{univ}\simeq 4.35\times 10^{17}$~s. On the other hand, the IceCube detector has reported the detection of three PeV neutrinos, a roughly $3\sigma$ excess above expected background rates~\cite{IceCube:2013cdw, IceCube:2013low, IceCube:2014stg}. The three highest energy events correspond to deposited 
energies of 1.04 PeV, 1.14 PeV and 2.0 PeV. 
Although the origin of these very high energy events is still unclear, it has been shown in Refs.~\cite{Bai:2013nga,Higaki:2014dwa,Esmaili:2014rma,Bhattacharya:2014vwa,Zavala:2014dla,Rott:2014kfa,Dudas:2014bca,Murase:2015gea,Anchordoqui:2015lqa,Ko:2015nma,Roland:2015yoa,Chianese:2016smc,ReFiorentin:2016rzn,Garcia:2020hyo,Dudas:2018npp} that such events could be sourced from the
decays of superheavy DM particles. The 
neutrino energy spectrum presents a high-energy cutoff
at $m_\text{DM}/2$~\cite{Higaki:2014dwa,Esmaili:2014rma} if 
two body decays including one neutrino are present. The total excess can be interpreted as high energy neutrinos resulting
from the decay of $N_1$ with
 $\tau_{N_1}\approx 10^{28}$ s for both normal
and inverted hierarchies~\cite{Higaki:2014dwa,Arguelles:2022nbl}. Given that the maximum energy of the IceCube events has been measured to be about 2 PeV,  the mass of the DM particle is constrained to be $\simeq$ 4 PeV. Moreover, the IceCube spectrum sets a lower bound on the DM lifetime of the order of $10^{28}$~s~\cite{Esmaili:2014rma,Arguelles:2022nbl}, which is approximately model-independent and orders of magnitude larger than the lifetime of the Universe. Thus, satisfying this bound automatically makes $N_1$ a nearly stable relic, and hence a good DM candidate. For $N_1\to\ell\,H$ decay, we find 
\begin{align}
& \tau\equiv\Gamma_{N_1\to\ell\,H}^{-1} 
\simeq \left( \frac{y_{N_1}^2 M_{N_1}}{8\,\pi} \right)^{-1}
\simeq 10^{28}~\text{s}\,\left(\frac{4\times 10^{-29}}{y_{N_1}}\right)^2\,\left(\frac{1\,\text{PeV}}{M_{N_1}}\right)\, ;   
\end{align}
that is, the Yukawa coupling $y_{N_1}$ must be highly suppressed. 

On the other hand, in order to satisfy the observed DM abundance via the freeze-in mechanism in the early Universe through inverse decay: $\nu\,,H\to N_1$ involving the same Yukawa, one needs~\cite{Hall:2009bx,Chianese:2016smc}
\begin{align}
& \Omega_{N_1}\,h^2\simeq 0.12\,\left(\frac{y_{N_1}}{1.2\times 10^{-12}}\right)^2\,\left(\frac{M_{N_1}}{1\,\text{PeV}}\right)\,. \end{align}
This means that the Yukawa required to interpret the PeV IceCube event from the decay of $N_1$, $y_{N_1}\sim 10^{-29}$, is far too small for the thermal bath to populate the Universe from the freeze in mechanism. Thus, if we are restricted to  dimension-four interactions involving RHN and the SM, it is not possible to simultaneously explain the DM relic density and the IceCube events. Alternatively, we may consider higher dimensional operators~\cite{Chianese:2016smc},  modified gravity/cosmology~\cite{Jizba:2022bfz} or some different production mechanism for DM~\cite{Ahmadvand:2021vxs}. 

\begin{figure}[htb!]
    \centering
    \includegraphics[width=0.495\linewidth]{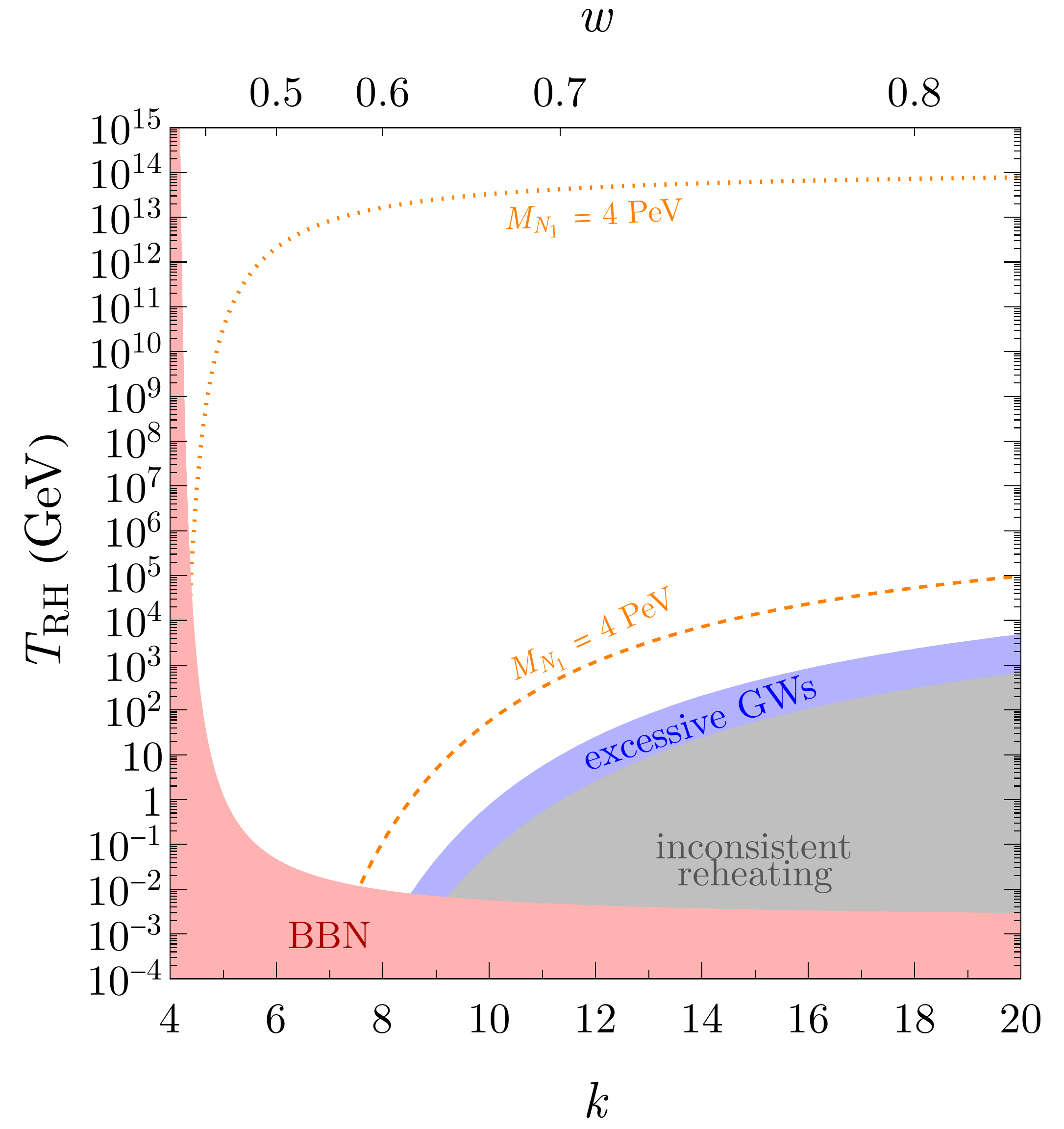}
    \caption{\it Contours of fixed relic density, $\Omega_{N_1} h^2 = 0.12$ for $M_{N_1}=4$ PeV. The upper dotted contour corresponds to production from gravitational scattering in the thermal bath (and requires a large value of $\xi_h$) and the lower dashed contour corresponds to production from inflaton scattering (and requires a relatively low value of $\xi_h$) Between the two contours $\Omega_{N_1} h^2 < 0.12$ for $M_{N_1}=4$ PeV.    
    }
    \label{fig:decay-icecube}
\end{figure}

Our minimalistic  framework contains a natural avenue
to reconcile both the DM abundance and IceCube events, 
through the gravitational production of decaying PeV neutrinos in the early Universe.
However, as discussed in Sec.~\ref{sec:grav-wave}, the case with minimal gravitational interactions is excluded by BBN for an excessive amount of gravitational waves as dark radiation.
Thus we then need to go (slightly) beyond the minimal setup and include non-minimal gravitational interactions. We show in Fig.~\ref{fig:decay-icecube} contours for $\Omega_{N_1} h^2 = 0.12$ for $M_{N_1} = 4$ PeV in the ($k$, $\Trh$) plane. The orange (dashed, dotted) lines correspond to the two dominant gravitational scattering processes involving the (inflaton, thermal particles) as discussed in the previous subsection. Note however that gravitational thermal production requires a high reheating temperature and is not compatible with the observed baryon asymmetry as can be understood from Fig.~\ref{fig:asym}. In contrast, at lower $\Trh$, the correct relic density can be produced from inflaton scattering with a lower value of $\xi_h \approx 2.5$.   In the left panel of Fig.~\ref{fig:allowed}, we show, by the black vertical line segment, the range in $M_{N_2}$ obtained from varying $k$ while fixing $M_{N_1} = 4$~PeV. Note that, since $N_1$ is a long-lived stable relic, it does not contribute to the generation of the baryon asymmetry as its decay takes place below the electroweak phase transition. In addition, because the Yukawa coupling of $N_1$ is extremely small, its interactions which violate lepton number will not be in equilibrium, and hence will not wash out any of the asymmetry produced by $N_2$. We summarize our analysis in the tables presented in the conclusion.

%%%%%%%%%%%%%%%%%%%%%%%%%%%%%%%
%%%%%%%%%%%%%%%%%%%%%%%%%%%%%%%

%%%%%%%%%%%%%%%%%%%%%%%
\section{Dark matter \& leptogenesis with a Majoron}
\label{sec:majoron}
\begin{figure}[htb!]
\centering
\includegraphics[scale=1.4]{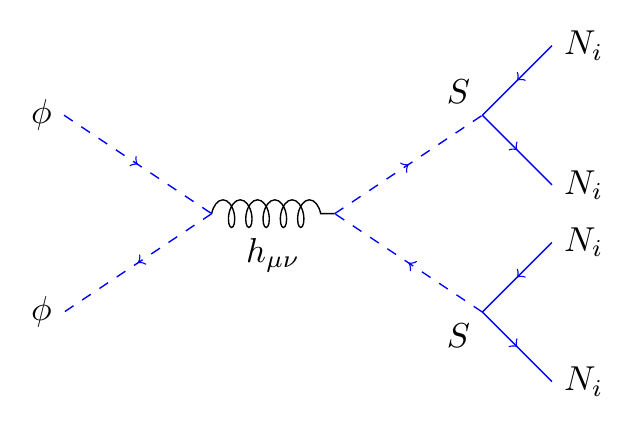}
\caption{\em Feynman diagram for the gravitational production of an on-shell scalar $S$ coupled to the heavy neutrinos. }
\label{Fig:feynman_Majoron}
\end{figure}

%%%%%%%%%%%%%%%%%%%%%%%
We have seen in the previous section that our result
was particularly constrained because of the strong
dependence of the production of the RHN on its mass
$M_{N_i}$, limiting our allowed region to masses 
above a PeV. In this section, we consider an alternative mechanism.  We include an additional complex scalar field, $\Phi$ containing the Majoron \cite{Chikashige:1980ui,Schechter:1981cv,Berezinsky:1993fm,Lattanzi:2007ux,Bazzocchi:2008fh,Lattanzi:2013uza,Dudas:2014bca,Lattanzi:2014mia,Gehrlein:2019iwl,Dudas:2020sbq}, that acts as an intermediate state in the interactions of the inflaton and RHNs.  This interaction is depicted in Fig.~\ref{Fig:feynman_Majoron}. 

%%%%%%%%%%%%%%%%%%%%%%%%%%%%%%
%%%%%%%%%%%%%%%%%%%%%%%%%%%%%%%

The relevant Lagrangian of this extension can be written as
\beq
{\cal L}_\Phi= (-y_R^i\,\Phi\,\overline{N_i^c}\,N_i + \text{h.c.}) + \frac{1}{2}\mu_\Phi^2\,\Phi^2 - \frac{1}{4}\lambda_\Phi\,\Phi^4\,.
\eeq
 After symmetry breaking, the real part of $\Phi$ acquires a non-zero vacuum expectation value, around which one can expand the field as: $\Phi=\frac{1}{\sqrt{2}}(S+v_S)e^{iJ/v_S}$, and $J$ is the Majoron.  This expectation value is the origin of the RHN Majorana masses, $M_{N_i}=y_R^i\,v_S/\sqrt{2}$. Then $m_S= \mu_\Phi < m_\phi$ and
 the gravitational production rate of the real scalar, $S$ is\footnote{As shown in Refs.~\cite{Clery:2021bwz,Haque:2021mab}, for the case of a scalar field, the gravitational thermal production is always negligible 
with respect to inflaton scattering.}
\begin{equation}
\label{Eq:ratephiS}
R^{\phi^k}_S=\frac{2 \times \rho_\phi^2}{16 \pi M_P^4} \Sigma_S^k \, ,
\end{equation}
where the factor of two accounts for the fact we produce two scalar particles per scattering, with \cite{Mambrini:2021zpp,Clery:2021bwz}
\begin{equation}
   \Sigma_S^k = \sum_{n = 1}^{\infty}  |{\cal P}^k_{2n}|^2\left[1+\frac{2\mu_\Phi^2}{E_{2n}^2}\right]^2 
\sqrt{1-\frac{4\mu_\Phi^2}{E_{2n}^2}} \, .
\label{Sigma0k}
\end{equation}
Since each scalar decays into 2 right-handed neutrinos, 
we obtain for the density of $N_i$ after integration of the
Boltzmann equation \cite{Clery:2021bwz},
\beq
n_{N_i}^{S\phi^k}(a_{\rm RH})\simeq\br_i \times
\frac{\sqrt{3}\rho_{\rm RH} ^{3/2}}{4 \pi M_P^3}
\frac{k+2}{6k-6}
\left(\frac{\rho_{\rm end}}{\rho_{\rm RH}}\right)^{1-\frac{1}{k}}\Sigma^k_S,
\label{nNiphi}
\eeq
where we assumed $a_{\rm RH} \gg a_{\rm end}$, and here $\br_i=\frac{(y_R^i)^2}{\sum (y_R^i)^2}$ so $\br_i = \frac{M_{N_i}^2}{M_{N_1}^2 + M_{N_2}^2+M_{N_3}^2}$ if $N_{1,2,3}$ are all lighter than $S$. The relic abundance 
of $N_1$ is then given by
\begin{eqnarray}
\frac{\Omega_{N_1}^{S\phi^k} h^2}{0.12}&\simeq& \br_1\times\left(\frac{\rho_{\rm end}}{10^{64} {\rm GeV}^4}\right)^{1-\frac{1}{k}}
\left(\frac{10^{40}{\rm GeV}^4}{\rho_{\rm RH}}\right)^{\frac{1}{4}-\frac{1}{k}}
\left(\frac{k+2}{6k-6}\right)
\nonumber
\\
&
\times&
\Sigma_S^k\times \frac{M_{N_1}}{ 2.5\times 10^{\frac{24}{k}-8} {\rm GeV}}\,,
\label{Eq:omegaN1}
\end{eqnarray}
whereas the baryon asymmetry follows from Eq.~\eqref{Eq:baryonassym}.
Note that so long as $M_{N_i} \ll \mu_\Phi \ll m_\phi$, the resulting dark matter abundance and baryon asymmetry will be independent of $m_S$.

\begin{figure}[htb!]
    \centering
\includegraphics[width=0.495\linewidth]{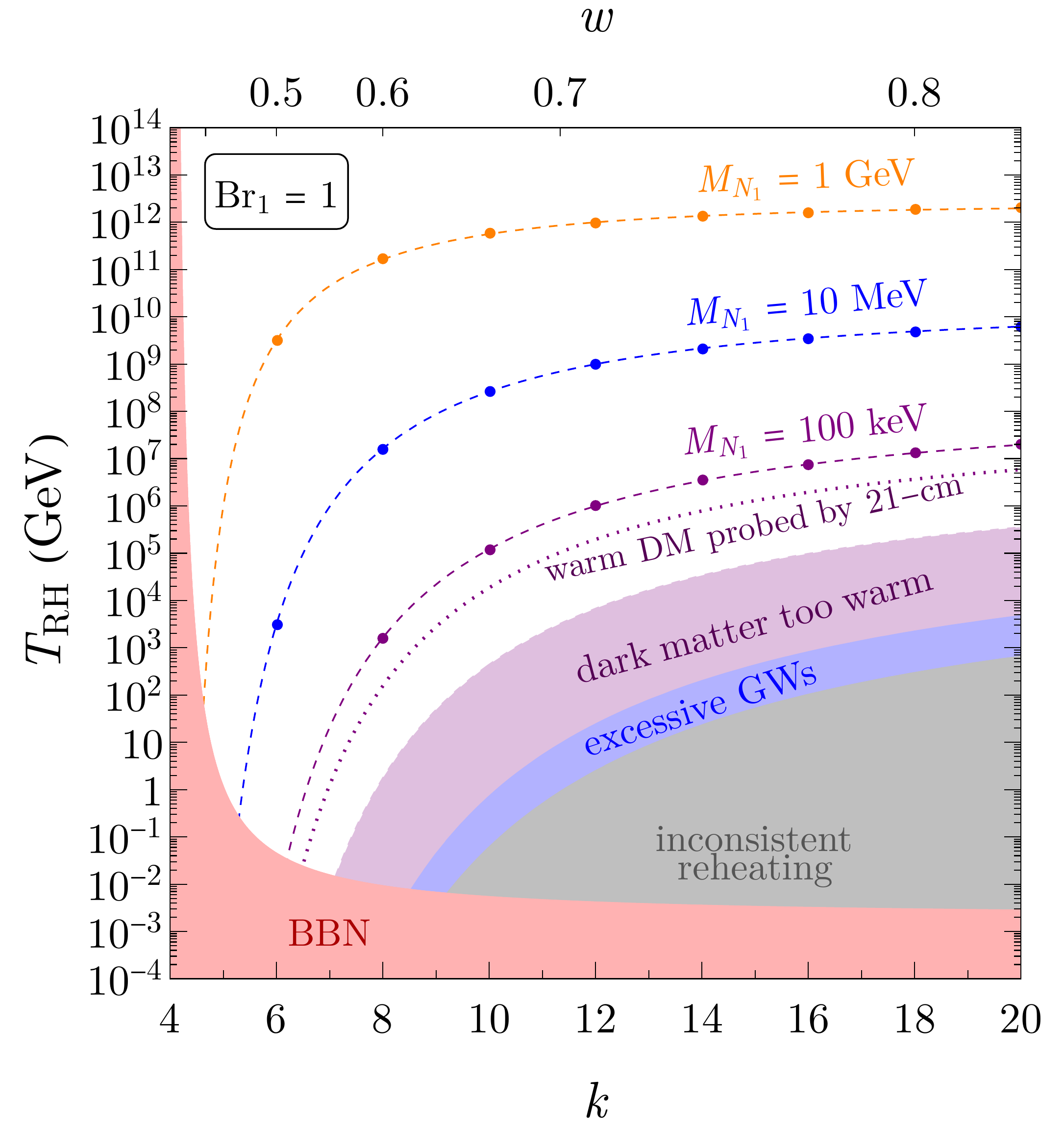}
\includegraphics[width=0.495\linewidth]{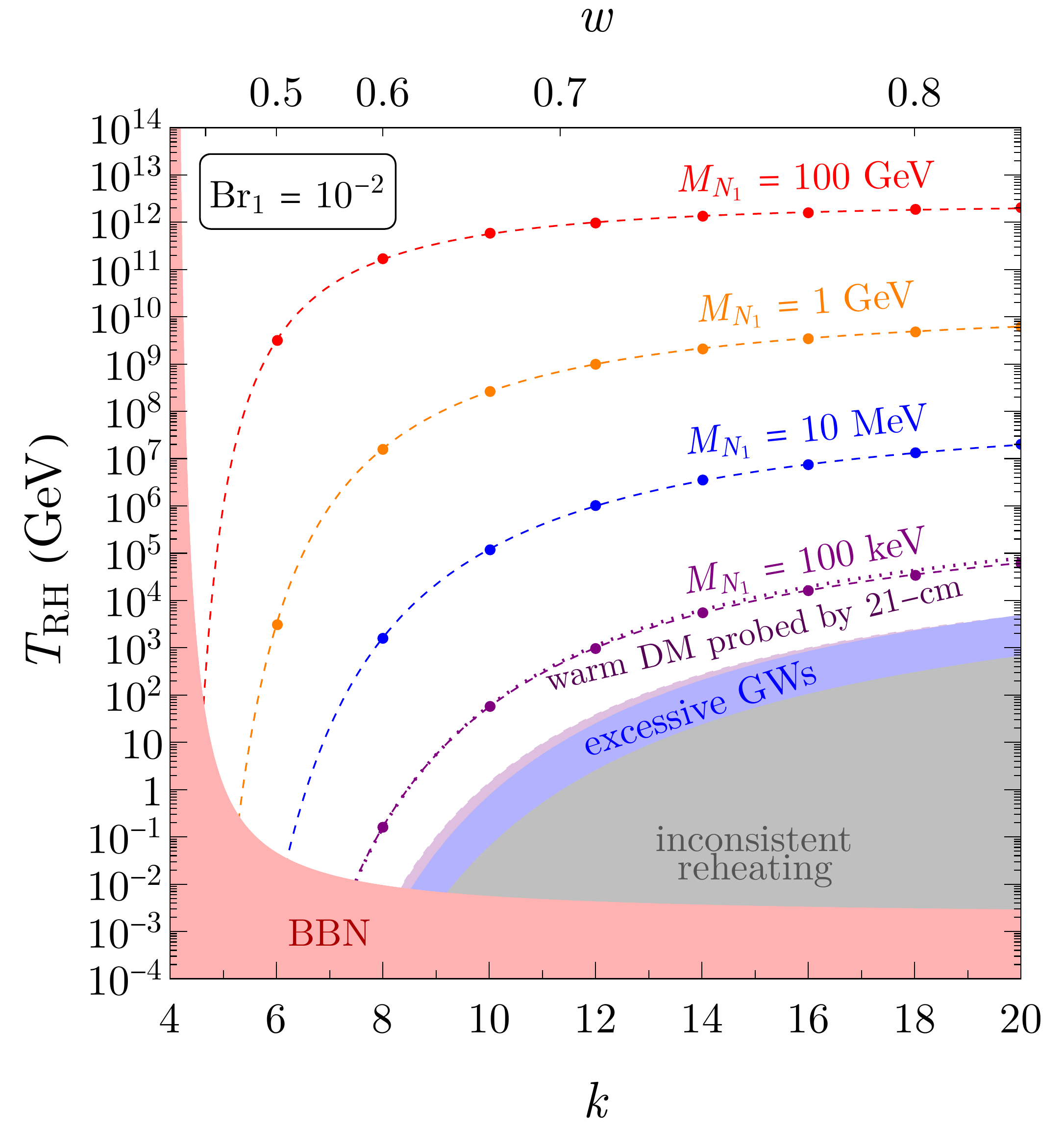}
    \caption{\it Contours of observed relic abundance assuming $\br_1=1$ (left) and $\br_1=10^{-2}$ (right) for different choices of the DM mass, considering only Majoron contribution. The purple-shaded region is disallowed from the warm DM limit (see text).}
    \label{fig:majrel}
\end{figure}

\begin{figure}[htb!]
    \centering
\includegraphics[width=0.495\linewidth]{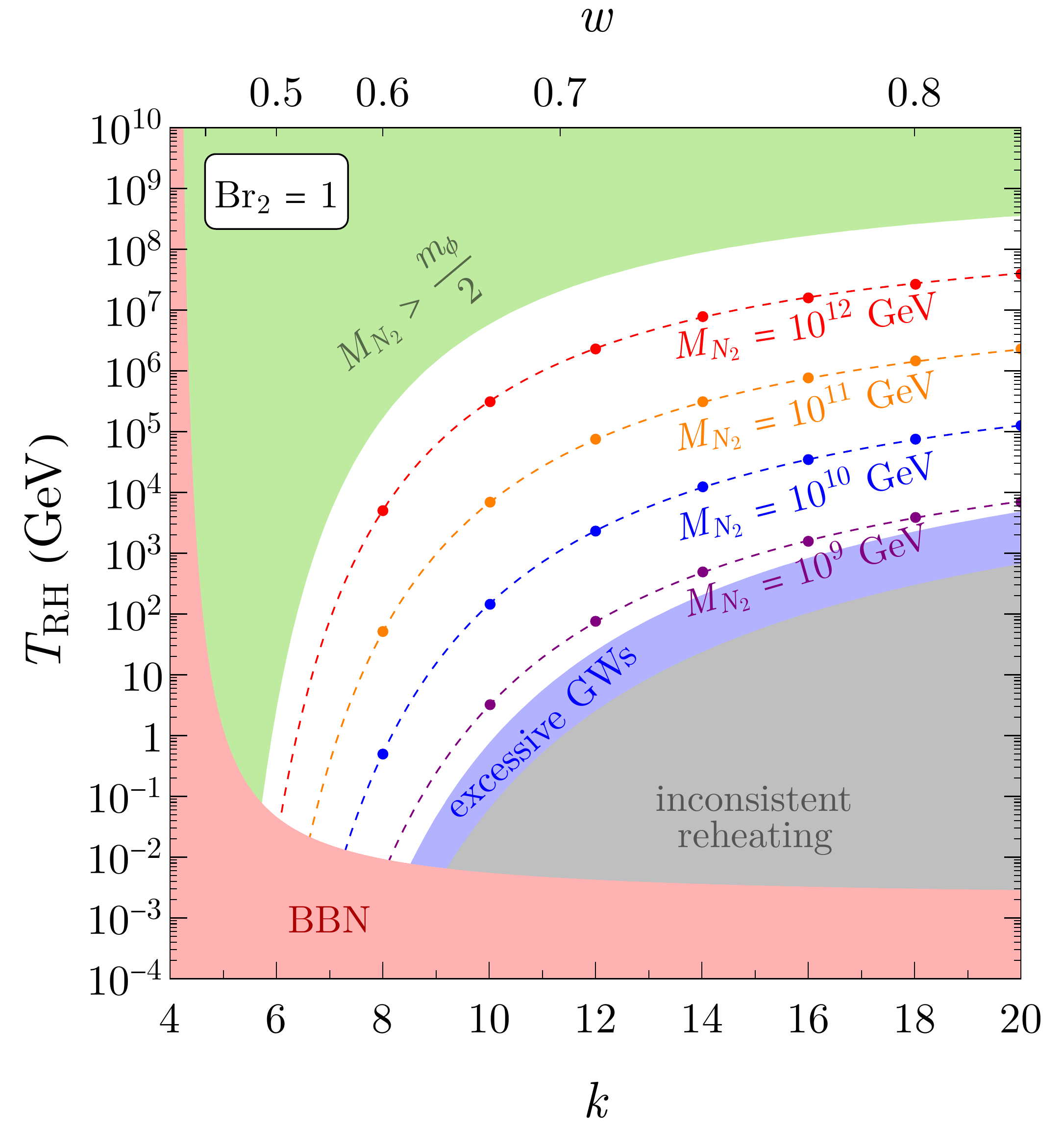}
\includegraphics[width=0.495\linewidth]{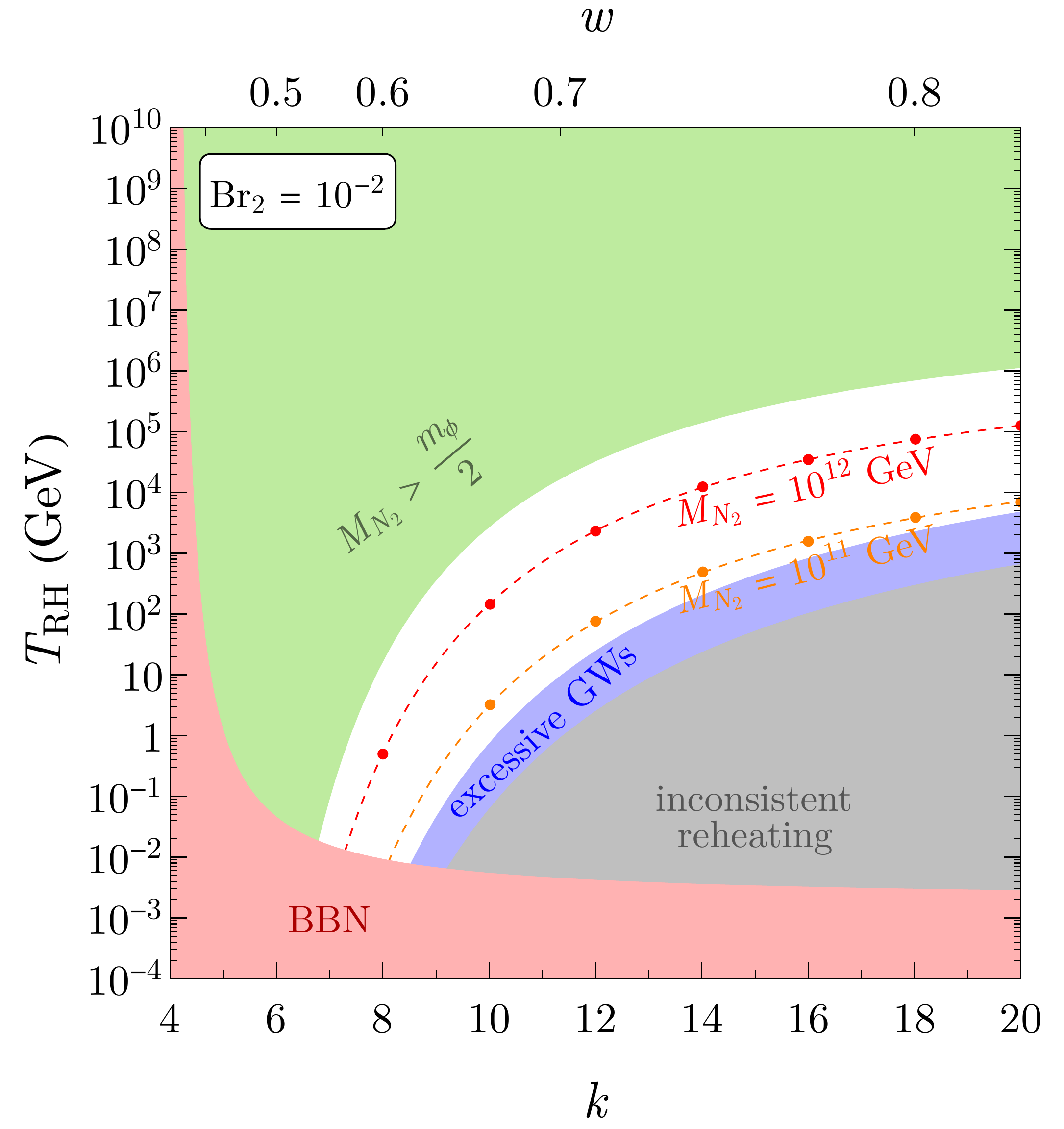}
    \caption{\it Contours of $N_{N_2}$ corresponding to the observed baryon asymmetry for $\br_2=1$ (left) and $\br_2=10^{-2}$ (right) in the $(k,\Trh)$ plane. Here only the contribution due to the intermediate scalar is included. The green-shaded region is kinematically inaccessible due to $M_{N_2}>m_\phi/2$ [cf. Fig.~\ref{Fig:feynman_Majoron}].}
    \label{fig:majlep}
\end{figure}

We show in Figs.~\ref{fig:majrel} and \ref{fig:majlep} respectively, the parameter space allowed by the relic abundance and the baryogenesis constraint in the ($k\,,\Trh$) plane.
Comparing Fig.~\ref{fig:majrel} and the dashed lines (from the inflaton scattering) in the bottom right panel of Fig.~\ref{fig:thermrelic}, we notice that the mass of the dark matter respecting Planck constraint is much lower, if the branching fraction to $N_1$ is large. For $\br_1=1$, the difference is about 8 orders of magnitude, and around 6 orders of magnitude for $\br_1=10^{-2}$. The reason is easy to understand:  the production rate through $S$ is boosted in comparison with the direct production, by a factor
\beq
\frac{R^{S \phi^k}_{N_1}}{R^{\phi^k}_{N_1}}\simeq \br_1\frac{m_\phi^2}{M_{N_1}^2}\,.
\eeq
For smaller branching fraction, the density of $N_1$ through this channel is suppressed and the effect is milder and proportional to $\br_1$, as one can see in Fig.~\ref{fig:majrel} right panel. 

%%%%%%%%%%%%%%%%%%%%%%%%%%%%%%%
%%%%%%%%%%%%%%%%%%%%%%%%%%%%%
%%%%%%%%%%%%%%%%%%%%%%%%%%%%%%%%%%%

Given that the required mass, $M_{N_1}$, can be much lower when we couple the RHNs to the scalar $S$, and $N_1$ is produced relativistically, $N_1$ dark matter may still be warm around the time of CMB decoupling. We derive the warmness constraint by redshifting the $N_1$ initial momentum of order $m_\phi$ at $\Tmax$ to the temperature $T\simeq 1$ eV and require that the velocity at $T\simeq 1$ eV is less than $2 \times 10^{-4}$. This bound on the velocity is obtained from translating the limit on the warm dark matter mass from the Lyman-$\alpha$ forest~\cite{Irsic:2017ixq} in the case where the abundance is generated thermally. The current warmness constraint is shown by the purple region, while the future sensitivity using cosmic 21-cm lines~\cite{Sitwell:2013fpa} is shown by the purple dotted curve. In summary, this mechanism interestingly allows for electroweak scale fermionic dark matter produced gravitationally, which is not possible by the direct scattering of the inflaton. We show in Fig.~\ref{fig:majlep} the parameter space allowed to obtain a sufficient amount of baryon asymmetry for the set of branching ratios $\br_2=1$ and $10^{-2}$. Comparing Figs.~\ref{fig:asym} and \ref{fig:majlep} left, we note that for $M_2=10^{13}$ GeV the situation is similar to the direct production because no real enhancement $\propto \frac{m_\phi^2}{M_{N_2}^2}$ exists. However, for $M_{N_2}=10^{11}$ GeV and large values of $k$, $\Trh$ should
be about 3 orders of magnitude larger to obtain the same asymmetry. The reason is that for a large value of $k$, $Y_B\propto \frac{1}{\Trh}$ when $S$ is produced
(combining Eqs.~\eqref{nNiphi} and \eqref{Eq:baryonassym}), and
$\propto \frac{M_{N_2}^2}{m_\phi^2 \Trh}$ when $N_2$ is produced directly. In other words, $\Trh$ should be compensated by a factor $\frac{m_\phi^2}{M_{N_2}^2}$ to avoid an excessive asymmetry. As in the case of dark matter, lowering the branching ratio dilutes the effect as one can see in the right panel of Fig.~\ref{fig:majlep}. 

\begin{figure}[htb!]
    \centering
    \includegraphics[width=0.495\linewidth]{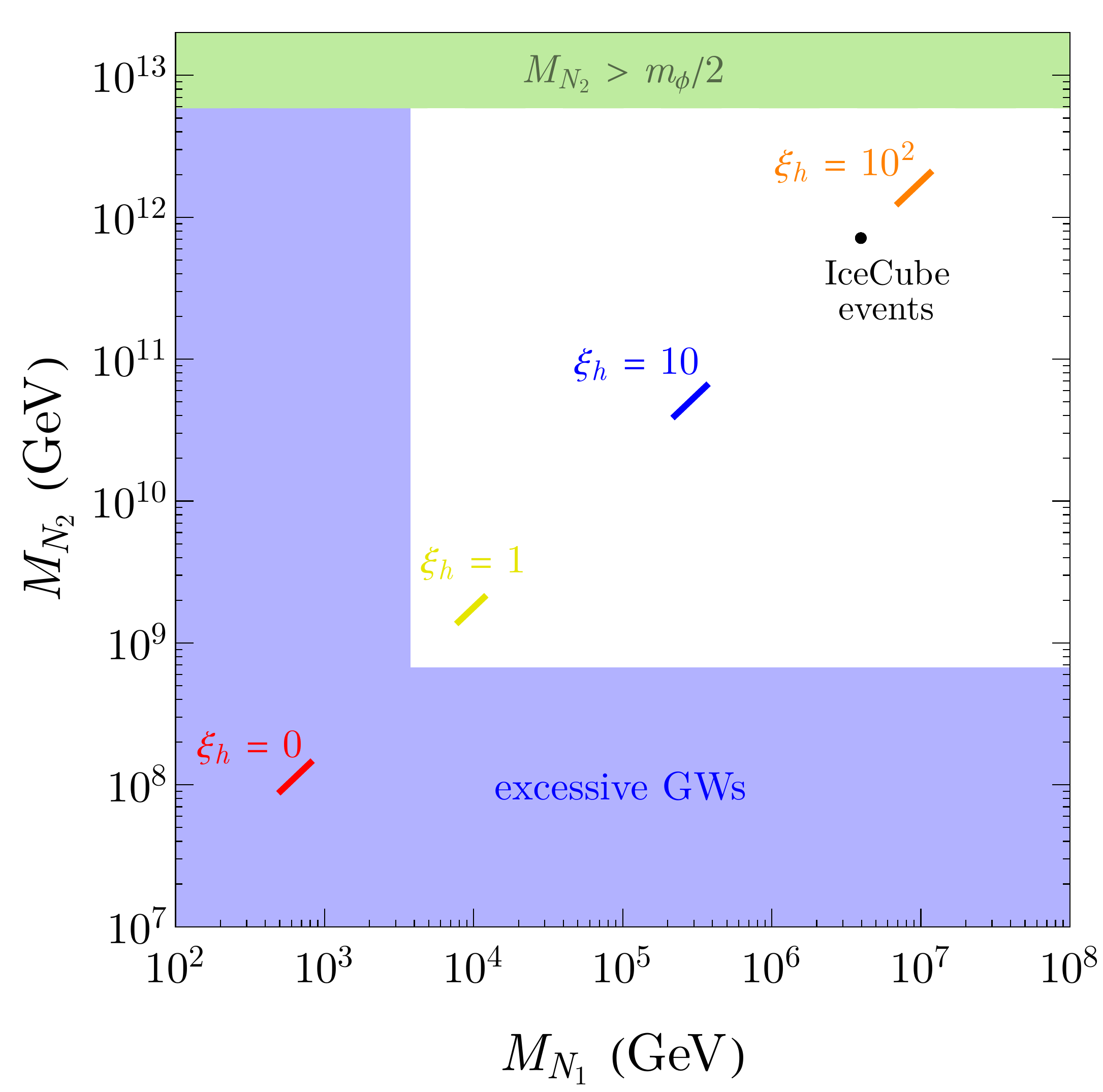}
    \caption{\it Parameter space satisfying the right dark matter relic abundance and baryon asymmetry, considering the production through $S$. The line colors correspond to different values of $\xi_h$, with $\xi_h=\{0,1,10,10^2\}$ from bottom to top, and $\xi_h=0$ corresponds to minimal graviton exchange. Each colored line segment shows the variation of the predicted masses with $k\in\left[6,\,20\right]$. The black dot marks the parameter point that can also explain the IceCube high-energy neutrino excess.}
    \label{fig:majparam}
\end{figure}

Finally, we can combine all of the preceding results, adding the possibility for a gravitational reheating with non-minimal coupling. We illustrate this in Fig.~\ref{fig:majparam}, which is the analogue of Fig.~\ref{fig:allowed} but with the scalar $S$ as an intermediate state. For demonstration purposes, here we suppose $M_{N_3} > m_\phi/2$ so that $N_3$ is not produced by the inflaton or $S$, resulting in $\br_2 = 1-\br_1 = 1-(M_{N_1}/M_{N_2})^2 \simeq 1$. As the branching ratios are completely determined by the masses $M_{N_1}$ and $M_{N_2}$, for a fixed $k$ and a fixed $\xi_h$, there will be again only one point in the $(M_{N_1}, M_{N_2})$ plane that could simultaneously obtain the CMB-determined DM relic abundance and the observed baryon asymmetry. Each color segment in Fig.~\ref{fig:majparam} assumes a fixed $\xi_h$ and allows values of $k \in [6,20]$ that are consistent with the BBN bound on $\Trh$. The black dot indicates the $M_{N_{1,2}}$ masses, independent of $k$, required to explain the IceCube high-energy neutrino excess.  The green region is inaccessible because $m_S > M_{N_2} > m_\phi/2$ forbids the production of $S$ via $\phi$ scattering.\footnote{Note that we have not included the effects of $\mu_\Phi$ in $\Sigma^k_S$. These start to play a role for large $\xi_h \gtrsim 10^2$ when $2 M_{N_2}$ becomes close to $m_\phi$, since we require $2m_{N_2} < m_S < m_\phi$.  } Compared to Fig.~\ref{fig:allowed}, we see that the effect of $S$ as an intermediate state expands the parameter space of allowed dark matter density and baryon asymmetry. Most notably, the parameter space opens up towards lower masses, and allows large values of $\xi_h$.  

%%%%%%%%%%%%%%%%%%%%%%%%%%%%%%%%

%%%%%%%%%%%%%%%%%%%%%%%
\section{Conclusions}
\label{sec:concl}
%%%%%%%%%%%%%%%%%%%%%%
In this paper, we have shown that there exists the possibility that inflationary reheating, dark matter and the baryon asymmetry can be generated solely gravitational interactions. The baryon asymmetry is produced through the decay of a right-handed neutrino $M_{N_2}$ (leading first to a non-zero lepton asymmetry). For minimal gravitational interactions, $\xi_h=0$, a large amount of dark radiation is created in the form of gravitational waves and is inconsistent with BBN.
Thus, we allow for a non-minimal gravitational coupling $\xi_h R H^2$ where $H$ the Standard Model Higgs field to enhance reheating, so that the ratio of gravitational wave energy density to the radiation is decreased. The lowest $\xi_h$ consistent with BBN is around 0.5. The range of the parameter space is 2-8 PeV for the dark matter mass $M_{N_1}$ and $0.3-1.7 \times 10^{12}$ GeV for the mass of the lepton number violating decaying RHN, $M_{N_2}$. The range corresponds to a scan over $\xi_h$. Our solution restricts $0.5 \lesssim \xi_h \lesssim 13.5$ and $\Trh < 3 \times 10^5$ GeV, where the maximum reheating temperature is attained with $\xi_h \simeq 4.7$. We summarize our results for fixing different values of $\xi_h$ or $M_{N_1}$ in the upper sections of Table~\ref{Tab:m1m2rng} and Table~\ref{Tab:m1m2xi} labeled ``Direct gravitational production." Primordial gravitational waves generated during inflation allow a large parameter space with $\Trh \lesssim 5 \times 10^6$ GeV to be probed in proposed gravitational wave detectors such as BBO, DECIGO, CE and ET.

% %%%%%%%%%%%%%%%%%%%%%%%%%%%%%%%%%%%%%%%%%%%%%%%
\begin{table}[htb!]
\centering
\begin{tabular}{|c|c|c|c|}
\hline
\multicolumn{4}{|c|}{\bf Direct gravitational production} \\
\hline
$\xi_h$ & $\Trh\,\text{[GeV]}$ & $M_{N_1}\,\text{[PeV]}$ & $M_{N_2}\,\text{[GeV]}$ \\
\hline
%0 & $\{0.057-630\}$ & $\{1.0-1.04\}$ & $\{1.8-1.9\}\times 10^{11}$ \\ \hline
1 & $\{5.6-1.8\times 10^4\}$  & $\{2.49-2.54\}$ & $\{4.5-4.8\}\times 10^{11}$ \\ \hline
$\{2.5-2.7\}$ & $\{0.11-9.8\times10^4\}$  & $4.0\star$ & $\{7.3-9.2\}\times 10^{11}$ \\ \hline
10 & $8.1$ & $8.1$ & $1.7\times 10^{12}$ \\
\hline\hline
\multicolumn{4}{|c|}{\bf Gravitational production via $S$} \\
\hline
$\xi_h$ & $\Trh\,\text{[GeV]}$ & $M_{N_1}$ & $M_{N_2}\,\text{[GeV]}$ \\
\hline
%0 & $\{0.057-630\}$ & $\{510-800\}~{\rm GeV}$ & $\{9.0-14\}\times 10^7$ \\ \hline
1 & $\{5.6-1.8\times 10^4\}$ & $\{7.9-12\}~{\rm TeV}$ & $\{1.4-2.1\}\times 10^9$ \\ \hline
$10$ & $\{8.1-1.3\times10^6\}$ & $\{220-360\}~{\rm TeV}$ & $\{4.0-6.5\} \times 10^{10}$ \\ \hline
$\{50-68\}$ & $\{2.6\times10^3-2.6\times10^7\}$ & $4.0~{\rm PeV} \star$ & $7.1 \times 10^{11}$ \\ \hline
$100$ & $\{8.1\times10^3-9.8\times10^7\}$ & $\{7.1-11\}~{\rm PeV}$ & $\{1.3-2.0\} \times 10^{12}$ \\ \hline
\end{tabular}
%%%%%
\caption{\it Ranges of $\Trh$, DM $(M_{N_1})$ and RHN $(M_{N_2})$ masses, over which baryon asymmetry and DM relic abundance are simultaneously satisfied via gravitational yield for different choices of the non-minimal coupling $\xi_h$; $\xi_h=0$ corresponds to minimal gravity, which is not shown since it is excluded by BBN for excessive GWs. The upper section assumes direction production of $N_{1,2}$, while the lower assumes production via Majoron's CP-even partner $S$ as an intermediate state. Here we allow $k\in [6,20]$ and omit points whose low $\Trh$ values are excluded by BBN. (For the direct gravitational production, a single value of $k=8$ is allowed for $\xi_h = 10$, while no points are allowed for $\xi_h \gtrsim 13.5 $.) The $\star$ entry corresponds to the DM mass that can explain the IceCube high-energy neutrino events.}
\label{Tab:m1m2rng}
\end{table}

%%%%%%%%%%%%%%%%%%%%%%%%%%%%%%%%%%%%%%%%%%%%%%
\begin{table}[htb!]
\centering
\begin{tabular}{|c|c|c|c|}
\hline 
\multicolumn{4}{|c|}{\bf Direct gravitational production} \\ 
\hline
$\xi_h$ & $\Trh\,\text{[GeV]}$ & $M_{N_1}\,\text{[PeV]}$ & $M_{N_2}\,\text{[GeV]}$ \\
\hline
%0 & $3.3 \times 10^{-5}$ (excluded) & -- & --  \\ \hline
1 & $0.0084$ (excluded) & -- & --  \\
\hline
2.5 & 0.11 & $4.0\star$ & $7.3 \times 10^{11}$ \\
\hline
10 & $8.1$ & $8.1$ & $1.7 \times 10^{12}$ \\
\hline\hline
\multicolumn{4}{|c|}{\bf Gravitational production via $S$} \\
\hline
$\xi_h$ & $\Trh\,\text{[GeV]}$ & $M_{N_1}$ & $M_{N_2}\,\text{[GeV]}$ \\
\hline 
%0 & $3.3 \times 10^{-5}$ (excluded) & -- & --  \\ \hline
1 & 0.0084 (excluded) & -- & -- \\
\hline
10 & $8.1$ & $220~{\rm TeV}$ & $4.0 \times 10^{10}$ \\
\hline
68 & $2.6 \times 10^3$ & $4.0\star$ & $7.1 \times 10^{11}$ \\
\hline
100 & $8.1\times10^3$ & $7.1~{\rm PeV}$ & $1.3 \times 10^{12}$ \\
\hline
\end{tabular} 
\caption{\it Same as Table~\ref{Tab:m1m2rng} but for a fixed $k=8$ as a benchmark.}
\label{Tab:m1m2xi}
\end{table}
%%%%%%%%%%%%%%%%%%%%%%%%%%%%%%%%%%%%%%%%%%%%%

We also showed that $N_1$, if unstable, can explain the recent IceCube PeV events through its decay into SM neutrinos. In this case, if we want to accommodate simultaneously the correct DM relic abundance, the observed baryon asymmetry, gravitational reheating and the IceCube events, the value of $\xi_h$ is fixed for a given $k$. We show this result in the second row of the upper section of each table where the assumed value of $m_{N_1} = 4$ PeV is marked by~$\star$. 

Finally, we proposed a new scenario where the RHN and the dark matter are produced through an intermediate scalar state $S$, the CP-even partner of the Majoron. In this case, the gravitational production of the scalar is not helicity suppressed by the mass of the final state fermions. As a result, the mass ranges for $N_1$ and $N_2$ are increased.  For $0.5 \lesssim \xi_h \lesssim 100$, the mass range for $N_1$ is $4$ TeV to $11$ PeV, and the range for $N_2$ is $7 \times 10^8 - 2 \times 10^{12}$ GeV. Finally, the IceCube events can be explained by appropriate choices of $\xi_h$ and $\Trh$ for each $k$, whereas $M_{N_2}$ is predicted to be $7.1 \times 10^{11}$ GeV for all $k$ when $M_{N_1}$ is fixed to 4 PeV. We summarize our overall results for fixing different values of $\xi_h$ or $M_{N_1}$ in the lower sections of Table~\ref{Tab:m1m2rng} and Table~\ref{Tab:m1m2xi} labeled ``Gravitational production via $S$".

%%%%%%%%%%%%%%%%
\section*{Acknowledgements}
%%%%%%%%%%%%%%%%
We would like to thank Essodjolo Kpatcha and Jong-Hyun Yoon for useful discussions. We thank Pedro Schwaller for bringing to our attention the constraint from the gravitational wave as dark radiation. BB received funding from the Patrimonio Autónomo - Fondo Nacional de Financiamiento para la Ciencia, la Tecnología y la Innovación Francisco José de Caldas (MinCiencias - Colombia) grant 80740-465-2020. This project has received funding /support from the European Union's Horizon 2020 research and innovation programme under the Marie Sklodowska-Curie grant agreement No 860881-HIDDeN, and the IN2P3 Master Projet UCMN. BB would also like to acknowledge the hospitality at the IJCLab during INVISIBLES'22 school+workshop, during which this work was initiated and fruitful discussions with Ashmita Das. The work of R.C.~and K.A.O.~was supported in part by DOE grant DE-SC0011842 at the University of Minnesota.

%%%%%%%%%%%%%%%%%%%%%
\bibliographystyle{JHEP}
\bibliography{Bibliography}

\providecommand{\href}[2]{#2}\begingroup\raggedright\begin{thebibliography}{100}

\bibitem{Poincare}
H.~Poincaré, \emph{{The Milky Way and the Theory of Gases}}, {\emph{Popular
  Astronomy, vol. 14, pp.475-488} {\bfseries 14} (1906) 475}.

\bibitem{Zwicky:1933gu}
F.~Zwicky, \emph{{Die Rotverschiebung von extragalaktischen Nebeln}},
  \href{https://doi.org/10.1007/s10714-008-0707-4}{\emph{Helv. Phys. Acta}
  {\bfseries 6} (1933) 110}.

\bibitem{Aghanim:2018eyx}
{\scshape Planck} collaboration, \emph{{Planck 2018 results. VI. Cosmological
  parameters}},  \href{https://arxiv.org/abs/1807.06209}{{\ttfamily
  1807.06209}}.

\bibitem{Gunn:1978gr}
J.E.~Gunn, B.W.~Lee, I.~Lerche, D.N.~Schramm and G.~Steigman, \emph{{Some
  Astrophysical Consequences of the Existence of a Heavy Stable Neutral
  Lepton}}, \href{https://doi.org/10.1086/156335}{\emph{Astrophys. J.}
  {\bfseries 223} (1978) 1015}.

\bibitem{Hut:1977zn}
P.~Hut, \emph{{Limits on Masses and Number of Neutral Weakly Interacting
  Particles}}, \href{https://doi.org/10.1016/0370-2693(77)90139-3}{\emph{Phys.
  Lett. B} {\bfseries 69} (1977) 85}.

\bibitem{Lee:1977ua}
B.W.~Lee and S.~Weinberg, \emph{{Cosmological Lower Bound on Heavy Neutrino
  Masses}}, \href{https://doi.org/10.1103/PhysRevLett.39.165}{\emph{Phys. Rev.
  Lett.} {\bfseries 39} (1977) 165}.

\bibitem{Ellis:1983ew}
J.R.~Ellis, J.S.~Hagelin, D.V.~Nanopoulos, K.A.~Olive and M.~Srednicki,
  \emph{{Supersymmetric Relics from the Big Bang}},
  \href{https://doi.org/10.1016/0550-3213(84)90461-9}{\emph{Nucl. Phys. B}
  {\bfseries 238} (1984) 453}.

\bibitem{Silveira:1985rk}
V.~Silveira and A.~Zee, \emph{{SCALAR PHANTOMS}},
  \href{https://doi.org/10.1016/0370-2693(85)90624-0}{\emph{Phys. Lett. B}
  {\bfseries 161} (1985) 136}.

\bibitem{McDonald:1993ex}
J.~McDonald, \emph{{Gauge singlet scalars as cold dark matter}},
  \href{https://doi.org/10.1103/PhysRevD.50.3637}{\emph{Phys. Rev.} {\bfseries
  D50} (1994) 3637} [\href{https://arxiv.org/abs/hep-ph/0702143}{{\ttfamily
  hep-ph/0702143}}].

\bibitem{Burgess:2000yq}
C.P.~Burgess, M.~Pospelov and T.~ter Veldhuis, \emph{{The Minimal model of
  nonbaryonic dark matter: A Singlet scalar}},
  \href{https://doi.org/10.1016/S0550-3213(01)00513-2}{\emph{Nucl. Phys. B}
  {\bfseries 619} (2001) 709}
  [\href{https://arxiv.org/abs/hep-ph/0011335}{{\ttfamily hep-ph/0011335}}].

\bibitem{Davoudiasl:2004be}
H.~Davoudiasl, R.~Kitano, T.~Li and H.~Murayama, \emph{{The New minimal
  standard model}},
  \href{https://doi.org/10.1016/j.physletb.2005.01.026}{\emph{Phys. Lett. B}
  {\bfseries 609} (2005) 117}
  [\href{https://arxiv.org/abs/hep-ph/0405097}{{\ttfamily hep-ph/0405097}}].

\bibitem{Cline:2013gha}
J.M.~Cline, K.~Kainulainen, P.~Scott and C.~Weniger, \emph{{Update on scalar
  singlet dark matter}},
  \href{https://doi.org/10.1103/PhysRevD.88.055025}{\emph{Phys. Rev. D}
  {\bfseries 88} (2013) 055025}
  [\href{https://arxiv.org/abs/1306.4710}{{\ttfamily 1306.4710}}].

\bibitem{Han:2015hda}
H.~Han and S.~Zheng, \emph{{New Constraints on Higgs-portal Scalar Dark
  Matter}}, \href{https://doi.org/10.1007/JHEP12(2015)044}{\emph{JHEP}
  {\bfseries 12} (2015) 044}
  [\href{https://arxiv.org/abs/1509.01765}{{\ttfamily 1509.01765}}].

\bibitem{Djouadi:2011aa}
A.~Djouadi, O.~Lebedev, Y.~Mambrini and J.~Quevillon, \emph{{Implications of
  LHC searches for Higgs--portal dark matter}},
  \href{https://doi.org/10.1016/j.physletb.2012.01.062}{\emph{Phys. Lett. B}
  {\bfseries 709} (2012) 65} [\href{https://arxiv.org/abs/1112.3299}{{\ttfamily
  1112.3299}}].

\bibitem{Lebedev:2011iq}
O.~Lebedev, H.M.~Lee and Y.~Mambrini, \emph{{Vector Higgs-portal dark matter
  and the invisible Higgs}},
  \href{https://doi.org/10.1016/j.physletb.2012.01.029}{\emph{Phys. Lett. B}
  {\bfseries 707} (2012) 570}
  [\href{https://arxiv.org/abs/1111.4482}{{\ttfamily 1111.4482}}].

\bibitem{Mambrini:2011ik}
Y.~Mambrini, \emph{{Higgs searches and singlet scalar dark matter: Combined
  constraints from XENON 100 and the LHC}},
  \href{https://doi.org/10.1103/PhysRevD.84.115017}{\emph{Phys. Rev. D}
  {\bfseries 84} (2011) 115017}
  [\href{https://arxiv.org/abs/1108.0671}{{\ttfamily 1108.0671}}].

\bibitem{Djouadi:2012zc}
A.~Djouadi, A.~Falkowski, Y.~Mambrini and J.~Quevillon, \emph{{Direct Detection
  of Higgs-Portal Dark Matter at the LHC}},
  \href{https://doi.org/10.1140/epjc/s10052-013-2455-1}{\emph{Eur. Phys. J. C}
  {\bfseries 73} (2013) 2455}
  [\href{https://arxiv.org/abs/1205.3169}{{\ttfamily 1205.3169}}].

\bibitem{Casas:2017jjg}
J.A.~Casas, D.G.~Cerde\~no, J.M.~Moreno and J.~Quilis, \emph{{Reopening the
  Higgs portal for single scalar dark matter}},
  \href{https://doi.org/10.1007/JHEP05(2017)036}{\emph{JHEP} {\bfseries 05}
  (2017) 036} [\href{https://arxiv.org/abs/1701.08134}{{\ttfamily
  1701.08134}}].

\bibitem{Arcadi:2014lta}
G.~Arcadi, Y.~Mambrini and F.~Richard, \emph{{Z-portal dark matter}},
  \href{https://doi.org/10.1088/1475-7516/2015/03/018}{\emph{JCAP} {\bfseries
  03} (2015) 018} [\href{https://arxiv.org/abs/1411.2985}{{\ttfamily
  1411.2985}}].

\bibitem{Escudero:2016gzx}
M.~Escudero, A.~Berlin, D.~Hooper and M.-X.~Lin, \emph{{Toward (Finally!)
  Ruling Out Z and Higgs Mediated Dark Matter Models}},
  \href{https://doi.org/10.1088/1475-7516/2016/12/029}{\emph{JCAP} {\bfseries
  12} (2016) 029} [\href{https://arxiv.org/abs/1609.09079}{{\ttfamily
  1609.09079}}].

\bibitem{Roszkowski:2017nbc}
L.~Roszkowski, E.M.~Sessolo and S.~Trojanowski, \emph{{WIMP dark matter
  candidates and searches—current status and future prospects}},
  \href{https://doi.org/10.1088/1361-6633/aab913}{\emph{Rept. Prog. Phys.}
  {\bfseries 81} (2018) 066201}
  [\href{https://arxiv.org/abs/1707.06277}{{\ttfamily 1707.06277}}].

\bibitem{Arcadi:2017kky}
G.~Arcadi, M.~Dutra, P.~Ghosh, M.~Lindner, Y.~Mambrini, M.~Pierre et~al.,
  \emph{{The waning of the WIMP? A review of models, searches, and
  constraints}},
  \href{https://doi.org/10.1140/epjc/s10052-018-5662-y}{\emph{Eur. Phys. J. C}
  {\bfseries 78} (2018) 203}
  [\href{https://arxiv.org/abs/1703.07364}{{\ttfamily 1703.07364}}].

\bibitem{Mambrini:2010dq}
Y.~Mambrini, \emph{{The Kinetic dark-mixing in the light of CoGENT and
  XENON100}}, \href{https://doi.org/10.1088/1475-7516/2010/09/022}{\emph{JCAP}
  {\bfseries 09} (2010) 022} [\href{https://arxiv.org/abs/1006.3318}{{\ttfamily
  1006.3318}}].

\bibitem{Lebedev:2014bba}
O.~Lebedev and Y.~Mambrini, \emph{{Axial dark matter: The case for an invisible
  Z{'}}}, \href{https://doi.org/10.1016/j.physletb.2014.05.025}{\emph{Phys.
  Lett. B} {\bfseries 734} (2014) 350}
  [\href{https://arxiv.org/abs/1403.4837}{{\ttfamily 1403.4837}}].

\bibitem{Alves:2015pea}
A.~Alves, A.~Berlin, S.~Profumo and F.S.~Queiroz, \emph{{Dark Matter
  Complementarity and the Z$^\prime$ Portal}},
  \href{https://doi.org/10.1103/PhysRevD.92.083004}{\emph{Phys. Rev. D}
  {\bfseries 92} (2015) 083004}
  [\href{https://arxiv.org/abs/1501.03490}{{\ttfamily 1501.03490}}].

\bibitem{Alves:2016cqf}
A.~Alves, G.~Arcadi, Y.~Mambrini, S.~Profumo and F.S.~Queiroz, \emph{{Augury of
  darkness: the low-mass dark Z' portal}},
  \href{https://doi.org/10.1007/JHEP04(2017)164}{\emph{JHEP} {\bfseries 04}
  (2017) 164} [\href{https://arxiv.org/abs/1612.07282}{{\ttfamily
  1612.07282}}].

\bibitem{Arcadi:2017jqd}
G.~Arcadi, P.~Ghosh, Y.~Mambrini, M.~Pierre and F.S.~Queiroz, \emph{{$Z'$
  portal to Chern-Simons Dark Matter}},
  \href{https://doi.org/10.1088/1475-7516/2017/11/020}{\emph{JCAP} {\bfseries
  1711} (2017) 020} [\href{https://arxiv.org/abs/1706.04198}{{\ttfamily
  1706.04198}}].

\bibitem{nos}
D.V.~Nanopoulos, K.A.~Olive and M.~Srednicki, \emph{{After Primordial
  Inflation}}, \href{https://doi.org/10.1016/0370-2693(83)91624-6}{\emph{Phys.
  Lett. B} {\bfseries 127} (1983) 30}.

\bibitem{Khlopov:1984pf}
M.Y.~Khlopov and A.D.~Linde, \emph{{Is It Easy to Save the Gravitino?}},
  \href{https://doi.org/10.1016/0370-2693(84)91656-3}{\emph{Phys. Lett. B}
  {\bfseries 138} (1984) 265}.

\bibitem{Olive:1984bi}
K.A.~Olive, D.N.~Schramm and M.~Srednicki, \emph{{Gravitinos as the Cold Dark
  Matter in an $\Omega$ = 1 Universe}},
  \href{https://doi.org/10.1016/0550-3213(85)90149-X}{\emph{Nucl. Phys. B}
  {\bfseries 255} (1985) 495}.

\bibitem{McDonald:2001vt}
J.~McDonald, \emph{{Thermally generated gauge singlet scalars as
  selfinteracting dark matter}},
  \href{https://doi.org/10.1103/PhysRevLett.88.091304}{\emph{Phys.Rev.Lett.}
  {\bfseries 88} (2002) 091304}
  [\href{https://arxiv.org/abs/hep-ph/0106249}{{\ttfamily hep-ph/0106249}}].

\bibitem{Hall:2009bx}
L.J.~Hall, K.~Jedamzik, J.~March-Russell and S.M.~West, \emph{{Freeze-In
  Production of FIMP Dark Matter}},
  \href{https://doi.org/10.1007/JHEP03(2010)080}{\emph{JHEP} {\bfseries 03}
  (2010) 080} [\href{https://arxiv.org/abs/0911.1120}{{\ttfamily 0911.1120}}].

\bibitem{Mambrini:2013iaa}
Y.~Mambrini, K.A.~Olive, J.~Quevillon and B.~Zaldivar, \emph{{Gauge Coupling
  Unification and Nonequilibrium Thermal Dark Matter}},
  \href{https://doi.org/10.1103/PhysRevLett.110.241306}{\emph{Phys. Rev. Lett.}
  {\bfseries 110} (2013) 241306}
  [\href{https://arxiv.org/abs/1302.4438}{{\ttfamily 1302.4438}}].

\bibitem{Bernal:2017kxu}
N.~Bernal, M.~Heikinheimo, T.~Tenkanen, K.~Tuominen and V.~Vaskonen, \emph{{The
  Dawn of FIMP Dark Matter: A Review of Models and Constraints}},
  \href{https://doi.org/10.1142/S0217751X1730023X}{\emph{Int. J. Mod. Phys.}
  {\bfseries A32} (2017) 1730023}
  [\href{https://arxiv.org/abs/1706.07442}{{\ttfamily 1706.07442}}].

\bibitem{Elahi:2014fsa}
F.~Elahi, C.~Kolda and J.~Unwin, \emph{{UltraViolet Freeze-in}},
  \href{https://doi.org/10.1007/JHEP03(2015)048}{\emph{JHEP} {\bfseries 03}
  (2015) 048} [\href{https://arxiv.org/abs/1410.6157}{{\ttfamily 1410.6157}}].

\bibitem{Mambrini:2015vna}
Y.~Mambrini, N.~Nagata, K.A.~Olive, J.~Quevillon and J.~Zheng, \emph{{Dark
  matter and gauge coupling unification in nonsupersymmetric SO(10) grand
  unified models}},
  \href{https://doi.org/10.1103/PhysRevD.91.095010}{\emph{Phys. Rev. D}
  {\bfseries 91} (2015) 095010}
  [\href{https://arxiv.org/abs/1502.06929}{{\ttfamily 1502.06929}}].

\bibitem{Bhattacharyya:2018evo}
G.~Bhattacharyya, M.~Dutra, Y.~Mambrini and M.~Pierre, \emph{{Freezing-in dark
  matter through a heavy invisible Z'}},
  \href{https://doi.org/10.1103/PhysRevD.98.035038}{\emph{Phys. Rev. D}
  {\bfseries 98} (2018) 035038}
  [\href{https://arxiv.org/abs/1806.00016}{{\ttfamily 1806.00016}}].

\bibitem{Chowdhury:2018tzw}
D.~Chowdhury, E.~Dudas, M.~Dutra and Y.~Mambrini, \emph{{Moduli Portal Dark
  Matter}}, \href{https://doi.org/10.1103/PhysRevD.99.095028}{\emph{Phys. Rev.
  D} {\bfseries 99} (2019) 095028}
  [\href{https://arxiv.org/abs/1811.01947}{{\ttfamily 1811.01947}}].

\bibitem{Benakli:2017whb}
K.~Benakli, Y.~Chen, E.~Dudas and Y.~Mambrini, \emph{{Minimal model of
  gravitino dark matter}},
  \href{https://doi.org/10.1103/PhysRevD.95.095002}{\emph{Phys. Rev. D}
  {\bfseries 95} (2017) 095002}
  [\href{https://arxiv.org/abs/1701.06574}{{\ttfamily 1701.06574}}].

\bibitem{Dudas:2017rpa}
E.~Dudas, Y.~Mambrini and K.~Olive, \emph{{Case for an EeV Gravitino}},
  \href{https://doi.org/10.1103/PhysRevLett.119.051801}{\emph{Phys. Rev. Lett.}
  {\bfseries 119} (2017) 051801}
  [\href{https://arxiv.org/abs/1704.03008}{{\ttfamily 1704.03008}}].

\bibitem{Dudas:2017kfz}
E.~Dudas, T.~Gherghetta, Y.~Mambrini and K.A.~Olive, \emph{{Inflation and
  High-Scale Supersymmetry with an EeV Gravitino}},
  \href{https://doi.org/10.1103/PhysRevD.96.115032}{\emph{Phys. Rev. D}
  {\bfseries 96} (2017) 115032}
  [\href{https://arxiv.org/abs/1710.07341}{{\ttfamily 1710.07341}}].

\bibitem{Dudas:2018npp}
E.~Dudas, T.~Gherghetta, K.~Kaneta, Y.~Mambrini and K.A.~Olive,
  \emph{{Gravitino decay in high scale supersymmetry with R -parity
  violation}}, \href{https://doi.org/10.1103/PhysRevD.98.015030}{\emph{Phys.
  Rev. D} {\bfseries 98} (2018) 015030}
  [\href{https://arxiv.org/abs/1805.07342}{{\ttfamily 1805.07342}}].

\bibitem{Kaneta:2019yjn}
K.~Kaneta, Y.~Mambrini, K.A.~Olive and S.~Verner, \emph{{Inflation and
  Leptogenesis in High-Scale Supersymmetry}},
  \href{https://doi.org/10.1103/PhysRevD.101.015002}{\emph{Phys. Rev. D}
  {\bfseries 101} (2020) 015002}
  [\href{https://arxiv.org/abs/1911.02463}{{\ttfamily 1911.02463}}].

\bibitem{Bernal:2018qlk}
N.~Bernal, M.~Dutra, Y.~Mambrini, K.~Olive, M.~Peloso and M.~Pierre,
  \emph{{Spin-2 Portal Dark Matter}},
  \href{https://doi.org/10.1103/PhysRevD.97.115020}{\emph{Phys. Rev. D}
  {\bfseries 97} (2018) 115020}
  [\href{https://arxiv.org/abs/1803.01866}{{\ttfamily 1803.01866}}].

\bibitem{Kaneta:2019zgw}
K.~Kaneta, Y.~Mambrini and K.A.~Olive, \emph{{Radiative production of
  nonthermal dark matter}},
  \href{https://doi.org/10.1103/PhysRevD.99.063508}{\emph{Phys. Rev. D}
  {\bfseries 99} (2019) 063508}
  [\href{https://arxiv.org/abs/1901.04449}{{\ttfamily 1901.04449}}].

\bibitem{Heurtier:2019eou}
L.~Heurtier and F.~Huang, \emph{{Inflaton portal to a highly decoupled EeV dark
  matter particle}},
  \href{https://doi.org/10.1103/PhysRevD.100.043507}{\emph{Phys. Rev. D}
  {\bfseries 100} (2019) 043507}
  [\href{https://arxiv.org/abs/1905.05191}{{\ttfamily 1905.05191}}].

\bibitem{Mambrini:2022uol}
Y.~Mambrini, K.A.~Olive and J.~Zheng, \emph{{Post-Inflationary Dark Matter
  Bremsstrahlung}},  \href{https://arxiv.org/abs/2208.05859}{{\ttfamily
  2208.05859}}.

\bibitem{Chung:1998rq}
D.J.~Chung, E.W.~Kolb and A.~Riotto, \emph{{Production of massive particles
  during reheating}},
  \href{https://doi.org/10.1103/PhysRevD.60.063504}{\emph{Phys. Rev. D}
  {\bfseries 60} (1999) 063504}
  [\href{https://arxiv.org/abs/hep-ph/9809453}{{\ttfamily hep-ph/9809453}}].

\bibitem{Giudice:2000ex}
G.F.~Giudice, E.W.~Kolb and A.~Riotto, \emph{{Largest temperature of the
  radiation era and its cosmological implications}},
  \href{https://doi.org/10.1103/PhysRevD.64.023508}{\emph{Phys. Rev. D}
  {\bfseries 64} (2001) 023508}
  [\href{https://arxiv.org/abs/hep-ph/0005123}{{\ttfamily hep-ph/0005123}}].

\bibitem{Garcia:2017tuj}
M.A.G.~Garcia, Y.~Mambrini, K.A.~Olive and M.~Peloso, \emph{{Enhancement of the
  Dark Matter Abundance Before Reheating: Applications to Gravitino Dark
  Matter}}, \href{https://doi.org/10.1103/PhysRevD.96.103510}{\emph{Phys. Rev.
  D} {\bfseries 96} (2017) 103510}
  [\href{https://arxiv.org/abs/1709.01549}{{\ttfamily 1709.01549}}].

\bibitem{Garcia:2020eof}
M.A.G.~Garcia, K.~Kaneta, Y.~Mambrini and K.A.~Olive, \emph{{Reheating and
  Post-inflationary Production of Dark Matter}},
  \href{https://doi.org/10.1103/PhysRevD.101.123507}{\emph{Phys. Rev. D}
  {\bfseries 101} (2020) 123507}
  [\href{https://arxiv.org/abs/2004.08404}{{\ttfamily 2004.08404}}].

\bibitem{Garcia:2020wiy}
M.A.G.~Garcia, K.~Kaneta, Y.~Mambrini and K.A.~Olive, \emph{{Inflaton
  Oscillations and Post-Inflationary Reheating}},
  \href{https://doi.org/10.1088/1475-7516/2021/04/012}{\emph{JCAP} {\bfseries
  04} (2021) 012} [\href{https://arxiv.org/abs/2012.10756}{{\ttfamily
  2012.10756}}].

\bibitem{Barman:2022tzk}
B.~Barman, N.~Bernal, Y.~Xu and O.~Zapata, \emph{{Ultraviolet freeze-in with a
  time-dependent inflaton decay}},
  \href{https://doi.org/10.1088/1475-7516/2022/07/019}{\emph{JCAP} {\bfseries
  07} (2022) 019} [\href{https://arxiv.org/abs/2202.12906}{{\ttfamily
  2202.12906}}].

\bibitem{Ema:2015dka}
Y.~Ema, R.~Jinno, K.~Mukaida and K.~Nakayama, \emph{{Gravitational Effects on
  Inflaton Decay}},
  \href{https://doi.org/10.1088/1475-7516/2015/05/038}{\emph{JCAP} {\bfseries
  05} (2015) 038} [\href{https://arxiv.org/abs/1502.02475}{{\ttfamily
  1502.02475}}].

\bibitem{Garny:2015sjg}
M.~Garny, M.~Sandora and M.S.~Sloth, \emph{{Planckian Interacting Massive
  Particles as Dark Matter}},
  \href{https://doi.org/10.1103/PhysRevLett.116.101302}{\emph{Phys. Rev. Lett.}
  {\bfseries 116} (2016) 101302}
  [\href{https://arxiv.org/abs/1511.03278}{{\ttfamily 1511.03278}}].

\bibitem{Garny:2017kha}
M.~Garny, A.~Palessandro, M.~Sandora and M.S.~Sloth, \emph{{Theory and
  Phenomenology of Planckian Interacting Massive Particles as Dark Matter}},
  \href{https://doi.org/10.1088/1475-7516/2018/02/027}{\emph{JCAP} {\bfseries
  02} (2018) 027} [\href{https://arxiv.org/abs/1709.09688}{{\ttfamily
  1709.09688}}].

\bibitem{Tang:2016vch}
Y.~Tang and Y.-L.~Wu, \emph{{Pure Gravitational Dark Matter, Its Mass and
  Signatures}},
  \href{https://doi.org/10.1016/j.physletb.2016.05.045}{\emph{Phys. Lett. B}
  {\bfseries 758} (2016) 402}
  [\href{https://arxiv.org/abs/1604.04701}{{\ttfamily 1604.04701}}].

\bibitem{Tang:2017hvq}
Y.~Tang and Y.-L.~Wu, \emph{{On Thermal Gravitational Contribution to Particle
  Production and Dark Matter}},
  \href{https://doi.org/10.1016/j.physletb.2017.10.034}{\emph{Phys. Lett. B}
  {\bfseries 774} (2017) 676}
  [\href{https://arxiv.org/abs/1708.05138}{{\ttfamily 1708.05138}}].

\bibitem{Ema:2016hlw}
Y.~Ema, R.~Jinno, K.~Mukaida and K.~Nakayama, \emph{{Gravitational particle
  production in oscillating backgrounds and its cosmological implications}},
  \href{https://doi.org/10.1103/PhysRevD.94.063517}{\emph{Phys. Rev. D}
  {\bfseries 94} (2016) 063517}
  [\href{https://arxiv.org/abs/1604.08898}{{\ttfamily 1604.08898}}].

\bibitem{Ema:2018ucl}
Y.~Ema, K.~Nakayama and Y.~Tang, \emph{{Production of Purely Gravitational Dark
  Matter}}, \href{https://doi.org/10.1007/JHEP09(2018)135}{\emph{JHEP}
  {\bfseries 09} (2018) 135}
  [\href{https://arxiv.org/abs/1804.07471}{{\ttfamily 1804.07471}}].

\bibitem{Ema:2019yrd}
Y.~Ema, K.~Nakayama and Y.~Tang, \emph{{Production of purely gravitational dark
  matter: the case of fermion and vector boson}},
  \href{https://doi.org/10.1007/JHEP07(2019)060}{\emph{JHEP} {\bfseries 07}
  (2019) 060} [\href{https://arxiv.org/abs/1903.10973}{{\ttfamily
  1903.10973}}].

\bibitem{Chianese:2020yjo}
M.~Chianese, B.~Fu and S.F.~King, \emph{{Impact of Higgs portal on
  gravity-mediated production of superheavy dark matter}},
  \href{https://doi.org/10.1088/1475-7516/2020/06/019}{\emph{JCAP} {\bfseries
  06} (2020) 019} [\href{https://arxiv.org/abs/2003.07366}{{\ttfamily
  2003.07366}}].

\bibitem{Chianese:2020khl}
M.~Chianese, B.~Fu and S.F.~King, \emph{{Interplay between neutrino and gravity
  portals for FIMP dark matter}},
  \href{https://doi.org/10.1088/1475-7516/2021/01/034}{\emph{JCAP} {\bfseries
  01} (2021) 034} [\href{https://arxiv.org/abs/2009.01847}{{\ttfamily
  2009.01847}}].

\bibitem{Redi:2020ffc}
M.~Redi, A.~Tesi and H.~Tillim, \emph{{Gravitational Production of a Conformal
  Dark Sector}}, \href{https://doi.org/10.1007/JHEP05(2021)010}{\emph{JHEP}
  {\bfseries 05} (2021) 010}
  [\href{https://arxiv.org/abs/2011.10565}{{\ttfamily 2011.10565}}].

\bibitem{Mambrini:2021zpp}
Y.~Mambrini and K.A.~Olive, \emph{{Gravitational Production of Dark Matter
  during Reheating}},
  \href{https://doi.org/10.1103/PhysRevD.103.115009}{\emph{Phys. Rev. D}
  {\bfseries 103} (2021) 115009}
  [\href{https://arxiv.org/abs/2102.06214}{{\ttfamily 2102.06214}}].

\bibitem{Barman:2021ugy}
B.~Barman and N.~Bernal, \emph{{Gravitational SIMPs}},
  \href{https://doi.org/10.1088/1475-7516/2021/06/011}{\emph{JCAP} {\bfseries
  06} (2021) 011} [\href{https://arxiv.org/abs/2104.10699}{{\ttfamily
  2104.10699}}].

\bibitem{Haque:2021mab}
M.R.~Haque and D.~Maity, \emph{{Gravitational dark matter: Free streaming and
  phase space distribution}},
  \href{https://doi.org/10.1103/PhysRevD.106.023506}{\emph{Phys. Rev. D}
  {\bfseries 106} (2022) 023506}
  [\href{https://arxiv.org/abs/2112.14668}{{\ttfamily 2112.14668}}].

\bibitem{Clery:2021bwz}
S.~Clery, Y.~Mambrini, K.A.~Olive and S.~Verner, \emph{{Gravitational portals
  in the early Universe}},
  \href{https://doi.org/10.1103/PhysRevD.105.075005}{\emph{Phys. Rev. D}
  {\bfseries 105} (2022) 075005}
  [\href{https://arxiv.org/abs/2112.15214}{{\ttfamily 2112.15214}}].

\bibitem{Clery:2022wib}
S.~Clery, Y.~Mambrini, K.A.~Olive, A.~Shkerin and S.~Verner,
  \emph{{Gravitational portals with nonminimal couplings}},
  \href{https://doi.org/10.1103/PhysRevD.105.095042}{\emph{Phys. Rev. D}
  {\bfseries 105} (2022) 095042}
  [\href{https://arxiv.org/abs/2203.02004}{{\ttfamily 2203.02004}}].

\bibitem{Ahmed:2022qeh}
A.~Ahmed, B.~Grzadkowski and A.~Socha, \emph{{Higgs Boson-Induced Reheating and
  Dark Matter Production}},
  \href{https://doi.org/10.3390/sym14020306}{\emph{Symmetry} {\bfseries 14}
  (2022) 306}.

\bibitem{Ahmed:2022tfm}
A.~Ahmed, B.~Grzadkowski and A.~Socha, \emph{{Higgs boson induced reheating and
  ultraviolet frozen-in dark matter}},
  \href{https://arxiv.org/abs/2207.11218}{{\ttfamily 2207.11218}}.

\bibitem{Fukugita:1986hr}
M.~Fukugita and T.~Yanagida, \emph{{Baryogenesis Without Grand Unification}},
  \href{https://doi.org/10.1016/0370-2693(86)91126-3}{\emph{Phys. Lett.}
  {\bfseries B174} (1986) 45}.

\bibitem{Kuzmin:1985mm}
V.A.~Kuzmin, V.A.~Rubakov and M.E.~Shaposhnikov, \emph{{On the Anomalous
  Electroweak Baryon Number Nonconservation in the Early Universe}},
  \href{https://doi.org/10.1016/0370-2693(85)91028-7}{\emph{Phys. Lett.}
  {\bfseries 155B} (1985) 36}.

\bibitem{Buchmuller:2002rq}
W.~Buchmuller, P.~Di~Bari and M.~Plumacher, \emph{{Cosmic microwave background,
  matter - antimatter asymmetry and neutrino masses}},
  \href{https://doi.org/10.1016/S0550-3213(02)00737-X}{\emph{Nucl. Phys. B}
  {\bfseries 643} (2002) 367}
  [\href{https://arxiv.org/abs/hep-ph/0205349}{{\ttfamily hep-ph/0205349}}].

\bibitem{Buchmuller:2003gz}
W.~Buchmuller, P.~Di~Bari and M.~Plumacher, \emph{{The Neutrino mass window for
  baryogenesis}},
  \href{https://doi.org/10.1016/S0550-3213(03)00449-8}{\emph{Nucl. Phys. B}
  {\bfseries 665} (2003) 445}
  [\href{https://arxiv.org/abs/hep-ph/0302092}{{\ttfamily hep-ph/0302092}}].

\bibitem{Chankowski:2003rr}
P.H.~Chankowski and K.~Turzynski, \emph{{Limits on T(reh) for thermal
  leptogenesis with hierarchical neutrino masses}},
  \href{https://doi.org/10.1016/j.physletb.2003.08.004}{\emph{Phys. Lett. B}
  {\bfseries 570} (2003) 198}
  [\href{https://arxiv.org/abs/hep-ph/0306059}{{\ttfamily hep-ph/0306059}}].

\bibitem{Giudice:2003jh}
G.~Giudice, A.~Notari, M.~Raidal, A.~Riotto and A.~Strumia, \emph{{Towards a
  complete theory of thermal leptogenesis in the SM and MSSM}},
  \href{https://doi.org/10.1016/j.nuclphysb.2004.02.019}{\emph{Nucl. Phys. B}
  {\bfseries 685} (2004) 89}
  [\href{https://arxiv.org/abs/hep-ph/0310123}{{\ttfamily hep-ph/0310123}}].

\bibitem{Davidson:2002qv}
S.~Davidson and A.~Ibarra, \emph{{A Lower bound on the right-handed neutrino
  mass from leptogenesis}},
  \href{https://doi.org/10.1016/S0370-2693(02)01735-5}{\emph{Phys. Lett.}
  {\bfseries B535} (2002) 25}
  [\href{https://arxiv.org/abs/hep-ph/0202239}{{\ttfamily hep-ph/0202239}}].

\bibitem{Giudice:1999fb}
G.~Giudice, M.~Peloso, A.~Riotto and I.~Tkachev, \emph{{Production of massive
  fermions at preheating and leptogenesis}},
  \href{https://doi.org/10.1088/1126-6708/1999/08/014}{\emph{JHEP} {\bfseries
  08} (1999) 014} [\href{https://arxiv.org/abs/hep-ph/9905242}{{\ttfamily
  hep-ph/9905242}}].

\bibitem{Asaka:1999yd}
T.~Asaka, K.~Hamaguchi, M.~Kawasaki and T.~Yanagida, \emph{{Leptogenesis in
  inflaton decay}},
  \href{https://doi.org/10.1016/S0370-2693(99)01020-5}{\emph{Phys. Lett. B}
  {\bfseries 464} (1999) 12}
  [\href{https://arxiv.org/abs/hep-ph/9906366}{{\ttfamily hep-ph/9906366}}].

\bibitem{Lazarides:1990huy}
G.~Lazarides and Q.~Shafi, \emph{{Origin of matter in the inflationary
  cosmology}}, \href{https://doi.org/10.1016/0370-2693(91)91090-I}{\emph{Phys.
  Lett. B} {\bfseries 258} (1991) 305}.

\bibitem{Campbell:1992hd}
B.A.~Campbell, S.~Davidson and K.A.~Olive, \emph{{Inflation, neutrino
  baryogenesis, and (S)neutrino induced baryogenesis}},
  \href{https://doi.org/10.1016/0550-3213(93)90619-Z}{\emph{Nucl. Phys. B}
  {\bfseries 399} (1993) 111}
  [\href{https://arxiv.org/abs/hep-ph/9302223}{{\ttfamily hep-ph/9302223}}].

\bibitem{Hahn-Woernle:2008tsk}
F.~Hahn-Woernle and M.~Plumacher, \emph{{Effects of reheating on
  leptogenesis}},
  \href{https://doi.org/10.1016/j.nuclphysb.2008.07.032}{\emph{Nucl. Phys. B}
  {\bfseries 806} (2009) 68} [\href{https://arxiv.org/abs/0801.3972}{{\ttfamily
  0801.3972}}].

\bibitem{Co:2022bgh}
R.T.~Co, Y.~Mambrini and K.A.~Olive, \emph{{Inflationary Gravitational
  Leptogenesis}},  \href{https://arxiv.org/abs/2205.01689}{{\ttfamily
  2205.01689}}.

\bibitem{Bernal:2021kaj}
N.~Bernal and C.S.~Fong, \emph{{Dark matter and leptogenesis from gravitational
  production}},
  \href{https://doi.org/10.1088/1475-7516/2021/06/028}{\emph{JCAP} {\bfseries
  06} (2021) 028} [\href{https://arxiv.org/abs/2103.06896}{{\ttfamily
  2103.06896}}].

\bibitem{Haque:2022kez}
M.R.~Haque and D.~Maity, \emph{{Gravitational Reheating}},
  \href{https://arxiv.org/abs/2201.02348}{{\ttfamily 2201.02348}}.

\bibitem{Figueroa:2018twl}
D.G.~Figueroa and E.H.~Tanin, \emph{{Inconsistency of an inflationary sector
  coupled only to Einstein gravity}},
  \href{https://doi.org/10.1088/1475-7516/2019/10/050}{\emph{JCAP} {\bfseries
  10} (2019) 050} [\href{https://arxiv.org/abs/1811.04093}{{\ttfamily
  1811.04093}}].

\bibitem{Kallosh:2013hoa}
R.~Kallosh and A.~Linde, \emph{{Universality Class in Conformal Inflation}},
  \href{https://doi.org/10.1088/1475-7516/2013/07/002}{\emph{JCAP} {\bfseries
  07} (2013) 002} [\href{https://arxiv.org/abs/1306.5220}{{\ttfamily
  1306.5220}}].

\bibitem{Choi:1994ax}
S.Y.~Choi, J.S.~Shim and H.S.~Song, \emph{{Factorization and polarization in
  linearized gravity}},
  \href{https://doi.org/10.1103/PhysRevD.51.2751}{\emph{Phys. Rev. D}
  {\bfseries 51} (1995) 2751}
  [\href{https://arxiv.org/abs/hep-th/9411092}{{\ttfamily hep-th/9411092}}].

\bibitem{Holstein:2006bh}
B.R.~Holstein, \emph{{Graviton Physics}},
  \href{https://doi.org/10.1119/1.2338547}{\emph{Am. J. Phys.} {\bfseries 74}
  (2006) 1002} [\href{https://arxiv.org/abs/gr-qc/0607045}{{\ttfamily
  gr-qc/0607045}}].

\bibitem{Ellis:2015pla}
J.~Ellis, M.A.G.~Garcia, D.V.~Nanopoulos and K.A.~Olive, \emph{{Calculations of
  Inflaton Decays and Reheating: with Applications to No-Scale Inflation
  Models}}, \href{https://doi.org/10.1088/1475-7516/2015/07/050}{\emph{JCAP}
  {\bfseries 07} (2015) 050}
  [\href{https://arxiv.org/abs/1505.06986}{{\ttfamily 1505.06986}}].

\bibitem{Planck:2018jri}
{\scshape Planck} collaboration, \emph{{Planck 2018 results. X. Constraints on
  inflation}}, \href{https://doi.org/10.1051/0004-6361/201833887}{\emph{Astron.
  Astrophys.} {\bfseries 641} (2020) A10}
  [\href{https://arxiv.org/abs/1807.06211}{{\ttfamily 1807.06211}}].

\bibitem{egnov}
J.~Ellis, M.A.G.~Garcia, D.V.~Nanopoulos, K.A.~Olive and S.~Verner,
  \emph{{BICEP/Keck constraints on attractor models of inflation and
  reheating}}, \href{https://doi.org/10.1103/PhysRevD.105.043504}{\emph{Phys.
  Rev. D} {\bfseries 105} (2022) 043504}
  [\href{https://arxiv.org/abs/2112.04466}{{\ttfamily 2112.04466}}].

\bibitem{MINKOWSKI1977421}
P.~Minkowski, \emph{$\mu \to e \gamma$ at a rate of one out of 109 muon
  decays?},
  \href{https://doi.org/https://doi.org/10.1016/0370-2693(77)90435-X}{\emph{Physics
  Letters B} {\bfseries 67} (1977) 421}.

\bibitem{GellMann:1980vs}
M.~Gell-Mann, P.~Ramond and R.~Slansky, \emph{{Complex Spinors and Unified
  Theories}}, {\emph{Conf. Proc.} {\bfseries C790927} (1979) 315}
  [\href{https://arxiv.org/abs/1306.4669}{{\ttfamily 1306.4669}}].

\bibitem{Yanagida:1979as}
T.~Yanagida, \emph{{HORIZONTAL SYMMETRY AND MASSES OF NEUTRINOS}}, {\emph{Conf.
  Proc.} {\bfseries C7902131} (1979) 95}.

\bibitem{Mohapatra:1979ia}
R.N.~Mohapatra and G.~Senjanovic, \emph{{Neutrino Mass and Spontaneous Parity
  Violation}}, \href{https://doi.org/10.1103/PhysRevLett.44.912}{\emph{Phys.
  Rev. Lett.} {\bfseries 44} (1980) 912}.

\bibitem{Schechter:1980gr}
J.~Schechter and J.W.F.~Valle, \emph{{Neutrino Masses in SU(2) x U(1)
  Theories}}, \href{https://doi.org/10.1103/PhysRevD.22.2227}{\emph{Phys. Rev.}
  {\bfseries D22} (1980) 2227}.

\bibitem{Schechter:1981cv}
J.~Schechter and J.W.F.~Valle, \emph{{Neutrino Decay and Spontaneous Violation
  of Lepton Number}},
  \href{https://doi.org/10.1103/PhysRevD.25.774}{\emph{Phys. Rev. D} {\bfseries
  25} (1982) 774}.

\bibitem{Frampton:2002qc}
P.H.~Frampton, S.L.~Glashow and T.~Yanagida, \emph{{Cosmological sign of
  neutrino CP violation}},
  \href{https://doi.org/10.1016/S0370-2693(02)02853-8}{\emph{Phys. Lett. B}
  {\bfseries 548} (2002) 119}
  [\href{https://arxiv.org/abs/hep-ph/0208157}{{\ttfamily hep-ph/0208157}}].

\bibitem{Raidal:2002xf}
M.~Raidal and A.~Strumia, \emph{{Predictions of the most minimal seesaw
  model}}, \href{https://doi.org/10.1016/S0370-2693(02)03124-6}{\emph{Phys.
  Lett. B} {\bfseries 553} (2003) 72}
  [\href{https://arxiv.org/abs/hep-ph/0210021}{{\ttfamily hep-ph/0210021}}].

\bibitem{Ibarra:2003up}
A.~Ibarra and G.G.~Ross, \emph{{Neutrino phenomenology: The Case of two
  right-handed neutrinos}},
  \href{https://doi.org/10.1016/j.physletb.2004.04.037}{\emph{Phys. Lett. B}
  {\bfseries 591} (2004) 285}
  [\href{https://arxiv.org/abs/hep-ph/0312138}{{\ttfamily hep-ph/0312138}}].

\bibitem{Rink:2017zrf}
T.~Rink, \emph{{Leptonic CP violation in the minimal type-I seesaw model:
  Bottom-up phenomenology $\&$ top-down model building}},  Master's thesis,
  Heidelberg U., 2017.

\bibitem{Ichikawa:2008ne}
K.~Ichikawa, T.~Suyama, T.~Takahashi and M.~Yamaguchi, \emph{{Primordial
  Curvature Fluctuation and Its Non-Gaussianity in Models with Modulated
  Reheating}}, \href{https://doi.org/10.1103/PhysRevD.78.063545}{\emph{Phys.
  Rev. D} {\bfseries 78} (2008) 063545}
  [\href{https://arxiv.org/abs/0807.3988}{{\ttfamily 0807.3988}}].

\bibitem{Kainulainen:2016vzv}
K.~Kainulainen, S.~Nurmi, T.~Tenkanen, K.~Tuominen and V.~Vaskonen,
  \emph{{Isocurvature Constraints on Portal Couplings}},
  \href{https://doi.org/10.1088/1475-7516/2016/06/022}{\emph{JCAP} {\bfseries
  06} (2016) 022} [\href{https://arxiv.org/abs/1601.07733}{{\ttfamily
  1601.07733}}].

\bibitem{book}
Y.~Mambrini, \emph{{Particles in the dark Universe}}, {\emph{Springer Ed., ISBN
  978-3-030-78139-2} (2021) }.

\bibitem{Lozanov:2016hid}
K.D.~Lozanov and M.A.~Amin, \emph{{Equation of State and Duration to Radiation
  Domination after Inflation}},
  \href{https://doi.org/10.1103/PhysRevLett.119.061301}{\emph{Phys. Rev. Lett.}
  {\bfseries 119} (2017) 061301}
  [\href{https://arxiv.org/abs/1608.01213}{{\ttfamily 1608.01213}}].

\bibitem{Lozanov:2017hjm}
K.D.~Lozanov and M.A.~Amin, \emph{{Self-resonance after inflation: oscillons,
  transients and radiation domination}},
  \href{https://doi.org/10.1103/PhysRevD.97.023533}{\emph{Phys. Rev. D}
  {\bfseries 97} (2018) 023533}
  [\href{https://arxiv.org/abs/1710.06851}{{\ttfamily 1710.06851}}].

\bibitem{Maity:2018qhi}
D.~Maity and P.~Saha, \emph{{(P)reheating after minimal Plateau Inflation and
  constraints from CMB}},
  \href{https://doi.org/10.1088/1475-7516/2019/07/018}{\emph{JCAP} {\bfseries
  07} (2019) 018} [\href{https://arxiv.org/abs/1811.11173}{{\ttfamily
  1811.11173}}].

\bibitem{Luty:1992un}
M.A.~Luty, \emph{{Baryogenesis via leptogenesis}},
  \href{https://doi.org/10.1103/PhysRevD.45.455}{\emph{Phys. Rev. D} {\bfseries
  45} (1992) 455}.

\bibitem{Flanz:1994yx}
M.~Flanz, E.A.~Paschos and U.~Sarkar, \emph{{Baryogenesis from a lepton
  asymmetric universe}},
  \href{https://doi.org/10.1016/0370-2693(94)01555-Q}{\emph{Phys. Lett. B}
  {\bfseries 345} (1995) 248}
  [\href{https://arxiv.org/abs/hep-ph/9411366}{{\ttfamily hep-ph/9411366}}].

\bibitem{Covi:1996wh}
L.~Covi, E.~Roulet and F.~Vissani, \emph{{CP violating decays in leptogenesis
  scenarios}}, \href{https://doi.org/10.1016/0370-2693(96)00817-9}{\emph{Phys.
  Lett. B} {\bfseries 384} (1996) 169}
  [\href{https://arxiv.org/abs/hep-ph/9605319}{{\ttfamily hep-ph/9605319}}].

\bibitem{Buchmuller:2004nz}
W.~Buchmuller, P.~Di~Bari and M.~Plumacher, \emph{{Leptogenesis for
  pedestrians}}, \href{https://doi.org/10.1016/j.aop.2004.02.003}{\emph{Annals
  Phys.} {\bfseries 315} (2005) 305}
  [\href{https://arxiv.org/abs/hep-ph/0401240}{{\ttfamily hep-ph/0401240}}].

\bibitem{Davidson:2008bu}
S.~Davidson, E.~Nardi and Y.~Nir, \emph{{Leptogenesis}},
  \href{https://doi.org/10.1016/j.physrep.2008.06.002}{\emph{Phys. Rept.}
  {\bfseries 466} (2008) 105}
  [\href{https://arxiv.org/abs/0802.2962}{{\ttfamily 0802.2962}}].

\bibitem{Asaka:2002zu}
T.~Asaka, H.~Nielsen and Y.~Takanishi, \emph{{Nonthermal leptogenesis from the
  heavier Majorana neutrinos}},
  \href{https://doi.org/10.1016/S0550-3213(02)00934-3}{\emph{Nucl. Phys. B}
  {\bfseries 647} (2002) 252}
  [\href{https://arxiv.org/abs/hep-ph/0207023}{{\ttfamily hep-ph/0207023}}].

\bibitem{Senoguz:2003hc}
V.N.~Senoguz and Q.~Shafi, \emph{{GUT scale inflation, nonthermal leptogenesis,
  and atmospheric neutrino oscillations}},
  \href{https://doi.org/10.1016/j.physletb.2003.12.020}{\emph{Phys. Lett. B}
  {\bfseries 582} (2004) 6}
  [\href{https://arxiv.org/abs/hep-ph/0309134}{{\ttfamily hep-ph/0309134}}].

\bibitem{Barman:2021tgt}
B.~Barman, D.~Borah and R.~Roshan, \emph{{Nonthermal leptogenesis and UV
  freeze-in of dark matter: Impact of inflationary reheating}},
  \href{https://doi.org/10.1103/PhysRevD.104.035022}{\emph{Phys. Rev. D}
  {\bfseries 104} (2021) 035022}
  [\href{https://arxiv.org/abs/2103.01675}{{\ttfamily 2103.01675}}].

\bibitem{Yeh:2022heq}
T.-H.~Yeh, J.~Shelton, K.A.~Olive and B.D.~Fields, \emph{{Probing Physics
  Beyond the Standard Model: Limits from BBN and the CMB Independently and
  Combined}},  \href{https://arxiv.org/abs/2207.13133}{{\ttfamily 2207.13133}}.

\bibitem{Co:2021lkc}
R.T.~Co, D.~Dunsky, N.~Fernandez, A.~Ghalsasi, L.J.~Hall, K.~Harigaya et~al.,
  \emph{{Gravitational wave and CMB probes of axion kination}},
  \href{https://doi.org/10.1007/JHEP09(2022)116}{\emph{JHEP} {\bfseries 09}
  (2022) 116} [\href{https://arxiv.org/abs/2108.09299}{{\ttfamily
  2108.09299}}].

\bibitem{Saikawa:2018rcs}
K.~Saikawa and S.~Shirai, \emph{{Primordial gravitational waves, precisely: The
  role of thermodynamics in the Standard Model}},
  \href{https://doi.org/10.1088/1475-7516/2018/05/035}{\emph{JCAP} {\bfseries
  05} (2018) 035} [\href{https://arxiv.org/abs/1803.01038}{{\ttfamily
  1803.01038}}].

\bibitem{Opferkuch:2019zbd}
T.~Opferkuch, P.~Schwaller and B.A.~Stefanek, \emph{{Ricci Reheating}},
  \href{https://doi.org/10.1088/1475-7516/2019/07/016}{\emph{JCAP} {\bfseries
  07} (2019) 016} [\href{https://arxiv.org/abs/1905.06823}{{\ttfamily
  1905.06823}}].

\bibitem{Giovannini:1998bp}
M.~Giovannini, \emph{{Gravitational waves constraints on postinflationary
  phases stiffer than radiation}},
  \href{https://doi.org/10.1103/PhysRevD.58.083504}{\emph{Phys. Rev. D}
  {\bfseries 58} (1998) 083504}
  [\href{https://arxiv.org/abs/hep-ph/9806329}{{\ttfamily hep-ph/9806329}}].

\bibitem{Crowder:2005nr}
J.~Crowder and N.J.~Cornish, \emph{{Beyond LISA: Exploring future gravitational
  wave missions}},
  \href{https://doi.org/10.1103/PhysRevD.72.083005}{\emph{Phys. Rev. D}
  {\bfseries 72} (2005) 083005}
  [\href{https://arxiv.org/abs/gr-qc/0506015}{{\ttfamily gr-qc/0506015}}].

\bibitem{Corbin:2005ny}
V.~Corbin and N.J.~Cornish, \emph{{Detecting the cosmic gravitational wave
  background with the big bang observer}},
  \href{https://doi.org/10.1088/0264-9381/23/7/014}{\emph{Class. Quant. Grav.}
  {\bfseries 23} (2006) 2435}
  [\href{https://arxiv.org/abs/gr-qc/0512039}{{\ttfamily gr-qc/0512039}}].

\bibitem{Harry:2006fi}
G.M.~Harry, P.~Fritschel, D.A.~Shaddock, W.~Folkner and E.S.~Phinney,
  \emph{{Laser interferometry for the big bang observer}},
  \href{https://doi.org/10.1088/0264-9381/23/15/008}{\emph{Class. Quant. Grav.}
  {\bfseries 23} (2006) 4887}.

\bibitem{Seto:2001qf}
N.~Seto, S.~Kawamura and T.~Nakamura, \emph{{Possibility of direct measurement
  of the acceleration of the universe using 0.1-Hz band laser interferometer
  gravitational wave antenna in space}},
  \href{https://doi.org/10.1103/PhysRevLett.87.221103}{\emph{Phys. Rev. Lett.}
  {\bfseries 87} (2001) 221103}
  [\href{https://arxiv.org/abs/astro-ph/0108011}{{\ttfamily
  astro-ph/0108011}}].

\bibitem{Kawamura:2006up}
S.~Kawamura et~al., \emph{{The Japanese space gravitational wave antenna
  DECIGO}}, \href{https://doi.org/10.1088/0264-9381/23/8/S17}{\emph{Class.
  Quant. Grav.} {\bfseries 23} (2006) S125}.

\bibitem{Yagi:2011wg}
K.~Yagi and N.~Seto, \emph{{Detector configuration of DECIGO/BBO and
  identification of cosmological neutron-star binaries}},
  \href{https://doi.org/10.1103/PhysRevD.83.044011}{\emph{Phys. Rev. D}
  {\bfseries 83} (2011) 044011}
  [\href{https://arxiv.org/abs/1101.3940}{{\ttfamily 1101.3940}}].

\bibitem{LIGOScientific:2016wof}
{\scshape LIGO Scientific} collaboration, \emph{{Exploring the Sensitivity of
  Next Generation Gravitational Wave Detectors}},
  \href{https://doi.org/10.1088/1361-6382/aa51f4}{\emph{Class. Quant. Grav.}
  {\bfseries 34} (2017) 044001}
  [\href{https://arxiv.org/abs/1607.08697}{{\ttfamily 1607.08697}}].

\bibitem{Reitze:2019iox}
D.~Reitze et~al., \emph{{Cosmic Explorer: The U.S. Contribution to
  Gravitational-Wave Astronomy beyond LIGO}}, {\emph{Bull. Am. Astron. Soc.}
  {\bfseries 51} (2019) 035}
  [\href{https://arxiv.org/abs/1907.04833}{{\ttfamily 1907.04833}}].

\bibitem{Punturo:2010zz}
M.~Punturo et~al., \emph{{The Einstein Telescope: A third-generation
  gravitational wave observatory}},
  \href{https://doi.org/10.1088/0264-9381/27/19/194002}{\emph{Class. Quant.
  Grav.} {\bfseries 27} (2010) 194002}.

\bibitem{Hild:2010id}
S.~Hild et~al., \emph{{Sensitivity Studies for Third-Generation Gravitational
  Wave Observatories}},
  \href{https://doi.org/10.1088/0264-9381/28/9/094013}{\emph{Class. Quant.
  Grav.} {\bfseries 28} (2011) 094013}
  [\href{https://arxiv.org/abs/1012.0908}{{\ttfamily 1012.0908}}].

\bibitem{Sathyaprakash:2012jk}
B.~Sathyaprakash et~al., \emph{{Scientific Objectives of Einstein Telescope}},
  \href{https://doi.org/10.1088/0264-9381/29/12/124013}{\emph{Class. Quant.
  Grav.} {\bfseries 29} (2012) 124013}
  [\href{https://arxiv.org/abs/1206.0331}{{\ttfamily 1206.0331}}].

\bibitem{Maggiore:2019uih}
M.~Maggiore et~al., \emph{{Science Case for the Einstein Telescope}},
  \href{https://doi.org/10.1088/1475-7516/2020/03/050}{\emph{JCAP} {\bfseries
  03} (2020) 050} [\href{https://arxiv.org/abs/1912.02622}{{\ttfamily
  1912.02622}}].

\bibitem{Schmitz:2020syl}
K.~Schmitz, \emph{{New Sensitivity Curves for Gravitational-Wave Experiments}},
   \href{https://arxiv.org/abs/2002.04615}{{\ttfamily 2002.04615}}.

\bibitem{IceCube:2013cdw}
{\scshape IceCube} collaboration, \emph{{First observation of PeV-energy
  neutrinos with IceCube}},
  \href{https://doi.org/10.1103/PhysRevLett.111.021103}{\emph{Phys. Rev. Lett.}
  {\bfseries 111} (2013) 021103}
  [\href{https://arxiv.org/abs/1304.5356}{{\ttfamily 1304.5356}}].

\bibitem{IceCube:2013low}
{\scshape IceCube} collaboration, \emph{{Evidence for High-Energy
  Extraterrestrial Neutrinos at the IceCube Detector}},
  \href{https://doi.org/10.1126/science.1242856}{\emph{Science} {\bfseries 342}
  (2013) 1242856} [\href{https://arxiv.org/abs/1311.5238}{{\ttfamily
  1311.5238}}].

\bibitem{IceCube:2014stg}
{\scshape IceCube} collaboration, \emph{{Observation of High-Energy
  Astrophysical Neutrinos in Three Years of IceCube Data}},
  \href{https://doi.org/10.1103/PhysRevLett.113.101101}{\emph{Phys. Rev. Lett.}
  {\bfseries 113} (2014) 101101}
  [\href{https://arxiv.org/abs/1405.5303}{{\ttfamily 1405.5303}}].

\bibitem{Bai:2013nga}
Y.~Bai, R.~Lu and J.~Salvado, \emph{{Geometric Compatibility of IceCube TeV-PeV
  Neutrino Excess and its Galactic Dark Matter Origin}},
  \href{https://doi.org/10.1007/JHEP01(2016)161}{\emph{JHEP} {\bfseries 01}
  (2016) 161} [\href{https://arxiv.org/abs/1311.5864}{{\ttfamily 1311.5864}}].

\bibitem{Higaki:2014dwa}
T.~Higaki, R.~Kitano and R.~Sato, \emph{{Neutrinoful Universe}},
  \href{https://doi.org/10.1007/JHEP07(2014)044}{\emph{JHEP} {\bfseries 07}
  (2014) 044} [\href{https://arxiv.org/abs/1405.0013}{{\ttfamily 1405.0013}}].

\bibitem{Esmaili:2014rma}
A.~Esmaili, S.K.~Kang and P.D.~Serpico, \emph{{IceCube events and decaying dark
  matter: hints and constraints}},
  \href{https://doi.org/10.1088/1475-7516/2014/12/054}{\emph{JCAP} {\bfseries
  12} (2014) 054} [\href{https://arxiv.org/abs/1410.5979}{{\ttfamily
  1410.5979}}].

\bibitem{Bhattacharya:2014vwa}
A.~Bhattacharya, M.H.~Reno and I.~Sarcevic, \emph{{Reconciling neutrino flux
  from heavy dark matter decay and recent events at IceCube}},
  \href{https://doi.org/10.1007/JHEP06(2014)110}{\emph{JHEP} {\bfseries 06}
  (2014) 110} [\href{https://arxiv.org/abs/1403.1862}{{\ttfamily 1403.1862}}].

\bibitem{Zavala:2014dla}
J.~Zavala, \emph{{Galactic PeV neutrinos from dark matter annihilation}},
  \href{https://doi.org/10.1103/PhysRevD.89.123516}{\emph{Phys. Rev. D}
  {\bfseries 89} (2014) 123516}
  [\href{https://arxiv.org/abs/1404.2932}{{\ttfamily 1404.2932}}].

\bibitem{Rott:2014kfa}
C.~Rott, K.~Kohri and S.C.~Park, \emph{{Superheavy dark matter and IceCube
  neutrino signals: Bounds on decaying dark matter}},
  \href{https://doi.org/10.1103/PhysRevD.92.023529}{\emph{Phys. Rev. D}
  {\bfseries 92} (2015) 023529}
  [\href{https://arxiv.org/abs/1408.4575}{{\ttfamily 1408.4575}}].

\bibitem{Dudas:2014bca}
E.~Dudas, Y.~Mambrini and K.A.~Olive, \emph{{Monochromatic neutrinos generated
  by dark matter and the seesaw mechanism}},
  \href{https://doi.org/10.1103/PhysRevD.91.075001}{\emph{Phys. Rev. D}
  {\bfseries 91} (2015) 075001}
  [\href{https://arxiv.org/abs/1412.3459}{{\ttfamily 1412.3459}}].

\bibitem{Murase:2015gea}
K.~Murase, R.~Laha, S.~Ando and M.~Ahlers, \emph{{Testing the Dark Matter
  Scenario for PeV Neutrinos Observed in IceCube}},
  \href{https://doi.org/10.1103/PhysRevLett.115.071301}{\emph{Phys. Rev. Lett.}
  {\bfseries 115} (2015) 071301}
  [\href{https://arxiv.org/abs/1503.04663}{{\ttfamily 1503.04663}}].

\bibitem{Anchordoqui:2015lqa}
L.A.~Anchordoqui, V.~Barger, H.~Goldberg, X.~Huang, D.~Marfatia,
  L.H.M.~da~Silva et~al., \emph{{IceCube neutrinos, decaying dark matter, and
  the Hubble constant}},
  \href{https://doi.org/10.1103/PhysRevD.94.069901}{\emph{Phys. Rev. D}
  {\bfseries 92} (2015) 061301}
  [\href{https://arxiv.org/abs/1506.08788}{{\ttfamily 1506.08788}}].

\bibitem{Ko:2015nma}
P.~Ko and Y.~Tang, \emph{{IceCube Events from Heavy DM decays through the
  Right-handed Neutrino Portal}},
  \href{https://doi.org/10.1016/j.physletb.2015.10.021}{\emph{Phys. Lett. B}
  {\bfseries 751} (2015) 81}
  [\href{https://arxiv.org/abs/1508.02500}{{\ttfamily 1508.02500}}].

\bibitem{Roland:2015yoa}
S.B.~Roland, B.~Shakya and J.D.~Wells, \emph{{PeV neutrinos and a 3.5 keV x-ray
  line from a PeV-scale supersymmetric neutrino sector}},
  \href{https://doi.org/10.1103/PhysRevD.92.095018}{\emph{Phys. Rev. D}
  {\bfseries 92} (2015) 095018}
  [\href{https://arxiv.org/abs/1506.08195}{{\ttfamily 1506.08195}}].

\bibitem{Chianese:2016smc}
M.~Chianese and A.~Merle, \emph{{A Consistent Theory of Decaying Dark Matter
  Connecting IceCube to the Sesame Street}},
  \href{https://doi.org/10.1088/1475-7516/2017/04/017}{\emph{JCAP} {\bfseries
  04} (2017) 017} [\href{https://arxiv.org/abs/1607.05283}{{\ttfamily
  1607.05283}}].

\bibitem{ReFiorentin:2016rzn}
M.~Re~Fiorentin, V.~Niro and N.~Fornengo, \emph{{A consistent model for
  leptogenesis, dark matter and the IceCube signal}},
  \href{https://doi.org/10.1007/JHEP11(2016)022}{\emph{JHEP} {\bfseries 11}
  (2016) 022} [\href{https://arxiv.org/abs/1606.04445}{{\ttfamily
  1606.04445}}].

\bibitem{Garcia:2020hyo}
M.A.G.~Garcia, Y.~Mambrini, K.A.~Olive and S.~Verner, \emph{{Case for decaying
  spin- 3/2 dark matter}},
  \href{https://doi.org/10.1103/PhysRevD.102.083533}{\emph{Phys. Rev. D}
  {\bfseries 102} (2020) 083533}
  [\href{https://arxiv.org/abs/2006.03325}{{\ttfamily 2006.03325}}].

\bibitem{Arguelles:2022nbl}
C.A.~Arg\"uelles, D.~Delgado, A.~Friedlander, A.~Kheirandish, I.~Safa,
  A.C.~Vincent et~al., \emph{{Dark Matter decay to neutrinos}},
  \href{https://arxiv.org/abs/2210.01303}{{\ttfamily 2210.01303}}.

\bibitem{Jizba:2022bfz}
P.~Jizba and G.~Lambiase, \emph{{Tsallis cosmology and its applications in dark
  matter physics with focus on IceCube high-energy neutrino data}},
  \href{https://arxiv.org/abs/2206.12910}{{\ttfamily 2206.12910}}.

\bibitem{Ahmadvand:2021vxs}
M.~Ahmadvand, \emph{{Filtered asymmetric dark matter during the Peccei-Quinn
  phase transition}},
  \href{https://doi.org/10.1007/JHEP10(2021)109}{\emph{JHEP} {\bfseries 10}
  (2021) 109} [\href{https://arxiv.org/abs/2108.00958}{{\ttfamily
  2108.00958}}].

\bibitem{Chikashige:1980ui}
Y.~Chikashige, R.N.~Mohapatra and R.D.~Peccei, \emph{{Are There Real Goldstone
  Bosons Associated with Broken Lepton Number?}},
  \href{https://doi.org/10.1016/0370-2693(81)90011-3}{\emph{Phys. Lett. B}
  {\bfseries 98} (1981) 265}.

\bibitem{Berezinsky:1993fm}
V.~Berezinsky and J.W.F.~Valle, \emph{{The KeV majoron as a dark matter
  particle}}, \href{https://doi.org/10.1016/0370-2693(93)90140-D}{\emph{Phys.
  Lett. B} {\bfseries 318} (1993) 360}
  [\href{https://arxiv.org/abs/hep-ph/9309214}{{\ttfamily hep-ph/9309214}}].

\bibitem{Lattanzi:2007ux}
M.~Lattanzi and J.W.F.~Valle, \emph{{Decaying warm dark matter and neutrino
  masses}}, \href{https://doi.org/10.1103/PhysRevLett.99.121301}{\emph{Phys.
  Rev. Lett.} {\bfseries 99} (2007) 121301}
  [\href{https://arxiv.org/abs/0705.2406}{{\ttfamily 0705.2406}}].

\bibitem{Bazzocchi:2008fh}
F.~Bazzocchi, M.~Lattanzi, S.~Riemer-S\o{}rensen and J.W.F.~Valle, \emph{{X-ray
  photons from late-decaying majoron dark matter}},
  \href{https://doi.org/10.1088/1475-7516/2008/08/013}{\emph{JCAP} {\bfseries
  08} (2008) 013} [\href{https://arxiv.org/abs/0805.2372}{{\ttfamily
  0805.2372}}].

\bibitem{Lattanzi:2013uza}
M.~Lattanzi, S.~Riemer-Sorensen, M.~Tortola and J.W.F.~Valle, \emph{{Updated
  CMB and x- and $\gamma$-ray constraints on Majoron dark matter}},
  \href{https://doi.org/10.1103/PhysRevD.88.063528}{\emph{Phys. Rev. D}
  {\bfseries 88} (2013) 063528}
  [\href{https://arxiv.org/abs/1303.4685}{{\ttfamily 1303.4685}}].

\bibitem{Lattanzi:2014mia}
M.~Lattanzi, R.A.~Lineros and M.~Taoso, \emph{{Connecting neutrino physics with
  dark matter}},
  \href{https://doi.org/10.1088/1367-2630/16/12/125012}{\emph{New J. Phys.}
  {\bfseries 16} (2014) 125012}
  [\href{https://arxiv.org/abs/1406.0004}{{\ttfamily 1406.0004}}].

\bibitem{Gehrlein:2019iwl}
J.~Gehrlein and M.~Pierre, \emph{{A testable hidden-sector model for Dark
  Matter and neutrino masses}},
  \href{https://doi.org/10.1007/JHEP02(2020)068}{\emph{JHEP} {\bfseries 02}
  (2020) 068} [\href{https://arxiv.org/abs/1912.06661}{{\ttfamily
  1912.06661}}].

\bibitem{Dudas:2020sbq}
E.~Dudas, L.~Heurtier, Y.~Mambrini, K.A.~Olive and M.~Pierre, \emph{{Model of
  metastable EeV dark matter}},
  \href{https://doi.org/10.1103/PhysRevD.101.115029}{\emph{Phys. Rev. D}
  {\bfseries 101} (2020) 115029}
  [\href{https://arxiv.org/abs/2003.02846}{{\ttfamily 2003.02846}}].

\bibitem{Irsic:2017ixq}
V.~Ir\v{s}i\v{c} et~al., \emph{{New Constraints on the free-streaming of warm
  dark matter from intermediate and small scale Lyman-$\alpha$ forest data}},
  \href{https://doi.org/10.1103/PhysRevD.96.023522}{\emph{Phys. Rev. D}
  {\bfseries 96} (2017) 023522}
  [\href{https://arxiv.org/abs/1702.01764}{{\ttfamily 1702.01764}}].

\bibitem{Sitwell:2013fpa}
M.~Sitwell, A.~Mesinger, Y.-Z.~Ma and K.~Sigurdson, \emph{{The Imprint of Warm
  Dark Matter on the Cosmological 21-cm Signal}},
  \href{https://doi.org/10.1093/mnras/stt2392}{\emph{Mon. Not. Roy. Astron.
  Soc.} {\bfseries 438} (2014) 2664}
  [\href{https://arxiv.org/abs/1310.0029}{{\ttfamily 1310.0029}}].

\end{thebibliography}\endgroup
%%%%%%%%%%%%%%%%%%

\end{document}